\documentclass[twocolumn,longauth]{aa}  

\usepackage{graphicx,lscape}
\usepackage{txfonts}
\usepackage{microtype}
\begin{document}

   \title{The Herschel Planetary Nebula Survey 
   (HerPlaNS)\thanks{{\sl Herschel} is an ESA space observatory with
   science instruments 
   provided by European-led Principal Investigator consortia and with
   important participation from NASA.}}  

   \subtitle{I.\ Data Overview and Analysis Demonstration with \object{NGC\,6781}}
   \author{%
   T.\,Ueta\inst{\ref{du},\ref{isas},\thanks{JSPS FY2013 Long-Term Invitation Fellow}\hspace{1pt}}
   \and
   D.\,Ladjal\inst{\ref{du},\thanks{The IAU Gruber
   Foundation Fellow 2014 at the Gemini South Observatory}\hspace{1.5pt}}
   \and
   K.\,M.\,Exter\inst{\ref{kul}}  
   \and
   M.\,Otsuka\inst{\ref{asiaa}} 
   \and
   R.\,Szczerba\inst{\ref{ncac}}  
   \and
   N.\,Si\'{o}dmiak\inst{\ref{ncac}}  
   \and
   I.\,Aleman\inst{\ref{leiden}}
   \and
   P.\,A.\,M.\,van Hoof\inst{\ref{rob}}
   \and
   J.\,H.\,Kastner\inst{\ref{rit}}  
   \and
   R.\,Montez\inst{\ref{vanderbilt}} 
   \and
   I.\,McDonald\inst{\ref{jodrell}}
   \and
   M.\,Wittkowski\inst{\ref{eso}} 
   \and
   C.\,Sandin\inst{\ref{aip}} 
   \and
   S.\,Ramstedt\inst{\ref{uppsala}}
   \and
   O.\,De\,Marco\inst{\ref{mac}}
   \and
   E.\,Villaver\inst{\ref{uam}} 
   \and
   Y.-H.\,Chu\inst{\ref{uiuc}}  
   \and
   W.\,Vlemmings\inst{\ref{chalmers}} 
   \and
   H.\,Izumiura\inst{\ref{oao}} 
   \and
   R.\,Sahai\inst{\ref{jpl}} 
   \and
   J.\,A.\,Lopez\inst{\ref{unam}} 
   \and
   B.\,Balick\inst{\ref{uw}}  
   \and
   A.\,Zijlstra\inst{\ref{jodrell}}
   \and
   A.\,G.\,G.\,M.\,Tielens\inst{\ref{leiden}}  
   \and
   R.\,E.\,Rattray\inst{\ref{du}}
   \and
   E.\,Behar\inst{\ref{technion}} 
   \and
   E.\,G.\,Blackman\inst{\ref{rochester}}  
   \and
   K.\,Hebden\inst{\ref{jodrell}}  
   \and
   J.\,L.\,Hora\inst{\ref{cfa}} 
   \and
   K.\,Murakawa\inst{\ref{reed}} 
   \and
   J.\,Nordhaus\inst{\ref{rit2}}
   \and
   R.\,Nordon\inst{\ref{mpe}} 
   \and
   I.\,Yamamura\inst{\ref{isas}}}

   \institute{%
   Department of Physics and Astronomy, University of Denver, 
   2112 E.\ Wesley Ave., Denver, CO 80210, USA
   \email{tueta@du.edu}\label{du}
   \and
   Institute of Space and Astronautical Science, Japan Aerospace
   Exploration Agency, 
   3-1-1 Yoshinodai, Chuo-ku, Sagamihara, Kanagawa, 252-5210, Japan\label{isas} %2 
   \and
   Instituut voor Sterrenkunde, KU Leuven, Celestijnenlaan 200D, 3001, Leuven, Belgium\label{kul} % 3
   \and
   Academia Sinica, Institute of Astronomy and Astrophysics, Taiwan\label{asiaa} % 4
   \and
   N.\ Copernicus Astronomical Center, Rabia\'{n}ska 8, 87-100 Toru\'{n}, Poland\label{ncac} % 5
   \and
   Leiden Observatory, Leiden University, PO Box 9513, 2300 RA Leiden, The Netherlands\label{leiden} % 6
   \and
   Royal Observatory of Belgium, Ringlaan 3, 1180 Brussels, Belgium\label{rob} % 7
   \and
   Rochester Institute of Technology, 54 Lomb Memorial Dr., Rochester,
   NY 14623, USA\label{rit} % 8
   \and
   Department of Physics and Astronomy, Vanderbilt University,
   Nashville, TN 37235, USA\label{vanderbilt} % 9
   \and
   Jodrell Bank Centre for Astrophysics, Alan Turing Building,
   Manchester M13 9PL, UK\label{jodrell} % 10
   \and
   ESO, Karl-Schwarzschild-Str. 2, 85748 Garching bei M\"{u}nchen, Germany\label{eso}  % 11
   \and
   Leibniz-Institut f\"{u}r Astrophysik Potsdam (AIP), An der Sternwarte 16,
   D-144 82 Potsdam, Germany\label{aip} % 16
   \and
   Department of Physics and Astronomy, Division of Astronomy \& Space
   Physics, Uppsala University, Box 515, 751 20 Uppsala, Sweden\label{uppsala} % 12
   \and
   Department of Physics \& Astronomy, Macquarie University, Sydney, NSW 2109, Australia\label{mac} % 13
   \and
   Departamento de F\'{\i}sica Te\'{o}rica, Universidad Aut\'{o}noma de Madrid, Cantoblanco 28049 Madrid, Spain\label{uam} % 14
   \and
   Department of Astronomy, University of Illinois at Urbana-Champaign, Urbana, IL 61801, USA\label{uiuc}  % 15
   \and
   Chalmers University of Technology, Onsala Space Observatory, 439 92,
   Onsala, Sweden\label{chalmers} % 17
   \and
   Okayama Astrophysical Observatory (OAO), National Astronomical
   Observatory of Japan (NAOJ), 3037-5 Honjo, Kamogata, Asakuchi,
   Okayama, 719-0232, Japan\label{oao} % 18
   \and
   Jet Propulsion Laboratory, MS 183-900, California Institute of
+   Technology, Pasadena, CA 91109, USA\label{jpl} % 19
   \and
   Instituto de Astronom\'{\i}a, Universidad Nacional Aut\'{o}noma de M\'{e}xico,
   Campus Ensenada, C.P. 22800, Baja California, M\'{e}xico\label{unam} % 20
   \and
   Department of Astronomy, University of Washington, Seattle, WA
   98195-1580, USA\label{uw} % 21
   \and
   Physics Department, Technion, Haifa 32000, Israel\label{technion} % 22
   \and
   Department of Physics and Astronomy, University of Rochester,
   Rochester, NY 14618, USA\label{rochester} % 23
   \and
   Center for Astrophysics, 60 Garden St., MS 65, Cambridge, MA 02138, USA\label{cfa} % 24
   \and
   School of Physics and Astronomy, EC Stoner Building, University of Leeds, Leeds LS2 9JT, UK\label{reed} % 25
   \and
   Center for Computational Relativity and Gravitation, Rochester
   Institute of Technology, Rochester, NY 14623, USA\label{rit2} % 26
   \and
   Max-Planck-Institut f\"{u}r Extraterrestrische Physik (MPE), Postfach, 1312 85741, Garching, Germany\label{mpe} % 27
   }

   \date{Received January 10, 2014; accepted March 11, 2014}

% \abstract{}{}{}{}{} 
% 5 {} token are mandatory
 
  \abstract
   {This is the first of a series of investigations into far-IR
   characteristics of 11 planetary nebulae (PNs) under the 
   {\sl Herschel Space Observatory} Open Time\,1 program, Herschel
   Planetary Nebula Survey (HerPlaNS).}
   {Using the HerPlaNS data set, we look into the PN energetics
   and variations of the physical conditions within the target nebulae.
   In the present work, we provide an overview of the survey, data
   acquisition and processing, and resulting data products.}
   {We perform
   (1) PACS/SPIRE broadband imaging to determine the spatial
   distribution of the cold dust component in the target PNs and
   (2) PACS/SPIRE spectral-energy-distribution (SED)
   and line spectroscopy to determine the spatial
   distribution of the gas component in the target PNs.}
   {For the case of NGC\,6781, the broadband maps confirm the nearly 
   pole-on barrel structure of the amorphous carbon-richdust shell and
   the surrounding halo having temperatures of 26--40\,K. 
   The PACS/SPIRE multi-position spectra show spatial variations of
   far-IR lines that reflect the physical stratification of the nebula. 
   We demonstrate that spatially-resolved far-IR line diagnostics yield
   the ($T_{\rm e}$, $n_{\rm e}$) profiles, from which distributions 
   of ionized, atomic, and molecular gases 
   can be determined.
   Direct comparison of the dust and gas column mass maps constrained 
   by the HerPlaNS data allows to construct an empirical gas-to-dust
   mass ratio map, which shows a range of ratios with the median of
   $195\pm110$. 
   The present analysis yields estimates of
   the total mass of the shell to be 0.86\,M$_{\odot}$, consisting of 
   0.54\,M$_{\odot}$ of ionized gas,
   0.12\,M$_{\odot}$ of atomic gas,
   0.2\,M$_{\odot}$ of molecular gas, and
   $4\times10^{-3}$\,M$_{\odot}$ of dust grains.
   These estimates also suggest that the central star of about 
   1.5\,M$_{\odot}$ initial mass is terminating its PN evolution onto the
   white dwarf cooling track.}      
   {The HerPlaNS data provide various diagnostics for both the
   dust and gas components in a spatially-resolved manner.
   In the forthcoming papers of the HerPlaNS series we will explore the 
   HerPlaNS data set fully for the entire sample of 11 PNs.} 

   \maketitle
%
%________________________________________________________________

 \section{Introduction}

 The planetary nebula (PN) phase marks the last throes of stellar
 evolution for low to intermediate initial mass stars (of about 
 0.8--8\,M$_{\odot}$, \citealt{kwokbook}).   
 During this phase, the circumstellar envelope of gas and
 dust, which is created by mass loss in the preceding asymptotic
 giant branch (AGB) and post-AGB phases, undergoes a dramatic 
 transformation (i.e., ionization, photo-dissociation, and dynamical
 shaping) 
 caused by the fast wind and the intense radiation from the central star
 and by the less powerful but often significant interstellar radiation
 field coming from the surrounding interstellar space.
 As a consequence, a wide variety of underlying physical conditions are
 showcased within PNs, from fully ionized hot plasma to dusty cold
 atomic/molecular clouds, which exist (at least to first order) in a
 stratified manner around the central star. 
 Therefore, PNs provide excellent astrophysical laboratories to test
 theories of stellar evolution as well as theories of gas-dust dynamical
 processes in interacting stellar winds that can also interact with the
 surrounding interstellar medium (ISM).

 While PN investigations have been traditionally done through
 diagnostics of optical emission lines, PNs are bright sources at a wide
 range of wavelengths from the radio through the UV, and in some cases,
 even in the X-ray (e.g., 
 \citealt{pottasch84,z89,st01,corradi03,schoen05,sandin08,sahai11,chanplans,gdm13}). 
 Investigations using far-infrared (far-IR) radiation are especially
 critical to comprehend PNs as complex physical systems in their
 entirety, because a large fraction of the nebula mass may reside
 outside the central ionized region 
 (e.g., \citealt{villaver02}).
 For example, up to about 4\,M$_{\odot}$ of matter has been found in
 the far-IR halo of \object{NGC\,650} \citep{ueta06,vanhoof13}. 
 However, according to the recent mass budget estimates based on the
 UV to mid-IR photometric survey of the Magellanic Clouds, the amount of
 circumstellar dust grains has been severely underestimated: only about
 3\% of the ISM dust grains is accounted for in the warm component of
 the circumstellar envelopes \citep{matsuura09,boyer12}. 
 What this implies is that the most extended cold regions
 of the circumstellar envelope could contain this ``missing mass''
 component, which can only be detected in the wavelength ranges in the
 far-IR and longer. 

 Recent opportunities provided by the
 {\sl Spitzer Space Telescope} ({\sl Spitzer\/}; \citealt{werner04}),
 {\sl AKARI Infrared Astronomy Satellite} ({\sl AKARI\/};
 \citealt{murakami07}), 
 and {\sl Herschel Space Observatory} ({\sl Herschel\/};
 \citealt{pilbratt10})
 have made it possible to probe the very extended, coldest parts of PN
 haloes at the highest spatial resolutions in the far-IR to date
 (the beam size of several to a few tens of arcsec;
 e.g., \citealt{ueta06,su07,vanhoof10,vanhoof13,cox11}). 
 The new far-IR window has not only given access to the bulk of the
 matter in the farthest reaches of PNs, 
 but also permitted us to probe the interacting boundary regions between 
 the PN haloes and ISM, spawning new insights into the
 processing of the mass loss ejecta as they merge into the
 ISM (e.g., \citealt{wareing06,sabin10,zhang12}). 

 Among these recent far-IR opportunities, those provided by 
 {\sl Herschel} are unique:
 {\sl Herschel} allows  
 simultaneous probing of the multiple phases of the gaseous components
 in PNs via far-IR ionic, atomic, and molecular line emission.
 The {\sl Infrared Space Observatory\/} ({\sl ISO\/};
 \citealt{iso}) made detections of far-IR lines from about two dozen
 PNs \citep{liu01} and another two dozen PN progenitors and other
 evolved stars \citep{fong01,cc01}.
 However, the {\sl ISO\/} apertures typically covered most of the
 optically-bright regions of the target objects,\footnote{The aperture
 size of the {\sl ISO} LWS detector in the spatial dimension is about
 $106^{\prime\prime}$, while the beam size is about
 $40^{\prime\prime}$ radius \citep{isohb}.} and 
 therefore, the previous {\sl ISO} spectroscopic analyses were 
 usually performed in a spatially-integrated manner. 
 
 {\sl Herschel\/}'s spectral mapping capabilities allow us to look for
 variations of line/continuum strengths as a function of location in the
 target nebulae, so that the spatially resolved energetics of the
 circumstellar envelope can be unveiled.  
 Far-IR line maps would help to trace the spatial variations of the 
 electron density, electron temperature, and relative elemental
 abundance, which may suggest how much of which material was ejected at
 what time over the course of the progenitor star's mass loss history.
 Also revealed is how PNs are influenced by the passage of the
 ionization front. 
 While such line diagnostics have been routinely performed in the optical
 line diagnostics in the far-IR can offer an alternative perspective,
 because 
 (1) far-IR line ratios are relatively insensitive to the electron 
 temperature due to smaller excitation energies of fine-structure
 transitions in the far-IR, 
 and
 (2) far-IR line and continuum measurements are often  
 extinction-independent, permitting probes into dusty PNs.
 Hence, PN investigations in the far-IR with {\sl Herschel\/} should have a
 bearing on abundance determinations and elemental column densities, and
 therefore can 
 heavily impact analyses in other wavelength regimes.

 With the foregoing as motivation, we have conducted
 a comprehensive far-IR imaging and spectroscopic survey of PNs, dubbed
 the Herschel Planetary Nebula Survey (HerPlaNS), using nearly 200\,hrs
 of {\sl Herschel\/} time by taking advantage of its mapping capabilities 
 -- broadband and spectral imaging as well as spatio-spectroscopy -- at 
 spatial resolutions made possible by its 3.3\,m effective aperture
 diameter.
 Our chief objective is to examine both the dust and gas components of the
 target PNs simultaneously in the far-IR at high spatial resolutions and 
 investigate the energetics of the entire gas-dust system as a function
 of location in the nebula. 
 In this first installment of the forthcoming HerPlaNS series of papers, we
 present an overview of the HerPlaNS survey by focusing on the data
 products and their potential. 
 Below we will describe the schemes of observations and data
 reduction (\S\,\ref{obs}), showcase the basic data characteristics 
 using the PN \object{NGC\,6781} as a representative sample (\S\,\ref{6781}),
 and summarize the potential of the data set (\S\,\ref{sum}) 
 to pave the way for more comprehensive and detailed analyses of the
 broadband mapping and spectroscopy data that will be presented in the 
 forthcoming papers of the series.

\section{The Herschel Planetary Nebula Survey (HerPlaNS)\label{obs}}

%_____________________________________________________________
% Table 1: Object List
%_____________________________________________________________
\begin{table*}
\caption{\label{objlist}  
 List of HerPlaNS Target PNs}
\centering                   
\begin{tabular}{lccccccccc}  
\hline\hline                 
 &  &  & D & R & Dyn.\,Age & T$_{\rm *}$ & & &
 X-Ray \\    % table heading 
Name & PN G & Morph\tablefootmark{a} & (kpc) & (pc) & (10$^3$ yr) &
			 (10$^3$ K) & Sp.\,Type & H$_{2}$ & Results\\
\hline                        % inserts single horizontal line
\object{NGC\,40}   & 120.0$+$09.8 & Bbsh    & 1.0 & 0.11 & \phantom{1}4 & \phantom{1}48 & [WC8] & Y & D \\      % inserting body of the table
\object{NGC\,2392} & 197.8$+$17.3 & Rsai    & 1.3 & 0.14 & \phantom{1}3 & \phantom{1}47 & Of(H) & N & D, P \\
\object{NGC\,3242} & 261.0$+$32.0 & Ecspaih & 1.0 & 0.10 & \phantom{1}4 & \phantom{1}89 & O(H)  & N & D \\
\object{NGC\,6445} & 008.0$+$03.9 & Mpi     & 1.4 & 0.14 & \phantom{1}3 & 170 & \dots   & Y & P \\
\object{NGC\,6543} & 096.4$+$29.9 & Mcspa   & 1.5 & 0.09 & \phantom{1}5 & \phantom{1}48 & Of-WR(H) & N & D, P \\ 
\object{NGC\,6720} & 063.1$+$13.9 & Ecsh    & 0.7 & 0.13 & \phantom{1}6 & 148 & hgO(H)   &  Y & N \\ 
\object{NGC\,6781} & 041.8$-$02.9 & Bth     & 1.0 & 0.32 & 38 & 110 & DAO & Y & N\\ 
\object{NGC\,6826} & 083.5$+$12.7 & Ecsah   & 1.3 & 0.08 & \phantom{1}5 & \phantom{1}50 & O3f(H) & N & D, P \\ 
\object{NGC\,7009} & 037.7$-$34.5 & Lbspa   & 1.5 & 0.09 & \phantom{1}3 & \phantom{1}87 & O(H) & N & D, P \\ 
\object{NGC\,7026}\tablefootmark{b} & 089.0$+$00.3 & Bs & 1.7 & 0.16 & <1 & \phantom{1}80 & [WC] & Y & D  \\ 
\object{Mz 3}\tablefootmark{c}  & 331.7$-$01.0 & Bps & 1--3 & 0.1--0.2 & 0.6--2 & \phantom{1}32: & \dots & N & D, P  \\ 
\hline                                   %inserts single line
\end{tabular}
\tablefoot{%
Adopted from ChanPlaNS Table\,1 (\citealt{chanplans}, 
and references therein), for which data are compiled from \citet{frew08}.
with additional information on H$_{2}$ \citep{hora99,smith03}.
X-ray results key: P = point source; D = diffuse source; N = not detected.
\tablefoottext{a}{According to the classification scheme by
 \citet{sahai11}; 
 B: bipolar, 
 E: elongated, 
 L: collimated lobe pair, 
 M: multipolar, 
 R: round,  
 a: ansae, 
 b: bright (barrel-shaped) central region, 
 c: closed outer lobes, 
 h: halo; 
 i: inner bubble, 
 p: point symmetry, 
 s: CSPN apparent, 
 t: bright central toroidal structure;} 
\tablefoottext{b}{Not a ChanPlaNS target PN; \object{NGC\,7026} from
 \citet{ggcw06}, and \object{Mz 3} from \citet{kastner03}.
The point spread function of XMM-Newton does not allow us to
 determine whether or not a point source is present.}
\tablefoottext{c}{Not a ChanPlaNS target PN; may be a symbiotic/PN mimic
 \citep{frew08}; data from \citet{kastner03}.}}
\end{table*}
%_____________________________________________________________

\subsection{Target Selection}

Our aim with HerPlaNS is to generate a comprehensive spatially-resolved
far-IR PN data resource which carries a rich and lasting legacy in the
follow-up investigations.  
As HerPlaNS was motivated partly by the 
Chandra Planetary Nebula Survey (ChanPlaNS; \citealt{chanplans})
conducted with 
the {\sl Chandra X-ray Observatory\/} \citep{cxc},
our target list is a subset of the initial ChanPlaNS sample (Cycle 12
plus archival). 
This sample is volume-limited, with an approximate cutoff distance of
1.5\,kpc, and is dominated by relatively high-excitation nebulae (see
\citealt{chanplans} for details).
Then, we took into account the far-IR detectability of the target
candidates based on the previous observations made with {\sl IRAS\/},
{\sl ISO\/}, {\sl Spitzer\/}, and {\sl AKARI\/}.  
Through this exercise,
we selected 11 PNs for comprehensive suites of observations with 
{\sl Herschel\/}, aiming to investigate
the potential effects of X-rays on the physics and chemistry of the
nebular gas and their manifestations in far-IR PN characteristics.
Table \ref{objlist} lists the whole HerPlaNS sample and its basic
characteristics. 

\subsection{Observing Modes and Strategies\label{modes}}

In executing the HerPlaNS survey, we used all observing modes available
with the photodetector array camera and spectrometer (PACS;
\citealt{pacs}) and the spectral and photometric imaging receiver
(SPIRE; \citealt{spire}). 
The log of observations is given in Table\,\ref{log}.

With PACS, we performed 
(1) dual-band imaging at 70\,$\mu$m (Blue band) 
and 160\,$\mu$m (Red band) 
with oversampling of the telescope point spread function
(PSF; diffraction/wavefront error limited) and  
(2) integral-field-unit (IFU) spectroscopy
by $5 \times 5$ spaxels (spectral-pixels), 
over 51--220\,$\mu$m.
For two targets (\object{NGC\,40} and \object{NGC\,6720}), 
an additional IFU spectroscopy was done at a higher spectral resolution 
with a $3 \times 3$ raster mapping (i.e., at higher spatial sampling)
for specific far-IR fine-structure lines.
With SPIRE, we carried out 
(1) triple-band imaging at 
250\,$\mu$m (PSW band), 
350\,$\mu$m (PMW band), and 
500\,$\mu$m (PLW band),
and
(2) Fourier-transform spectrometer (FTS) spectroscopy in two
overlapping bands to cover 
194--672\,$\mu$m (SSW band over
194--313\,$\mu$m with 35 detectors and 
SLW band over
303--672\,$\mu$m with 19 detectors).  

%-----------------------------------------------------------------
% Figure 1: detector footprints
%-----------------------------------------------------------------
   \begin{figure}[h]
    \centering
    \includegraphics[width=0.74\hsize]{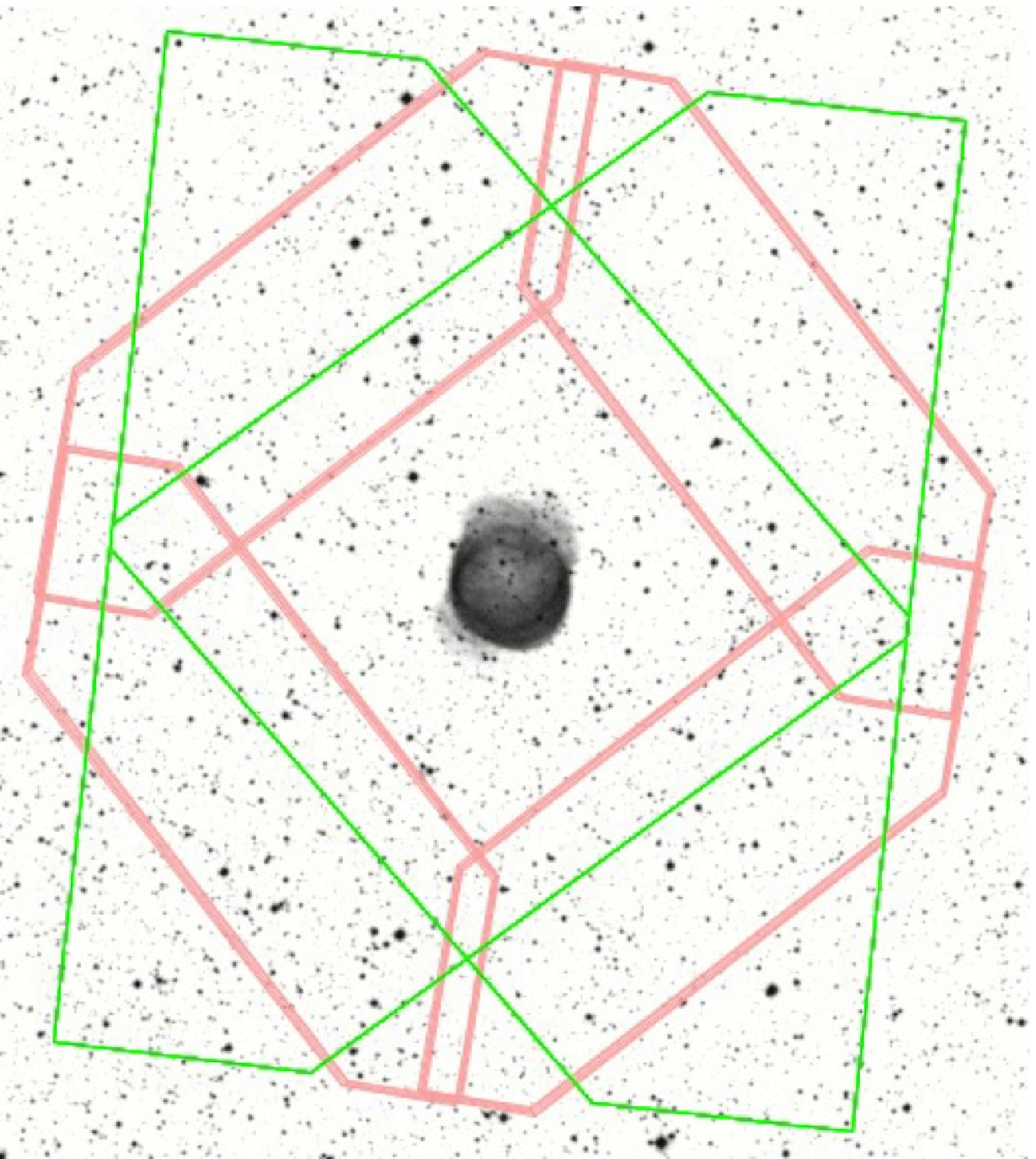}
    \includegraphics[width=0.245\hsize]{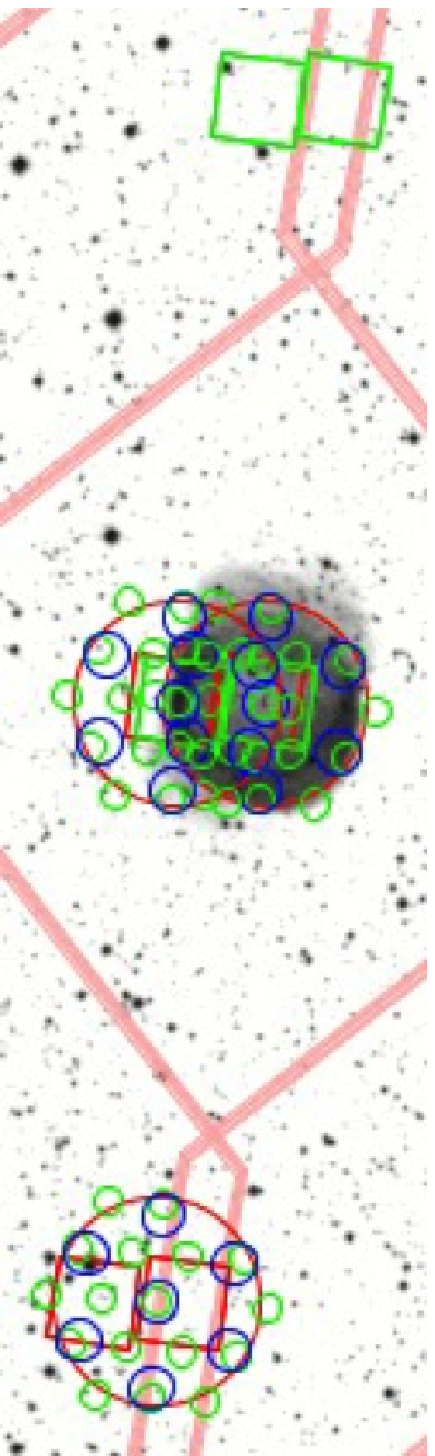}
    \caption{\label{6781example}
    HerPlaNS spatial coverage of \object{NGC\,6781}.
    The footprint of each instrument/observing mode is overlaid
    with the Digitized Sky Survey POSS2 Red map.
    [Left] Broadband imaging: 
    (1) PACS scan-mapping -- each pink polygon corresponds to
    a single medium-speed scan, delineating the four
    sides of the total $8^{\prime} \times 8^{\prime}$ region of mapping, 
    (2) SPIRE scan-mapping -- each green polygon corresponds to
    a single scan; two orthogonal scans define the total coverage.
    [Right] Spectroscopy (blow-up of the central region):
    (3) PACS spectroscopy -- a pair of red/green squares, each
    corresponding to chop/nod exposures (about $0\fdg1$ away in the N
    and S) of the $5 \times 5$ IFU field 
    of $47^{\prime\prime} \times 47^{\prime\prime}$, pointed at the
    center and eastern rim of the nebula,
    (4) SPIRE spectroscopy -- groups of green/blue/red circles,
    each corresponding to a detector feedhorn of the SSW and SLW bands 
    and the unvignetted 2\farcm6 field of view of the FTS bolometer
    array pointed at the same target locations as the PACS spectroscopy
    apertures with one off-source 
    pointing in the S.}  
   \end{figure}
%_________________________________________________________________

Using these capabilities, we obtained 
(1) broadband images in the above five bands
and 
(2) IFU spectral cubes in the PACS band,
FTS sparsely-sampled spectral array in the SPIRE SSW and SLW bands, and
at multiple locations in the target nebulae (``pointings'' hereafter).
From these IFU spectra it is also possible to extract spectral images
over a certain wavelength range (e.g., over a particular line or a
continuum) to recover the 
spatial extent of a specific emission.
Fig.\,\ref{6781example} shows footprints of detector apertures
for \object{NGC\,6781}, the target PN we consider in detail in this
paper, to illustrate how each of these data sets was obtained. 
In the forthcoming papers of the HerPlaNS series, we will discuss the
broadband mapping and spectroscopic data separately for the entire
HerPlaNS sample of 11 PNs plus others in the archive.

%_____________________________________________________________
% Table 2: Observing Log - landscape, longtable
%-------------------------------------------------------------
\begin{longtab}
%\begin{landscape}
\scriptsize
\centering
\begin{longtable}{lcccccccc}
\caption{Log of HerPlaNS Observations\label{log}}\\
\hline\hline             
Object & OD & Pointing & RA (2000) & DEC (2000) & AOT & Duration & Start
 Date \& Time & Obs.\ Id \\
 &  &  & (h m s) & (d m s) &  & (Sec) & (UTC) &  \\
\hline
\endfirsthead
\caption{Continued.}\\
\hline\hline
Object & OD & Pointing & RA (2000) & DEC (2000) & AOT & Duration & Start
 Date \& Time & Obs. Id \\
 &  &  & (h m s) & (d m s) &  & (Sec) & (UTC) &  \\
\hline
\endhead
\hline
\endfoot
NGC\,40 & \phantom{1}788 & Center & 00 13 01.010 & $+$72 31 19.10 & PacsPhoto & 15038 & 2011-07-10T18:34:30Z & 1342223905 \\
 & \phantom{1}788 & Center & 00 13 01.010 & $+$72 31 19.10 & PacsPhoto & 15038 & 2011-07-10T22:46:11Z & 1342223906 \\
 & \phantom{1}970 & Center & 00 13 01.010 & $+$72 31 19.10  & PacsRangeSpec & 13435 & 2012-01-08T14:30:35Z & 1342236879 \\
 & \phantom{1}970 & Center & 00 13 01.010 & $+$72 31 19.10 & PacsRangeSpec & \phantom{1}6673 & 2012-01-08T18:16:42Z & 1342236880 \\
 & 1118 & Center & 00 13 01.010 & $+$72 31 19.10 & PacsLineSpec & 21733 & 2012-06-04T19:48:53Z & 1342246640 \\
 & 1118 & Center & 00 13 01.010 & $+$72 31 19.10 & PacsLineSpec & 33903 & 2012-06-05T01:53:19Z & 1342246641 \\
 & \phantom{1}862 & Center & 00 13 01.010 & $+$72 31 19.10 & SpirePhoto & \phantom{11}721 & 2011-09-23T08:38:43Z & 1342229623 \\
 & 1054 & Center & 00 13 01.010 & $+$72 31 19.10 & SpireSpectrometer & \phantom{1}2936 & 2012-04-02T19:10:28Z & 1342243640 \\
 & 1079 & West Rim & 00 12 57.580 & $+$72 31 26.70 & SpireSpectrometer & \phantom{1}2936 & 2012-04-27T13:26:53Z & 1342245122 \\
 & 1054 & Off Sky & 00 12 55.200 & $+$72 25 30.00 & SpireSpectrometer & \phantom{11}502 & 2012-04-02T19:59:50Z & 1342243641 \\
NGC\,2392 & 888 & Center & 07 29 10.770 & $+$20 54 42.50 & PacsPhoto & 12486 & 2011-10-18T20:07:55Z & 1342231154 \\
 & \phantom{1}888 & Center & 07 29 10.770 & $+$20 54 42.50 & PacsPhoto & 12486 & 2011-10-18T23:37:04Z & 1342231155 \\
 & \phantom{1}866 & Center & 07 29 10.770 & $+$20 54 42.50 & PacsRangeSpec & 16685 & 2011-09-26T16:04:42Z & 1342229792 \\
 & \phantom{1}867 & Center & 07 29 10.770 & $+$20 54 42.50 & PacsRangeSpec & 17895 & 2011-09-27T15:04:58Z & 1342229816 \\
 & \phantom{1}862 & Center & 07 29 10.770 & $+$20 54 42.50 & SpirePhoto & \phantom{11}721 & 2011-09-22T15:32:21Z & 1342229466 \\
 & 1079 & Center & 07 29 10.770 & $+$20 54 42.50 & SpireSpectrometer & \phantom{1}4288 & 2012-04-27T12:02:25Z & 1342245121 \\
 & 1079 & South Rim & 07 29 10.500 & $+$20 54 27.00 & SpireSpectrometer & \phantom{1}3070 & 2012-04-27T11:11:02Z & 1342245120 \\
 & 1079 & Off Sky & 07 29 12.900 & $+$20 57 31.60 & SpireSpectrometer & \phantom{11}502 & 2012-04-27T11:02:19Z & 1342245119 \\
NGC\,3242 & \phantom{1}782 & Center & 10 24 46.090 & $-$18 38 28.30 & PacsPhoto & 15038 & 2011-07-04T17:56:50Z & 1342223696 \\
 & \phantom{1}782 & Center & 10 24 46.090 & $-$18 38 28.30 & PacsPhoto & 15038 & 2011-07-04T22:08:31Z & 1342223697 \\
 & \phantom{1}912 & Center & 10 24 46.090 & $-$18 38 28.30 & PacsRangeSpec & \phantom{1}6673 & 2011-11-12T10:42:02Z & 1342232278 \\
 & \phantom{1}912 & Center & 10 24 46.090 & $-$18 38 28.30 & PacsRangeSpec & \phantom{1}8963 & 2011-11-12T12:35:24Z & 1342232279 \\
 & \phantom{1}948 & Center & 10 24 46.090 & $-$18 38 28.30 & SpirePhoto & \phantom{1}1135 & 2011-12-18T00:36:09Z & 1342234839 \\
 & 1098 & Center & 10 24 46.090 & $-$18 38 28.30 & SpireSpectrometer & \phantom{1}2936 & 2012-05-16T01:00:55Z & 1342245847 \\
 & 1098 & South Rim & 10 24 46.520 & $-$18 38 47.30 & SpireSpectrometer & \phantom{1}2936 & 2012-05-16T00:11:43Z & 1342245846 \\
 & 1098 & Off Sky & 10 24 55.510 & $-$18 33 04.90 & SpireSpectrometer & \phantom{11}502 & 2012-05-16T00:02:55Z & 1342245845 \\
NGC\,6445 & \phantom{1}889 & Center & 17 49 15.210 & $-$20 00 34.50 & PacsPhoto & 10076 & 2011-10-19T19:20:14Z & 1342231253 \\
 & \phantom{1}889 & Center & 17 49 15.210 & $-$20 00 34.50 & PacsPhoto & 10076 & 2011-10-19T22:09:13Z & 1342231254 \\
 & 1047 & Center & 17 49 15.210 & $-$20 00 34.50 & PacsRangeSpec & \phantom{1}6731 & 2012-03-26T02:12:26Z & 1342242440 \\
 & 1047 & Center & 17 49 15.210 & $-$20 00 34.50 & PacsRangeSpec & \phantom{1}3351 & 2012-03-26T04:06:49Z & 1342242441 \\
 & 861 & Center & 17 49 15.210 & $-$20 00 34.50 & SpirePhoto & \phantom{11}307 & 2011-09-22T08:54:34Z & 1342229195 \\
 & 1054 & Center & 17 49 15.210 & $-$20 00 34.50 & SpireSpectrometer & \phantom{1}2800 & 2012-04-02T03:33:34Z & 1342243630 \\
 & 1054 & West Lobe & 17 49 14.290 & $-$20 00 21.80 & SpireSpectrometer & \phantom{1}2800 & 2012-04-02T02:46:41Z & 1342243629 \\
 & 1054 & Off Sky & 17 49 14.400 & $-$20 06 39.60 & SpireSpectrometer & \phantom{11}502 & 2012-04-02T02:37:53Z & 1342243628 \\
NGC\,6543 & \phantom{1}991 & Center & 17 58 33.240 & $+$66 37 58.80 & PacsRangeSpec & \phantom{1}6731 & 2012-01-29T14:25:29Z & 1342238388 \\
 & \phantom{1}991 & Center & 17 58 33.240 & $+$66 37 58.80 & PacsRangeSpec & \phantom{1}3351 & 2012-01-29T16:19:52Z & 1342238389 \\
 & \phantom{1}837 & Center & 17 58 33.240 & $+$66 37 58.80 & SpireSpectrometer\tablefootmark{a} & \phantom{1}7027 & 2011-08-29T06:52:20Z & 1342227789 \\
 & \phantom{1}972 & Center & 17 58 33.240 & $+$66 37 58.80 & SpireSpectrometer & \phantom{1}2664 & 2012-01-11T07:28:42Z & 1342237028 \\
 & \phantom{1}958 & West Knot & 17 58 17.350 & $+$66 38 08.30 & PacsRangeSpec & \phantom{1}6731 & 2011-12-27T14:58:17Z & 1342235679 \\
 & \phantom{1}958 & West Knot & 17 58 17.350 & $+$66 38 08.30 & PacsRangeSpec & \phantom{1}3351 & 2011-12-27T16:52:40Z & 1342235680 \\
 & \phantom{1}837 & West Knot & 17 58 17.350 & $+$66 38 08.30 & SpireSpectrometer\tablefootmark{a} & \phantom{1}7027 & 2011-08-29T04:43:45Z & 1342227787 \\
 & \phantom{1}972 & West Knot & 17 58 17.350 & $+$66 38 08.30 & SpireSpectrometer & \phantom{1}2664 & 2012-01-11T06:43:58Z & 1342237027 \\
 & \phantom{1}837 & Off Sky & 17 58 30.710 & $+$66 43 49.70 & SpireSpectrometer & \phantom{11}636 & 2011-08-29T06:41:18Z & 1342227788 \\
NGC\,6720 & \phantom{1}938 & Off Center & 18 53 34.500 & $+$33 01 57.40 & PacsRangeSpec & \phantom{1}6731 & 2011-12-07T12:57:48Z & 1342233716 \\
 & \phantom{1}938 & Off Center & 18 53 34.500 & $+$33 01 57.40 & PacsRangeSpec & \phantom{1}8893 & 2011-12-07T14:52:11Z & 1342233717 \\
 & \phantom{1}920 & Center & 18 53 35.100 & $+$33 01 44.90 & PacsLineSpec & 33903 & 2011-11-19T15:41:38Z & 1342232561 \\
 & \phantom{1}920 & Center & 18 53 35.100 & $+$33 01 44.90 & PacsLineSpec & 36925 & 2011-11-20T01:08:48Z & 1342232562 \\
 & 1053 & Center & 18 53 35.080 & $+$33 01 45.00 & SpireSpectrometer & \phantom{1}4288 & 2012-04-01T14:40:15Z & 1342243616 \\
 & 1053 & North Rim & 18 53 34.280 & $+$33 02 09.00 & SpireSpectrometer & \phantom{1}3070 & 2012-04-01T15:52:02Z & 1342243617 \\
 & 1053 & Off Sky & 18 53 34.200 & $+$32 56 04.00 & SpireSpectrometer & \phantom{11}502 & 2012-04-01T16:43:38Z & 1342243618 \\
NGC\,6781 & \phantom{1}887 & Center & 19 18 28.090 & $+$6 32 19.30 & PacsPhoto & 12681 & 2011-10-17T21:52:48Z & 1342231099 \\
 & \phantom{1}887 & Center & 19 18 28.090 & $+$6 32 19.30 & PacsPhoto & 12681 & 2011-10-18T01:25:12Z & 1342231100 \\
 & \phantom{1}884 & Center & 19 18 28.090 & $+$6 32 19.30 & PacsRangeSpec & 13435 & 2011-10-14T20:01:18Z & 1342230999 \\
 & \phantom{1}884 & Center & 19 18 28.090 & $+$6 32 19.30 & PacsRangeSpec & \phantom{1}9995 & 2011-10-14T23:47:25Z & 1342231000 \\
 & \phantom{1}884 & Rim & 19 18 31.520 & $+$6 32 19.80 & PacsRangeSpec & 17899 & 2011-10-15T02:36:09Z & 1342231001 \\
 & \phantom{1}884 & Rim & 19 18 31.520 & $+$6 32 19.80 & PacsRangeSpec & \phantom{1}9995 & 2011-10-15T07:36:40Z & 1342231002 \\
 & \phantom{1}880 & Center & 19 18 28.090 & $+$6 32 19.30 & SpirePhoto & \phantom{11}583 & 2011-10-11T06:57:06Z & 1342230841 \\
 & 1053 & Center & 19 18 28.090 & $+$6 32 19.30 & SpireSpectrometer & \phantom{1}2800 & 2012-04-01T12:21:55Z & 1342243612 \\
 & 1053 & Rim & 19 18 31.520 & $+$6 32 19.80 & SpireSpectrometer & \phantom{1}2800 & 2012-04-01T11:34:58Z & 1342243611 \\
 & 1053 & Off Sky & 19 18 32.390 & $+$6 26 32.10 & SpireSpectrometer & \phantom{11}502 & 2012-04-01T11:26:11Z & 1342243610 \\
NGC\,6826 & \phantom{1}771 & Center & 19 44 48.150 & $+$50 31 30.30 & PacsPhoto & 10076 & 2011-06-23T22:08:04Z & 1342223188 \\
 & \phantom{1}771 & Center & 19 44 48.150 & $+$50 31 30.30 & PacsPhoto & 10076 & 2011-06-24T00:57:03Z & 1342223189 \\
 & 1003 & Center & 19 44 48.150 & $+$50 31 30.30 & PacsRangeSpec & \phantom{1}8963 & 2012-02-10T15:02:19Z & 1342238926 \\
 & 1003 & Center & 19 44 48.150 & $+$50 31 30.30 & PacsRangeSpec & 13338 & 2012-02-10T17:33:54Z & 1342238927 \\
 & \phantom{1}962 & Rim & 19 44 52.990 & $+$50 31 44.60 & PacsRangeSpec & 17899 & 2011-12-31T16:28:58Z & 1342235850 \\
 & \phantom{1}962 & Rim & 19 44 52.990 & $+$50 31 44.60 & PacsRangeSpec & 23333 & 2011-12-31T21:29:29Z & 1342235851 \\
 & \phantom{1}862 & Center & 19 44 48.150 & $+$50 31 30.30 & SpirePhoto & \phantom{11}997 & 2011-09-23T07:04:09Z & 1342229608 \\
 & \phantom{1}941 & Center & 19 44 48.150 & $+$50 31 30.30 & SpireSpectrometer\tablefootmark{b} & \phantom{1}2936 & 2011-12-10T13:12:34Z & 1342233822 \\
 & \phantom{1}971 & Center & 19 44 48.150 & $+$50 31 30.30 & SpireSpectrometer & \phantom{1}2936 & 2012-01-10T09:38:19Z & 1342237006 \\
 & \phantom{1}741 & Rim & 19 44 51.230 & $+$50 31 39.20 & SpireSpectrometer\tablefootmark{a} & \phantom{1}8611 & 2011-05-25T16:39:54Z & 1342221687 \\
 & \phantom{1}741 & Off Sky & 19 44 27.380 & $+$50 37 08.90 & SpireSpectrometer & \phantom{11}735 & 2011-05-25T16:27:12Z & 1342221686 \\
NGC\,7009 & \phantom{1}894 & Center & 21 04 10.820 & $-$11 21 48.60 & PacsPhoto & 14900 & 2011-10-24T18:05:45Z & 1342231400 \\
 & \phantom{1}894 & Center & 21 04 10.820 & $-$11 21 48.60 & PacsPhoto & 14900 & 2011-10-24T22:15:08Z & 1342231401 \\
 & \phantom{1}913 & Center & 21 04 10.820 & $-$11 21 48.60 & PacsRangeSpec & \phantom{1}4499 & 2011-11-13T07:27:59Z & 1342232300 \\
 & \phantom{1}913 & Center & 21 04 10.820 & $-$11 21 48.60 & PacsRangeSpec & \phantom{1}5571 & 2011-11-13T08:45:10Z & 1342232301 \\
 & 1064 & Center & 21 04 10.820 & $-$11 21 48.60 & SpirePhoto & \phantom{11}583 & 2012-04-11T23:12:25Z & 1342244153 \\
 & \phantom{1}741 & Center & 21 04 10.820 & $-$11 21 48.60 & SpireSpectrometer\tablefootmark{a} & \phantom{1}7651 & 2011-05-25T11:41:29Z & 1342221681 \\
 & 1080 & Center & 21 04 10.820 & $-$11 21 48.20 & SpireSpectrometer & \phantom{1}4288 & 2012-04-27T22:17:32Z & 1342245079 \\
 & 1080 & East Flier & 21 04 12.440 & $-$11 21 43.90 & SpireSpectrometer & \phantom{1}2936 & 2012-04-27T21:28:22Z & 1342245078 \\
 & \phantom{1}741 & Off Sky & 21 04 04.280 & $-$11 16 13.80 & SpireSpectrometer & \phantom{11}675 & 2011-05-25T11:29:48Z & 1342221680 \\
NGC\,7026 & \phantom{1}788 & Center & 21 06 18.570 & $+$47 51 06.90 & PacsPhoto & 12486 & 2011-07-11T06:43:28Z & 1342223919 \\
 & \phantom{1}788 & Center & 21 06 18.570 & $+$47 51 06.90 & PacsPhoto & 12486 & 2011-07-11T10:12:37Z & 1342223920 \\
 & \phantom{1}936 & Center & 21 06 18.570 & $+$47 51 06.90 & PacsRangeSpec & \phantom{1}8959 & 2011-12-06T07:53:03Z & 1342234268 \\
 & \phantom{1}936 & Center & 21 06 18.570 & $+$47 51 06.90 & PacsRangeSpec & \phantom{1}6669 & 2011-12-06T10:24:34Z & 1342234269 \\
 & \phantom{1}722 & Center & 21 06 18.570 & $+$47 51 06.90 & SpirePhoto & \phantom{11}445 & 2011-05-06T00:55:57Z & 1342219975 \\
 & 1011 & Center & 21 06 18.570 & $+$47 51 06.90 & SpireSpectrometer & \phantom{1}2936 & 2012-02-19T02:50:17Z & 1342239347 \\
 & 1011 & North Lobe & 21 06 18.520 & $+$47 51 19.70 & SpireSpectrometer & \phantom{1}2936 & 2012-02-19T02:01:01Z & 1342239346 \\
 & 1011 & Off Sky & 21 06 05.180 & $+$47 52 59.50 & SpireSpectrometer & \phantom{11}502 & 2012-02-19T01:52:18Z & 1342239345 \\
PN Mz 3 & 1042 & Center & 16 17 13.400 & $-$51 59 10.60 & PacsRangeSpec & \phantom{1}5567 & 2012-03-21T17:58:45Z & 1342243109 \\
 & 1042 & Center & 16 17 13.400 & $-$51 59 10.60 & PacsRangeSpec & \phantom{1}4495 & 2012-03-21T19:33:41Z & 1342243110 \\
 & 1229 & Center & 16 17 13.400 & $-$51 59 10.60 & SpireSpectrometer & \phantom{1}2800 & 2012-09-24T02:21:17Z & 1342251318 \\
 & 1229 & South Lobe & 16 17 13.240 & $-$51 59 25.50 & SpireSpectrometer & \phantom{1}2800 & 2012-09-24T01:34:21Z & 1342251317 \\
 & 1229 & Off Sky & 16 17 08.620 &
   $-$52 01 59.80 & SpireSpectrometer &
    \phantom{11}502 & 2012-09-24T01:25:36Z &
     1342251316 \\
\hline
\end{longtable}
\tablefoot{%
\tablefoottext{a}{These SPIRE spectral-mapping were done originally with 
 full-spatial-sampling, resulting in an insufficient signal: hence the
 observations were executed again with sparse-spatial-sampling.  
 The rest of the ``SpireSpectrometer'' observations were done with
 sparse-spatial-sampling after an AOT revision.}
\tablefoottext{b}{Data failure: observation repeated.}}
%\end{landscape}
\end{longtab}
%_____________________________________________________________

\subsection{Data Reduction}

%_____________________________________________________________
% Table 3: Obs Summary
%_____________________________________________________________
\begin{table*}
 \footnotesize
\caption{\label{datasummary}  
 Summary of HerPlaNS Data Products and their Characteristics}
\centering                   
\begin{tabular}{lcccccccccc}  
\hline\hline                 
Observing Mode & Instrument/Band & $\lambda (\Delta\lambda)$ & \multicolumn{2}{c}{Data Characteristics} \\
& & ($\mu$m) & & \\
\hline                        % inserts single horizontal line
Imaging & PACS/Blue & 70 (25) & 
scan map (5\farcs6 beam at 1$^{\prime\prime}$\,pix$^{-1}$) & 
$2\farcm5 \times 2\farcm5$ to $7^{\prime} \times 7^{\prime}$ field of view \\
& PACS/Green & 110 (45) & scan map (6\farcs8 beam at
	     1$^{\prime\prime}$\,pix$^{-1}$) & by 2 orthogonal scans\\
& PACS/Red & 160 (85) & scan map (11\farcs4 beam at 2$^{\prime\prime}$\,pix$^{-1}$) &\\
& SPIRE/PSW & 250 (76) & scan map (18\farcs2 beam at 6$^{\prime\prime}$\,pix$^{-1}$) &
$4^{\prime} \times 8^{\prime}$ field of view \\
& SPIRE/PMW & 350 (103) & scan map (24\farcs9 beam at
	     9$^{\prime\prime}$\,pix$^{-1}$) & by 2 orthogonal scans \\
& SPIRE/PLW & 500 (200) & scan map (36\farcs3 beam at
	     14$^{\prime\prime}$\,pix$^{-1}$) & \\
Spectroscopy & 
   PACS/B2A & 51--72 & spectral cube ($R\approx4000$ at 9\farcs6\,spaxel$^{-1}$) &
 $\sim 50^{\prime\prime} \times 50^{\prime\prime}$ field of view \\
 & PACS/B2B & 70--105 &  spectral cube ($R\approx2000$ at
	     $10^{\prime\prime}$\,spaxel$^{-1}$) & by $5 \times 5$ IFU spaxels\\
 & PACS/R1 & 103--145 &  spectral cube ($R\approx1500$ at 11\farcs6\,spaxel$^{-1}$)\\
 & PACS/R1 & 140--220 &  spectral cube ($R\approx1000$ at 13\farcs2\,spaxel$^{-1}$) \\
& SPIRE/SSW & 194--342 & spectral array
($R\approx1000$ at $17^{\prime\prime}$--$21^{\prime\prime}$\,beam$^{-1}$) & 
$4^{\prime}$ diameter field of view \\
&&&&  by a 35-bolometer array \\
& SPIRE/SLW & 316--672 & spectral array
($R\approx500$ at $29^{\prime\prime}$--$42^{\prime\prime}$\,beam$^{-1}$) & 
 $4^{\prime}$ diameter field of view \\
&&&& by a 19-bolometer array  \\
\hline
\end{tabular}
\tablefoot{%
See Fig.\,\ref{specposmaps} for relative placements of PACS and
 SPIRE spectroscopic apertures.
The outermost SPIRE bolometers (16 for SSW and 12 for SLW) are
 located outside of the unvignetted 2\farcm6 field of view.}
\end{table*}
%_____________________________________________________________

Here, we briefly summarize the data reduction steps we adopted.
Complete accounts of reduction processes will be presented
in the forthcoming papers of the series (D.\,Ladjal et al.\ 
{\sl in prep}; K.\ M.\ Exter et al.\ {\sl in prep}).
A summary of the HerPlaNS data products and their characteristics is
given in Table\,\ref{datasummary}.

\subsubsection{Broadband Imaging}

To generate broadband images, we used the Herschel interactive
processing environment (HIPE, version 11; \citealt{hipe}) and 
Scanamorphos data reduction tool (Scanamorphos, version 21;
\citealt{scana}).
First, the raw scan map data were processed with HIPE from level 0 to 
level 1.  
During this stage, basic pipeline reduction steps were applied while the
data were corrected for instrumental effects. 
The level 1 data were then ingested into Scanamorphos, which corrects
for brightness drifts and signal jumps caused by electronic
instabilities and performs deglitching, flux calibration, and map 
projection. 
Scanamorphos was chosen as our map-making engine over other choices 
-- photoproject (the default HIPE mapper) and MADmap \citep{madmap} -- 
because it reconstructs
surface brightness maps of extended sources with the lowest noise, which
is of great importance for our purposes.
After processing with HIPE and Scanamorphos, we obtained far-IR surface
brightness maps at 5 bands (70, 160, 250, 350, and 500\,$\mu$m) for 11
PNs, each covering at most
$7^{\prime} \times 7^{\prime}$ unvignetted field centered at the target
source.

\subsubsection{PACS Spectroscopy}

We used HIPE track 11 with the calibration release version 44 to reduce
all of the PACS spectroscopy data of HerPlaNS.
Within HIPE, we selected the background normalization PACS spectroscopy
pipeline script for long range and SEDs 
to reduce the range scan data and 
the same pipeline script for line scans to reduce the line scan data. 
Our reduction steps follow those described in the PACS
Data Reduction Guide: Spectroscopy.\footnote{%
\url{http://herschel.esac.esa.int/hcss-doc-9.0/load/pacs_spec/html/pacs_spec.html}
(Version 1, Aug.\ 2012)}

In the range scan mode, we used the blue bands B2A (51--72\,$\mu$m)
and B2B (70--105\,$\mu$m) and each time we also got simultaneous
spectra in the R1 band (103--145\,$\mu$m and 140--220\,$\mu$m),
achieving the full spectral coverage from 51--220\,$\mu$m.
Each observation results in simultaneous spatial coverage of a
$\sim50^{\prime\prime} \times 50^{\prime\prime}$ field by a set of 
$5 \times 5$ spaxels of the IFU (each spaxel covering
roughly a $10^{\prime\prime} \times 10^{\prime\prime}$ field).
The PACS IFU $5\times5$ data cubes can also be integrated over a
specific wavelength range to generate a 2-D line map. 
This process can be done for any line detected at a reasonable S/N.

\subsubsection{SPIRE Spectroscopy}

We used the standard HIPE-SPIRE spectroscopy data reduction pipeline for 
the single-pointing mode (version 11 with SPIRE calibration tree
version 11) to reduce all of the SPIRE spectroscopy data of HerPlaNS,
but with the following three major modifications; 
(1) we extracted and reduced signal from each bolometer individually
instead of signal from only the central bolometer as nominally done for
single-pointing observations; 
(2) we applied the extended source flux calibration correction
to our data; and 
(3) we used our own dedicated off-target sky observations for the
background subtraction 
(Fig.\,\ref{6781example}).
Besides these extra steps, our reduction steps basically copy those
described in the  
SPIRE Data Reduction Guide.\footnote{%
\url{http://herschel.esac.esa.int/hcss-doc-9.0/load/spire_drg/html/spire_drg.html}
(version 2.1, Document Number: SPIRE-RAL-DOC 003248, 06 July 2012)}
The standard apodization function was applied to the data to minimize
the ringing in the instrument line shape wings at the expense of
spectral resolution. 

At the end of these processes, each of the on-source (center and
off-center) and off-sky pointings would yield 35 short-band
spectra\footnote{In the SLW band, 2 bolometers out of the total of 37
are blind.} from individual hexagonal bolometer positions
($33^{\prime\prime}$ spacing between bolometers)
for 194--342\,$\mu$m and 
19 long-band spectra from individual hexagonal bolometer positions
($51^{\prime\prime}$ spacing between bolometers)
for 316--672\,$\mu$m.
The bolometer beams for the short and long band arrays overlap
spatially at about a dozen positions, from which the full range spectrum
(194--672\,$\mu$m) can be constructed. 

We created an off-sky spectrum by taking a median of spectra taken from
the detectors located within the unvignetted field (i.e., all but the
outermost bolometers) of each off-sky position and 
subtracting the off-sky spectrum from each on-source spectrum taken from
the unvignetted field of the bolometer array.
Data from the vignetted outermost bolometers are not included for
the present science analyses because these bolometers are not
sufficiently calibrated for their uncertainties and long term
stabilities by the instrument team. 
Because of the large data volume collected and redundant spatial
coverage by center and off-center pointings for some of the target
sources, it is possible to self-calibrate data from the outermost
bolometers.
However, this is beyond the scope of the present overview and
hence will be discussed in the forthcoming papers of the series.

\section{HerPlaNS Data: A Case Study of NGC\,6781\label{6781}}

In the following, we showcase the wealth of the HerPlaNS data
set by revealing the far-IR characteristics of \object{NGC\,6781}.
\object{NGC\,6781} is an evolved PN\footnote{Here, this PN is described
as ``evolved'' to mean that its central star has started descending
along the white dwarf cooling track in the  
Hertzsprung-Russell (H-R) diagram (e.g., \citealt{vw94}).}
whose central star has an effective temperature of
110\,kK (DAO spectral type; \citealt{frew08})
and a luminosity of 
385\,L$_{\odot}$ for our adopted distance of $950 \pm 143$\,pc
(based on iterative photo-ionization model fitting constrained 
by various optical line maps; \citealt{sm06}).\footnote{%
In the present work, all distance-dependent quantities are derived
from the adopted distance of $950\pm143$\,pc \citep{sm06}, and hence,
its uncertainty propagates in subsequent calculations.}
The initial and present masses of the central star
are estimated to be 
$1.5 \pm 0.5$ and
$0.60 \pm 0.03$\,M$_{\odot}$, respectively, via comparison between 
evolutionary tracks of \citet{vw94} with photo-ionization model 
parameters (Fig.\,6 of \citealt{sm06}).
This comparison with evolutionary tracks also suggests that the age of
the PN since the AGB turn-off is (2--4)\,$\times 10^{4}$\,yr 
\citep{vw94}.

%_____________________________________________________________
% Figure 2: 3-D schematic of the cylindrical barrel structure
%-----------------------------------------------------------
   \begin{figure}
   \centering
   \includegraphics[width=\hsize]{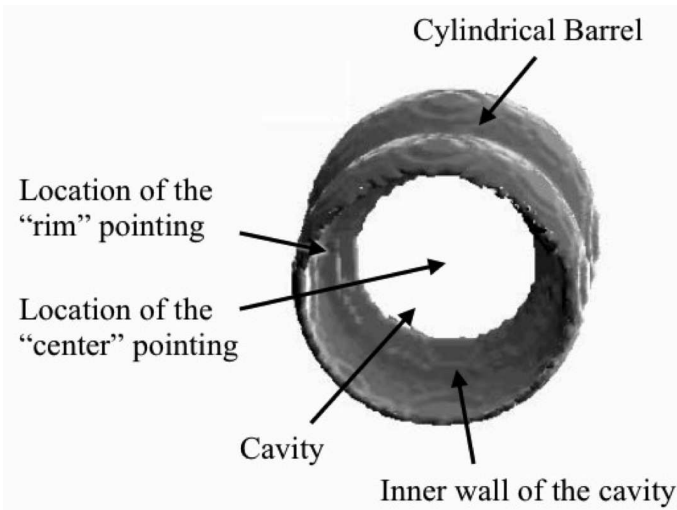}
    \caption{\label{schematic}
    3-D schematic of the central ``ring'' region of NGC\,6781 showing its
    orientation with respect to us,
    in which the cylindrical barrel, barrel cavity, and inner wall of
    the cavity, as well as the locations of the multi-point observations,
    are identified as a visual guide for readers.
    Note that the gray-scale image represents an isodensity surface of
    the barrel structure and that the density distribution is NOT
    necessarily confined within the surface shown here.
    This image is 
    reproduced from the work by \citet{sm06} with permission of the
    American Astronomical Society.} 
   \end{figure}
%______________________________________________________________

The surface brightness of \object{NGC\,6781} in the optical is known to
be very low and rather uniform, indicative of its relatively evolved
state (i.e., the stellar ejecta have been expanding for 
(2--4)\,$\times 10^{4}$\,yr).  
Previous optical imaging revealed the object's signature appearance of a
bright ring of about $130^{\prime\prime}$ diameter, which consists of
two separate rings in some parts.
The ring emission is embedded in faint extended lobes that are elongated
along the NNW-SSE direction \citep{mpp01,phillips11}. 
Morpho-kinematic observations in molecular lines
\citep{z90,bachiller93,hiriart05} and photo-ionization models \citep{sm06}
indicated that the density distribution in the nebula was 
cylindrical with an equatorial enhancement (i.e., a cylindrical
barrel) oriented at nearly pole-on.

The axis of the cylindrical barrel is thought to be inclined roughly at
$\sim23^{\circ}$ to the line of sight, with its south side pointed to us 
(Fig.\,\ref{schematic}). 
Optical emission line diagnostics yielded a relatively low electron
density of about 130--210\,cm$^{-3}$ and an electron temperature of
about $10^{4}$\,K \citep{liuy04,liu04}.
The dynamical age of the object is at least $3 \times 10^4$\,yr,
based on the observed extent of the faintest optical nebula
($\sim108^{\prime\prime}$; \citealt{mpp01}) and the shell expansion
velocity of 15\,km\,s$^{-1}$ 
(the average of expansion velocities measured from optical lines and
molecular radio emission; e.g., \citealt{weinberger98,bachiller93}) at
the adopted 950\,pc \citep{sm06}.
This age is consistent with the theoretical estimates mentioned above. 

\object{NGC\,6781} is representative of the class of
axisymmetric, dusty, and molecule-rich PNs (such as \object{NGC
6720} and \object{NGC\,6445} in the HerPlaNS sample).
These nebulae appear to be distinct from H$_2$-poor objects
(such as \object{NGC\,2392}, \object{NGC\,6543}, and others in the HerPlaNS
sample; Table\,\ref{objlist}) in terms of
progenitor mass, structure, and evolutionary history. 
It is therefore our aim in the forthcoming papers in
the HerPlaNS series to shed light on similarities and differences of
the far-IR PN characteristics in the HerPlaNS sample to enhance our
understanding of the physical properties of PNs. 

\subsection{Broadband Imaging\label{bbandimg}}

%_____________________________________________________________
% Figure 3: Broadband Maps
%-------------------------------------------------------------
   \begin{figure*}
    \centering
    \includegraphics[width=\hsize]{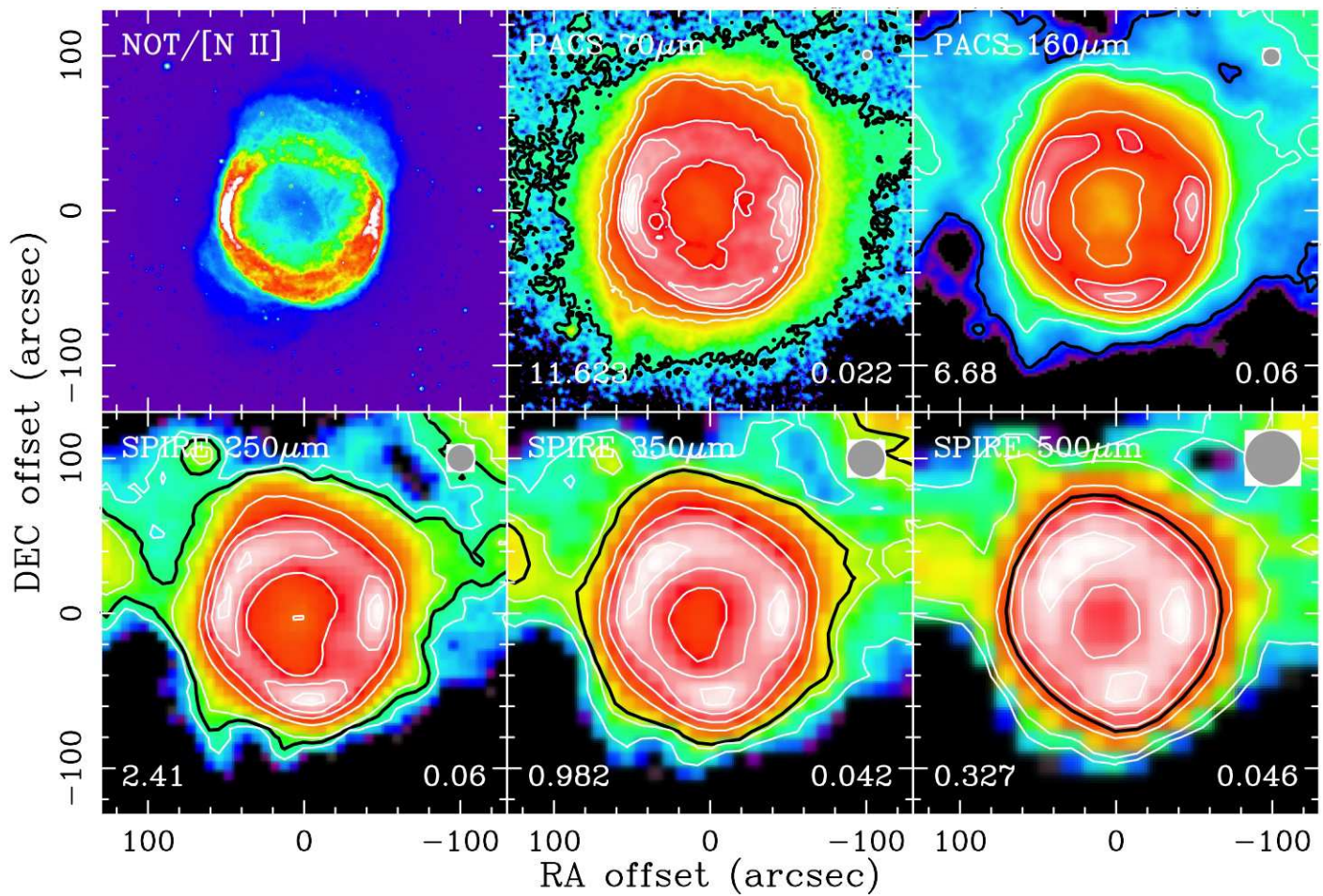}
    \caption{\label{broadmaps}
    HerPlaNS PACS/SPIRE broadband images of 
    \object{NGC\,6781} at 70, 160, 250, 350, and 500\,$\mu$m 
    in a $300^{\prime\prime}\times300^{\prime\prime}$ field
    centered at the position of the central star, 
    ($\alpha$, $\delta$) = (19:18:28.085, $+$06:32:19.29) 
    \citep{kerber03}, along with 
    an [\ion{N}{II}] image taken at the NOT \citep{phillips11}.
    The far-IR peak surface brightness is indicated at the bottom-left
    corner, 
    while the one-$\sigma_{\rm sky}$ noise is shown at the bottom-right corner
    (in mJy arcsec$^{-2}$) in each frame.
    At the top-right corner, the beam size for the band is indicated by
    a gray circle (5\farcs6, 11\farcs4, 18\farcs2, 24\farcs9, and
    36\farcs3, respectively).
    White contours represent 90, 70, 50, 30, 10, and 5\% of the peak,
    respectively, and the black contour indicates 3-$\sigma_{\rm sky}$ detection.   
    The pixel scales are 1, 1, 2,
    6, 9, and 14 arcsec\,pix$^{-1}$, respectively from upper left to
    lower right.
    To convert from mJy\,arcsec$^{-2}$ to MJy sr$^{-1}$, multiply by
    42.5.}
   \end{figure*}
%_____________________________________________________________

PACS/SPIRE broadband images of \object{NGC\,6781} are presented in 
Fig.\,\ref{broadmaps}, along with an optical image in the
[\ion{N}{II}]\,$\lambda$6584 band taken at the Nordic Optical Telescope
(NOT) for comparison \citep{phillips11}. 
These images reveal the far-IR structures of \object{NGC\,6781}, 
which are comparable to those in the optical.
The signature ``ring'' appearance of the near pole-on cylindrical barrel
structure is clearly resolved in all five far-IR bands, even at 500\,$\mu$m. 

The detected far-IR emission is dominated by thermal dust 
continuum: the degree of line contamination is determined to be at most 
8--20\% based on the HerPlaNS PACS/SPIRE spectroscopy data (see below).
The strength of the surface brightness is indicated by the color scale
and contours in Fig.\,\ref{broadmaps} (the band-specific values are
indicated at the bottom corners of each frame).  
The background root-mean-square (rms) noise ($= \sigma_{\rm sky}$),
determined using the off-source background sky regions,
is measured to be
between 0.023 and  
0.18\,mJy\,arcsec$^{-2}$ (0.97 and 7.66\,MJy\,sr$^{-1}$, respectively)
in these bands 
(the band-specific value is indicated at the bottom right in each
frame).   
The black contour is drawn to mark the 3-$\sigma_{\rm sky}$ detection level. 

\begin{figure}
    \centering
    \includegraphics[width=\hsize]{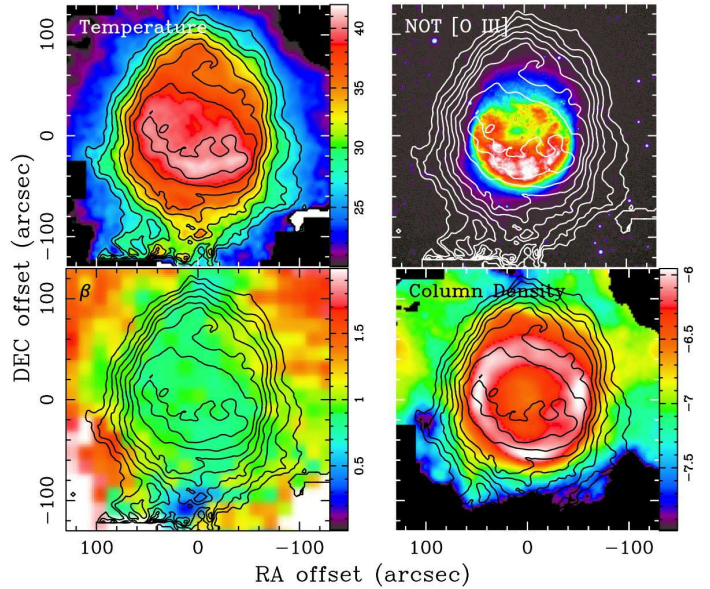}
    \caption{\label{tempmaps}
    [Top Left]
    The dust temperature ($T_{\rm dust}$) map of NGC\,6781 
    at $11\farcs4$ resolution
    derived by
    fitting the PACS/SPIRE maps with a power-law dust emissivity,
    $I_{\nu} \propto \lambda^{-\beta} B_{\nu}(T_{\rm dust})$
    in the same
    $300^{\prime\prime}\times300^{\prime\prime}$ field.
    The peak is 41.3\,K, with contours indicating temperatures from 40 to
    26\,K at 2\,K intervals (a linear color-scale wedge is also shown on the
    right).  
    [Top Right]
    $T_{\rm dust}$ contours overlaid with the 
    [\ion{O}{III}]\,$\lambda$5007 map taken at the NOT \citep{phillips11}.
    [Bottom Left]
    The power-law index ($\beta$) map with a linear color-scale wedge at
    the $36\farcs3$ resolution. 
    [Bottom Right]
    The dust column mass density ($\rho$ in ${\rm M}_{\odot}$\,pix$^{-1}$) map
    in log-scale at $11\farcs4$ resolution.  
    The peak is  
    $1.3 \times 10^{-6}$\,M$_{\odot}$\,pix$^{-1}$.}
   \end{figure}

In the continuum maps, far-IR emission from \object{NGC\,6781} is
detected from the 
$240^{\prime\prime} \times 200^{\prime\prime}$ region encompassing the
entire optical ring structure. 
While the optical ring appears somewhat incomplete due to a relatively
smaller surface brightness in the north side, the far-IR ring looks more
complete with a smoother surface brightness distribution, especially
in bands at longer wavelengths ($>\,160$\,$\mu$m).
This difference is most likely due to extinction of the optical line
emission emanating from the inner surface of the northern cylindrical
barrel by the dusty column of the inclined barrel wall that lies in 
front of the optical emission regions along the line of sight.
Our interpretation is consistent with the optical extinction map
presented by \citet{mpp01} (their Fig.\,3).
We show below that the dust column mass density is roughly constant 
all around the ring structure ($\sim 10^{-6}$\,M$_{\odot}$\,pix$^{-1}$ at
the 2$^{\prime\prime}$\,pix$^{-1}$ scale; Fig.\,\ref{tempmaps}).

The total extent of the far-IR emission (especially at 70\,$\mu$m)
encompasses that of the 
deep exposure [\ion{N}{II}] image (about 100$^{\prime\prime}$ radius at 
five-$\sigma_{\rm sky}$; Fig.\,2 of \citealt{mpp01}): hence, we infer that the diffuse
extended line emission is most likely caused by scattering of line
emission emanating from the central ionized region by dust grains in the
dusty extended part of the nebula.  
Given that the highly-ionized region is restricted within the ring
structure (e.g., \citealt{mpp01}), we can conclude that the total
extent of the nebula (both in the optical and far-IR) is 
sensitivity limited.
Adopting the constant expansion velocity of $\sim 15$\,km\,s$^{-1}$
as above, we confirm that the dynamical age of the observed far-IR
nebula is at least (3--4)\,$\times 10^{4}$\,yr.

At 70\,$\mu$m, the distribution of thermal dust continuum is 
very similar to what is seen in the [\ion{N}{II}]\,$\lambda$6584 
image \citep{phillips11}.
This indicates that the lateral density gradient in the barrel
wall along the equatorial plane
is very steep\footnote{\citet{mpp01} reported about a
$10^{\prime\prime}$ offset along the E-W direction between the
[\ion{N}{II}] and [\ion{O}{III}] profile peaks in the optical, 
which is barely resolvable at the {\sl Herschel} bands.}
and that the temperature stratification along this direction occurs 
in a physically very restricted region, i.e., on scales smaller than
{\sl Herschel}'s far-IR spatial resolution.
Hence, the surface brightness peaks on the eastern and western rims 
represent the pivot points of the inclined barrel about
20$^{\circ}$ tilted to the south with respect to the line of sight
\citep{bachiller93,hiriart05,sm06}.
By the same token, the southern rim of the ring represents the interior
wall of the cylindrical barrel on the far side seen through the cavity,
while the northern rim is the exterior wall of the barrel on the near
side.

At longer wavelengths, the locations of the surface brightness peaks 
change: the surface brightness becomes brighter on the NW and SE
sides of the ring structure.
Similar surface brightness characteristics were seen in the radio
emission map in the CO J=2--1 line \citep{bachiller93}.
The overall appearance of the far-IR emission regions is more circular
at longer wavelengths.
This is partly due to spatial resolution, but is also due to the fact
that the redder far-IR images in the SPIRE bands probe the colder part
of the shell which gives about the same column density around the ring
structure.

%_____________________________________________________________
% Table 4: Photometry Table
%_____________________________________________________________ 
\begin{table*}
 \caption{\label{photom}
 Far-IR Image Characteristics and Photometry of \object{NGC\,6781}}
 \centering
 \begin{tabular}{lccccc}
  \hline
  Band & $\lambda$ & $\Delta\lambda$ & $I_{\rm peak}$ & 
  $\sigma_{\rm sky}$ & $F_{\nu}$ \\
  & ($\mu$m) & ($\mu$m) & (mJy arcsec$^{-2}$) &
		      (mJy arcsec$^{-2}$) & (Jy) \\  
  \hline
  PACS Blue & \phantom{1}70 & \phantom{1}25 & 11.623 &
		  0.022  & $65.42 \pm 3.28$\\
  PACS Red  & 160 & \phantom{1}85 & \phantom{1}6.68\phantom{1} &
		  0.06\phantom{0} 
		      & $64.88 \pm 3.28$\\
  SPIRE PSW & 250 & \phantom{0}76 & \phantom{0}2.41\phantom{0} &
		  0.06\phantom{0} 
		      & $30.04 \pm 4.60$\\
  SPIRE PMW & 350 & 103 & \phantom{0}0.982 &
		  0.042 
		      & $14.56 \pm 2.25$\\
  SPIRE PLW & 500 & 200 & \phantom{0}0.327 &
		  0.046 
		      & $\phantom{0}6.41 \pm 1.02$\\
  \hline
 \end{tabular}
 \tablefoot{%
 The total specific flux, $F_{\nu}$, is measured above the
 three-$\sigma_{\rm sky}$.}
\end{table*}
%_____________________________________________________________

The total fluxes of the target at the PACS/SPIRE wavebands ($F_{\nu}$) 
are computed by aperture photometry: 
we adopted the three-$\sigma_{\rm sky}$ detection contour (the black
contour in Fig.\,\ref{broadmaps}) of the 70\,$\mu$m map (of the best S/N
among all) as the photometry aperture and summed up pixel values within
the aperture in all five bands.  
Upon running the aperture photometry routine, we first subtracted
background point sources using the IRAF {\sl daophot} routine built into
HIPE to make background-point-source-free maps.
The uncertainty of the total flux is set to be the combined
uncertainties of the sky scatter/variation ($\sigma_{\rm sky}$) and
the absolute flux calibration error (which is as high as 5\% for PACS and
15\% for SPIRE maps according to the Herschel PACS/SPIRE documentation).  
Color correction was applied using correction factors derived for the
appropriate temperature of the SED based on the color correction tables
provided in Table 3 of the PACS release note PICC-ME-TN-038 and 
Table 5.3 of the SPIRE Observing Manual v2.4. 
Table\,\ref{photom} summarizes photometric measurements for 
\object{NGC\,6781} made from the broadband images.

%_____________________________________________________________
% Figure 4: SED fit
%-------------------------------------------------------------
   \begin{figure}
    \centering
    \includegraphics[width=\hsize]{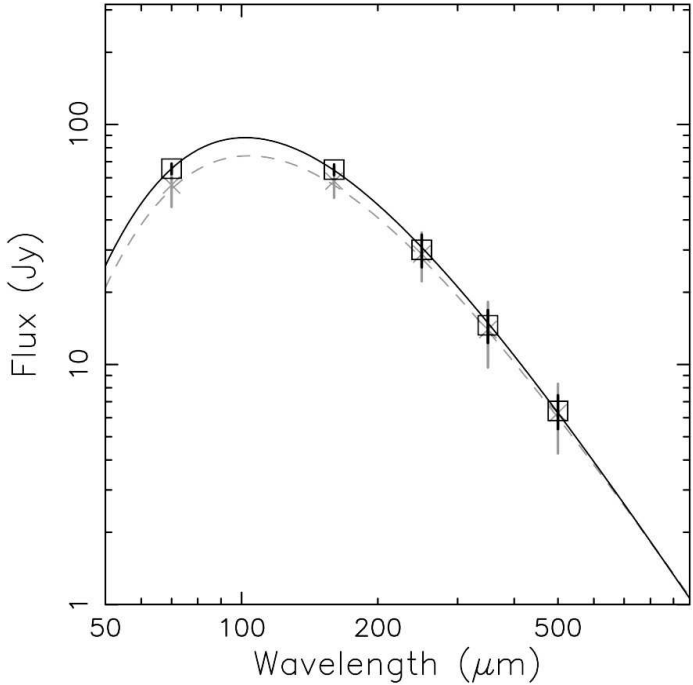}
    \caption{\label{sedfit}
    Far-IR SED of NGC\,6781 as fit by the HerPlaNS data.
    Squares indicate the measured photometry
    (with line emission)
    with vertical lines corresponding to uncertainties. 
    The black solid curve is the best-fit SED (i.e., with line
    contamination) with  
    $T_{\rm dust} = 36 \pm 2$\,K and $\beta = 1.0 \pm 0.1$.
    Crosses show the measured photometry (but line emission
    contribution subtracted) with vertical lines corresponding to
    uncertainties.
    The gray dashed curve is the best-fit SED (i.e., without line
    contamination) with
    $T_{\rm dust} = 37 \pm 5$\,K and $\beta = 0.9 \pm 0.3$.}  
   \end{figure}
%_____________________________________________________________

As seen from the photometry (Table \ref{photom}), the present 
{\sl Herschel}
observations cover the Rayleigh-Jeans shoulder of the thermal dust
emission component of the spectral energy distribution (SED) of 
\object{NGC\,6781}.
The dust temperature, $T_{\rm dust}$, therefore, can be estimated
by fitting the far-IR SED with a power-law dust emissivity, 
$I_{\nu} \propto \lambda^{-\beta} B_{\nu}(T_{\rm dust})$,
where $\beta$ defines the emissivity characteristics of the far-IR
emitting dust grains.
Typically, the value of $\beta$ is roughly 
2 for silicate dust grains
and 
graphite grains
and close to 1 for amorphous carbon grains 
(e.g., \citealt{bh83,dl84,rm91,mcb95} for theoretical/lab studies,
and \citealt{ksr93,kbyp94,gby02} for observational studies).  

%_____________________________________________________________
% Figure 5: Temperature Maps
%-------------------------------------------------------------
%_____________________________________________________________

Using the integrated fluxes in Table\,\ref{photom}, we obtain 
$T_{\rm dust} = 36 \pm 2$\,K and $\beta = 1.0 \pm 0.1$
(Fig.\,\ref{sedfit}).
Note, however, that these broadband fluxes include some line emission. 
Thus, we assessed the amount of line contamination in the broadband
fluxes using spectra taken by individual spaxels.
While the amount of line emission is spatially variable, we found the
level of line contamination to be 9--20\% at 70\,$\mu$m and 8--16\% at
160\,$\mu$m. 
As demonstrated by Fig.\,\ref{sedfit}, the line contamination
contributes negligibly to the uncertainties
in fitting $T_{\rm dust}$ and $\beta$.
Therefore, we concluded that for dust-rich objects direct fitting of
broadband fluxes with the modified blackbody yielded acceptable
$T_{\rm dust}$ and other derivatives.

From spatially resolved PACS/SPIRE maps (Fig.\,\ref{broadmaps}), 
we can recover $T_{\rm dust}$ and $\beta$ maps at the spatial
resolution of the 160\,$\mu$m map.
First, we performed five-point fitting of the modified blackbody curve
at the 
spatial resolution of the SPIRE 500\,$\mu$m map (FWHM of $36\farcs3$).
The derived $\beta$ map was fed back into the surface brightness ratio
map to solve for the dust temperature at the spatial resolution of the
160\,$\mu$m map (FWHM of $11\farcs4$) via the relation
\begin{eqnarray} \label{eq:fnumeric}
\frac{B_{\nu (70\mu m)}(T_{\rm dust})}{B_{\nu (160\mu m)}(T_{\rm dust})}
 &=&
\frac{F_{\nu (70\mu m)}}{F_{\nu (160\mu m)}} \times 
\frac{\nu (70\mu m)^{\beta}}{\nu (160\mu m)^{\beta}}. 
\end{eqnarray}
Upon being fed into the above relation, the $\beta$ map was
re-gridded at the pixel scale of the 160\,$\mu$m map 
($2^{\prime\prime}$ pix$^{-1}$) by 2-D linear interpolation.
We consider this approximation reasonable 
and better than the brute-force five-point SED fitting at the pixel scale of 
the 70\,$\mu$m map, which involves a substantial amount of interpolation,
because the range of $\beta$ is not large  
(between -0.5 and 2.5) and so there were no strong gradients over which  
values had to be spatially resampled.

The derived $T_{\rm dust}$ and $\beta$ maps, along with the
dust column density ($\rho$) map, are presented in Fig.\,\ref{tempmaps},
together with the [\ion{O}{III}]\,$\lambda$5007 map taken at the NOT 
for comparison \citep{phillips11}.
As shown in the top frames of Fig.\,\ref{tempmaps}, the highest dust
temperature region 
($T_{\rm dust} \gtrsim 40$\,K within the dust ring of $\sim 36$\,K) is 
spatially coincident with the region of highly-ionized optical line
emission (e.g., [\ion{O}{III}]\,$\lambda$5007 and H$\alpha$;
\citealt{mpp01,phillips11}), delineating the interior walls of
the barrel cavity directly visible to us through the polar opening.
While the median uncertainties of $\beta$ and $T_{\rm dust}$ in 
fitting pixel-wise surface brightnesses are as large as those in 
fitting integrated fluxes ($\pm 0.2$ for $\beta$ and $\pm 5$\,K 
for $T_{\rm dust}$), these pixel-wise uncertainties are not completely
independent because of the nature of dust heating (i.e., the radiative
equilibrium is achieved in an optically-thin medium) and the
differential spatial resolution over one decade of wavelengths.
Hence, the $\beta$ and $T_{\rm dust}$ distributions are as continuous 
as the dust distribution.

Therefore, we conclude that the 
five-point SED fitting of dust temperature was successful
and that the gradient of dust temperature within the dust ring is real.
The value of $\beta$ is close to unity around the central ionized
region (Fig.\,\ref{tempmaps}, bottom-left), suggesting that the major
component of the far-IR emitting dust 
is likely carbon-based \citep{vk88}.
This is consistent with previous chemical abundance analyses performed
with optical line measurements \citep{liu04,mk09}
as well as with the absence of silicate dust features in mid-IR spectra
taken by the {\sl Spitzer} IRS (e.g., \citealt{phillips11}).

The $\rho$ map can be derived from the observed
surface brightness maps and the derived $T_{\rm dust}$ map via
\begin{eqnarray}\label{dustmass}
 M_{\rm dust} = \frac{I_{\nu}D^2}{\kappa_{\nu} B_{\nu}(T_{\rm dust})},
\end{eqnarray}
where $D$ is the distance to the object and $\kappa_{\nu}$ is the
opacity of the dust grains.
Based on the 160\,$\mu$m map and adopting the dust opacity 
of $\kappa_{160\mu m} = 23$\,cm$^{2}$\,g$^{-1}$ at
160\,$\mu$m\footnote{We computed the dust opacity following the Mie 
formulation with the optical constants of amorphous carbon grains
determined by \citet{rm91}, assuming spherical grains having the size 
distribution of the ``MRN'' type \citep{mrn}, from 0.01--1\,$\mu$m.}, 
the $\rho$ map yields up to  
$1.6 \times 10^{-6}$\,M$_{\odot}$ per pixel (Fig.\,\ref{tempmaps}, bottom right).
As mentioned above, the $\rho$ map delineates the relatively uniform
distribution of dust grains in the cylindrical barrel.
Because the far-IR thermal dust continuum is optically thin all around the
ring structure  
($\tau_{160\mu{\rm m}} = 10^{-5}$--$10^{-6}$ on the ring), 
the dust column mass density map probes the whole depth of the
inclined nebula along the line of sight, corroborating
the pole-on cylindrical barrel structure that was previously only
inferred from optical images.
By integrating over the entire nebula, we determined that the
total amount of far-IR emitting dust is
$M_{\rm dust} = 4 \times 10^{-3}$\,M$_{\odot}$,
of which roughly 50\% appears to be contained in 
the cylindrical barrel (defined to be the region where $\rho$ is more
than 40\% of the peak). 

Fig.\,\ref{tempmaps} shows that there is still a substantial amount of
dust column along lines of sight toward the inner cavity of the barrel,
even though this region is expected to be filled with highly-ionized gas
as seen from emission maps in high-excitation optical lines such as
\ion{He}{II}$\lambda$4846 and [\ion{O}{III}]\,$\lambda$5007
\citep{mpp01}, and in mid-IR lines such as [\ion{O}{IV}]\,25.8\,$\mu$m and
\ion{He}{II} 24.3/25.2\,$\mu$m (which dominate most of the emission
detected in the archived {\sl WISE} 24\,$\mu$m map of the object). 
Moreover, the relatively high dust continuum emission detected toward the
inner cavity in the SPIRE range (both in images and spectra; see below)
suggests the presence of colder material toward this direction, as has
been implied by the previous detection of H$_{2}$ emission in v=0--0
S(2) to S(7) transitions in the bipolar lobes \citep{phillips11}.  
These pieces of evidence indicate that there are distributions of cold
dust (and gas) in front of and behind the highly-ionized central cavity
along the inclined polar axis (i.e., polar caps). 

While a high degree of symmetry is exhibited by the $\rho$ map by the
 cylindrical barrel structure, a highly lopsided structure -- warmer
 dust grains are concentrated along the surface of the S side of the
 barrel wall -- is presented by the $T_{\rm dust}$ map
 (Fig.\,\ref{schematic}, bottom right and top left, respectively). 
 Given that the dusty cylindrical barrel is optically thin in
 the far-IR, the $T_{\rm dust}$ distribution is not
 caused by the
 inclination of the cylindrical barrel (i.e., the inner surface of the S
 side of the barrel is directly seen from us, while the inner surface of
 the N side is obscured from our direct view by the barrel wall itself).
 Hence, we conclude that the barrel is not completely symmetric and its
 S side is somewhat closer than the N side (both
 the gas and dust components; Fig.\,\ref{tempmaps}, top frames). 

\subsection{Spatio-Spectroscopy}

%____________________________________________________________
% Figure 6: Spectroscopic aperture position maps
%-------------------------------------------------------------
   \begin{figure}
    \centering
    \includegraphics[width=\hsize]{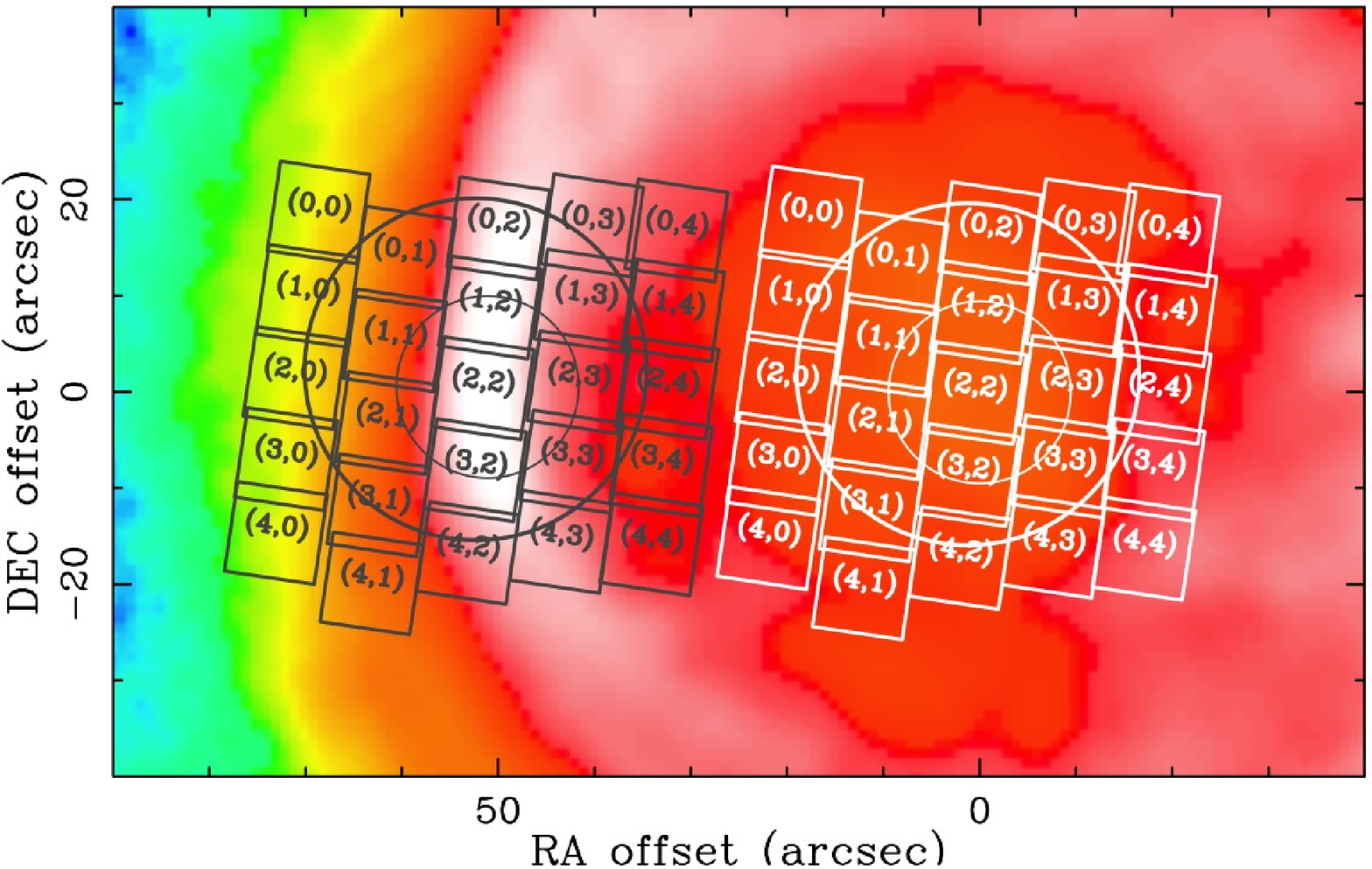}\\
    \includegraphics[width=\hsize]{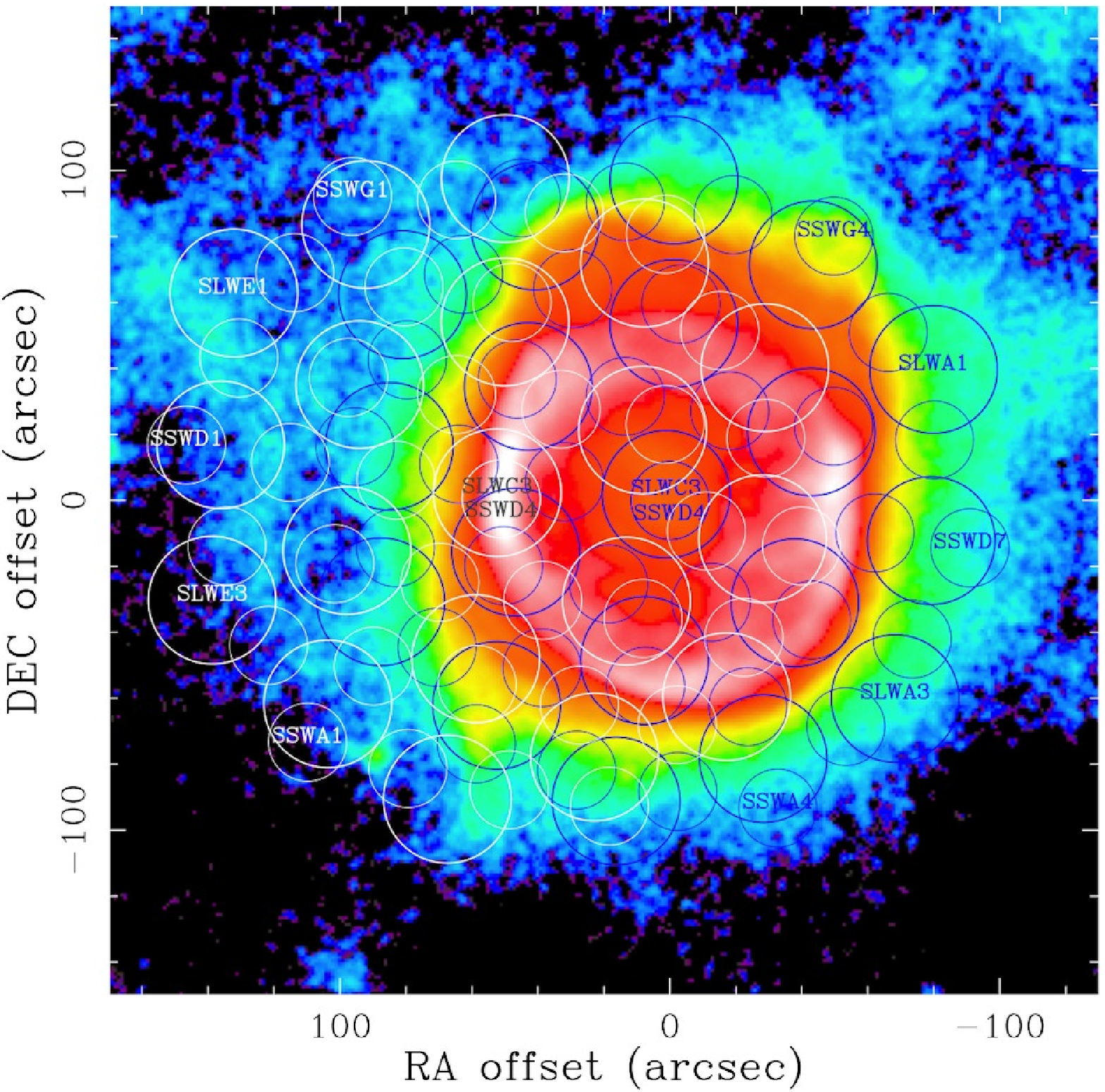}
    \caption{\label{specposmaps}
    [Top] Individual PACS spaxels and the central SPIRE SSW/SLW 
    bolometers (the smaller circles represent SSW) for
    each of the two pointings (center in white, rim in gray) are
    overlaid on the central   
    $130^{\prime\prime} \times 80^{\prime\prime}$ region of the 
    PACS 70\,$\mu$m image.
    Labels indicate specific spaxels.
    [Bottom] Individual SPIRE SSW/SLW bolometers for each of the
    two pointings (center in blue, rim in white; the smaller circles
    represent SSW) are overlaid on
    the central 
    $290^{\prime\prime} \times 290^{\prime\prime}$ region of the 
    PACS 70\,$\mu$m image.
    To specify the instrument orientation, certain bolometers are
    identified by their identifiers.}
   \end{figure}
%_____________________________________________________________

To investigate the spatial variations of the spectral characteristics 
in \object{NGC\,6781}, we extracted individual spectra from each of
the PACS spaxels and SPIRE bolometers at two distinct pointings
within the target nebula.
While the center pointing was directed toward the middle of the cavity of
the ring structure, the rim pointing was aimed at the continuum surface
brightness peak in the eastern rim of the ring structure.
Fig.\,\ref{specposmaps} displays the complete two-pointing footprints of
the PACS spaxels and SPIRE bolometers with respect to the structures of
\object{NGC\,6781} seen at 70\,$\mu$m (see also Fig.\,\ref{schematic}).
With these pointings, we obtained 50, 70, and 38 separate PACS,
SPIRE/SSW, and SPIRE/SLW spectra, respectively, each probing a specific
line of sight in each band.
In the HerPlaNS data set, the spectroscopic flux units are set to be those
of surface brightness (mJy\,arcsec$^{-2}$ per wavelength bin, whose
size is optimized roughly to 0.013\,$\mu$m for PACS and 
0.037--0.45\,$\mu$m for SPIRE). 
The flux density in Jy for an emitting region can be obtained by
integrating the surface brightness within the area of that
region.\footnote{To compute practically the flux for extended sources,
the spectroscopic surface brightness at each 
wavelength bin must be multiplied by the {\sl apparent} area of 
the aperture or the region of interest  
(cf.\ the spaxel/bolometer aperture size varies with wavelength from 
$9\farcs6$--$13\farcs2$ side for PACS, from 17--21$^{\prime\prime}$
diameter for SPIRE/SSW, and from 29--42$^{\prime\prime}$ diameter for
SPIRE/SLW).} 

\begin{figure*}
    \centering
    \includegraphics[width=\hsize]{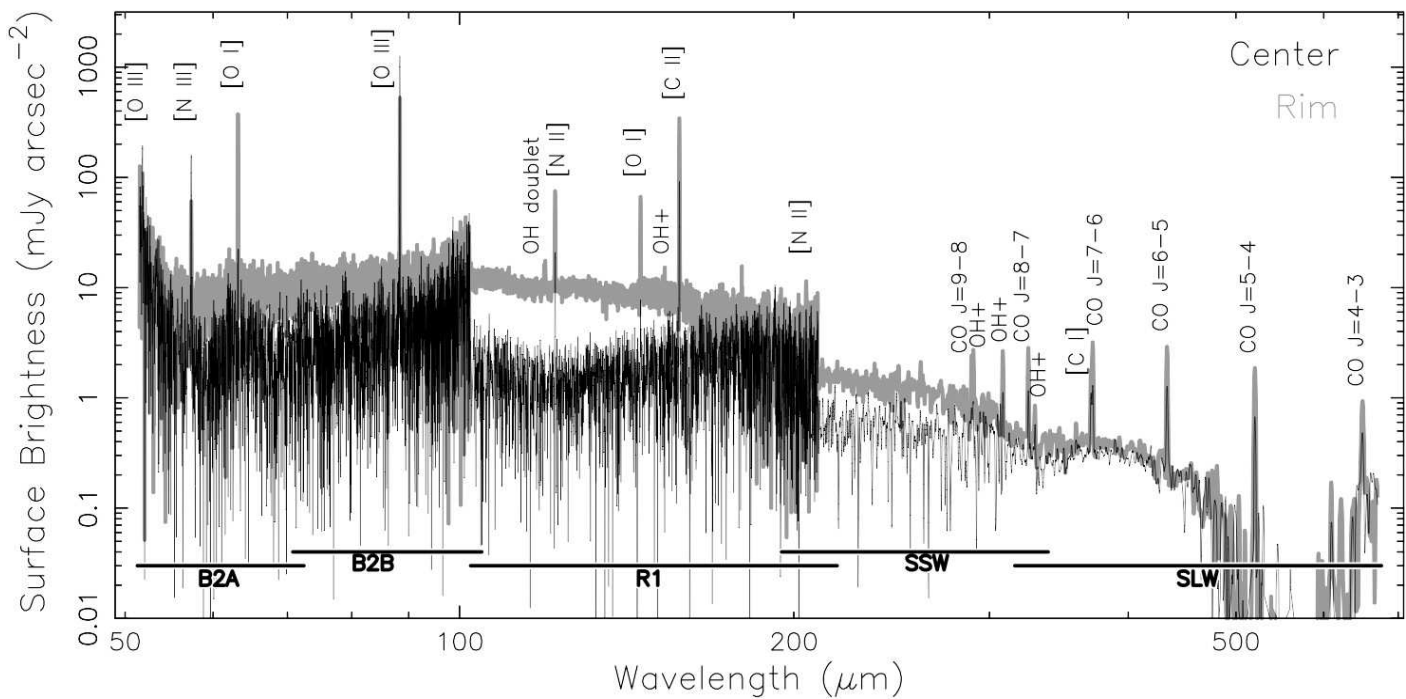}
    \caption{\label{wholespec}
    Spectra of \object{NGC\,6781} over the complete PACS/SPIRE 
    spectral coverage (51--672\,$\mu$m) extracted from the PACS
    central spaxel and the SPIRE central
    bolometers, taken at the ``center'' of the nebula (black line)
    and on the eastern ``rim'' (gray line).
    The range of each spectral band is indicated by a horizontal bar
    near the bottom edge of the plot. 
    The flux units are set to be those of the surface brightness 
    (mJy\,arcsec$^{-2}$) at the specific position in the nebula.
    Various ionic, atomic, and molecular lines are detected and 
    labeled as identified.}
   \end{figure*}

Far-IR spectra of \object{NGC\,6781} for the complete PACS/SPIRE
spectral range (51--672\,$\mu$m) at each of the two pointings
are presented in Fig.\,\ref{wholespec}:
the black (gray) spectrum is taken at the center (rim) pointing.
These spectra are constructed by combining 
a PACS spectrum from the central (2,2) spaxel and SPIRE SSW/SLW spectra
from the central C3/D4 bolometers, respectively
(see Fig.\,\ref{specposmaps} for their spatial relationship).  
For the present analysis, we did not take into account the difference in
the actual area of the sky subtended by the central PACS spaxel and
SPIRE bolometers as well as the beam dilution effect).  
The peak (median) sensitivities in the PACS B2A/B2B and R1 bands
and in the SPIRE SSW and SLW bands are
0.56 (3.74)\,mJy\,arcsec$^{-2}$,
0.24 (1.53)\,mJy\,arcsec$^{-2}$,
0.004 (0.101)\,mJy\,arcsec$^{-2}$, and
0.001 (0.023)\,mJy\,arcsec$^{-2}$,
respectively.

In Table \ref{linefluxes}, flux measurements for the presently
confirmed lines are summarized. 
Note that these measurements are based on the total PACS spectrum (i.e.,
all 25 spaxels integrated) and SPIRE spectra from the central D4/C3
bolometers.
In general, the measured fluxes appear to be consistent with those
obtained with {\sl ISO\/} LWS \citep{liu01}, 
given the different aperture size of {\sl ISO\/} LWS (about 
$40^{\prime\prime}$ radius).
However, calibration of the [\ion{O}{III}]\,51.8\,$\mu$m line flux at
the PACS B2A band edge may be uncertain due to known spectral 
leakage (see Sect.\,\ref{linediag}).

%_____________________________________________________________
% Figure 7: Whole Spectral Range
%-------------------------------------------------------------
%_____________________________________________________________

At both pointings, we detected continuum emission ranging from a few to 
10\,mJy\,arcsec$^{-2}$ in the PACS bands ($< 210\,\mu$m) and from a few
tenths to a few\,mJy\,arcsec$^{-2}$ in the SPIRE bands ($> 210\,\mu$m).
Thermal dust emission in the PACS bands and the SPIRE SSW band is
stronger at the eastern rim than at the nebula center, while it is about
the same at both pointings in the SPIRE SLW band.
This indicates that 
(1) dust grains having temperatures less than about 10\,K (corresponding
to those emitting at $\gtrsim 300$\,$\mu$m) are distributed uniformly in
the nebula,
and
(2) dust grains having temperatures more than about 10\,K (corresponding
to those emitting at $\lesssim 300$\,$\mu$m) are more abundant along the
columns toward the rim. 
Hence, the dust component emitting at $\gtrsim$\,300\,$\mu$m probably
represents the part of the nebula surrounding the central highly-ionized
regions, corroborating the presence of the polar caps as suggested by
the analysis of the broadband images. 
The equation of thermal balance between radiation and the cold dust
component in the polar caps under the $\lambda^{-1}$ dust emissivity
assumption suggests that dust grains of 10\,K would
be located at 50$^{\prime\prime}$ away from the star at 950\,pc.
As the radius of the dust barrel is about 40$^{\prime\prime}$, 
this simple calculation suggests that the inner cavity of
\object{NGC\,6781} is slightly elongated along the polar axis.

%_____________________________________________________________
% Table 5: Line Fluxes Table
%_____________________________________________________________
\begin{table}
\footnotesize
\caption{\label{linefluxes}  
 Line fluxes measured at two positions in NGC\,6781}         % title of Table
\centering                 
\begin{tabular}{lccc}     
\hline\hline               
Line & $\lambda$ & \multicolumn{2}{c}{Flux (10$^{-16}$ W/m$^{2}$)} \\
  & ($\mu$m) & Center & Rim \\ 
\hline
{[}\ion{O}{III}]  & \phantom{1}51.8 & 248\phantom{.00}$\pm$10\phantom{.00} & 112\phantom{.00}$\pm$13\phantom{.00} \\
{[}\ion{N}{III}]   & \phantom{1}57.3 & 237\phantom{.00}$\pm$\phantom{0}2\phantom{.00} & 103\phantom{.00}$\pm$\phantom{0}1\phantom{.00} \\
{[}\ion{O}{I}] & \phantom{1}63.2 & \phantom{0}48.7\phantom{0}$\pm$\phantom{0}0.9\phantom{0} & 104\phantom{.00}$\pm$\phantom{0}1\phantom{.00} \\
{[}\ion{O}{III}] & \phantom{1}88.4 & 573\phantom{.00}$\pm$\phantom{0}1\phantom{.00} & 252\phantom{.00}$\pm$\phantom{0}1\phantom{.00} \\
OH  & 119.2 & \phantom{00}0.91$\pm$\phantom{0}0.16 & \phantom{00}1.64$\pm$\phantom{0}0.17  \\
OH  & 119.4 & \phantom{00}0.99$\pm$\phantom{0}0.18 & \phantom{00}1.84$\pm$\phantom{0}0.18  \\
{[}\ion{N}{II}]  & 121.9 & \phantom{0}15.6\phantom{0}$\pm$\phantom{0}0.3\phantom{0} & \phantom{0}18.8\phantom{0}$\pm$\phantom{0}0.4\phantom{0} \\
{[}\ion{O}{I}]  & 145.6 & \phantom{00}4.12$\pm$\phantom{0}0.15 & \phantom{00}9.50$\pm$\phantom{0}0.17 \\
OH$^{+}$  & 153.0 & \phantom{00}0.49$\pm$\phantom{0}0.12 & \phantom{00}0.79$\pm$\phantom{0}0.14 \\
{[}\ion{C}{II}] & 157.8 & \phantom{0}36.8\phantom{0}$\pm$\phantom{0}0.17 & \phantom{0}48.1\phantom{0}$\pm$\phantom{0}0.1\phantom{0} \\
{[}\ion{N}{II}] &  205.2 & \phantom{00}1.08$\pm$\phantom{0}0.05 &  \phantom{00}1.84$\pm$\phantom{0}0.06\\
CO J=9--8 &  289.1 & \phantom{00}0.25$\pm$\phantom{0}0.12 & \phantom{00}0.71$\pm$\phantom{0}0.11 \\
OH$^{+}$ &  290.2 & \phantom{00}0.26$\pm$\phantom{0}0.12 & \phantom{00}0.72$\pm$\phantom{0}0.11 \\
OH$^{+}$ &  308.4 & \phantom{00}0.47$\pm$\phantom{0}0.04 & \phantom{00}0.43$\pm$\phantom{0}0.01 \\
CO J=8--7 & 325.3 & \phantom{00}0.40$\pm$\phantom{0}0.04 & \phantom{00}0.46$\pm$\phantom{0}0.02 \\
OH$^{+}$   & 329.7 & \phantom{00}0.10$\pm$\phantom{0}0.05 & \phantom{00}0.05$\pm$\phantom{0}0.04 \\
{[}\ion{C}{I}] & 370.3 & \phantom{00}0.86$\pm$\phantom{0}0.03 & \phantom{00}0.67$\pm$\phantom{0}0.02 \\
CO J=7--6 & 371.6 & \phantom{00}0.86$\pm$\phantom{0}0.03 & \phantom{00}0.90$\pm$\phantom{0}0.01 \\
CO J=6--5 & 433.5 & \phantom{00}0.53$\pm$\phantom{0}0.03 & \phantom{00}0.51$\pm$\phantom{0}0.01 \\
CO J=5--4 & 520.3 & \phantom{00}0.30$\pm$\phantom{0}0.03 & \phantom{00}0.38$\pm$\phantom{0}0.13 \\
CO J=4--3 & 650.3 & \phantom{00}0.16$\pm$\phantom{0}0.03 & \phantom{00}0.17$\pm$\phantom{0}0.01 \\
\hline
\end{tabular}
\tablefoot{Flux values are integrated over the entire
 PACS aperture up to 200\,$\mu$m, 
 and measured only from the central bolometer 
 at the SPIRE SSW/SLW bands beyond 200\,$\mu$m,
 without considering the beam dilution effect.
 See the top frame of Fig.\,\ref{specposmaps} for the relative
 placements of the PACS spaxels and SPIRE central bolometers.}
\end{table}
%_____________________________________________________________

Besides the continuum, 
detected are a number of ionic and atomic emission lines such as 
[\ion{O}{III}]\,52, 88\,$\mu$m,
[\ion{N}{III}]\,57\,$\mu$m,
[\ion{O}{I}]\,63, 146\,$\mu$m,
[\ion{N}{II}]\,122, 205\,$\mu$m, and
[\ion{C}{II}]\,158\,$\mu$m
in the PACS bands
and a number of CO rotational lines in the SPIRE bands.
In the center pointing spectra, high-excitation ionic lines are stronger 
whereas low-excitation ionic, atomic, and molecular lines are weaker.
In the rim spectra, however, we observe the opposite.
The different relative strengths of these lines at the two positions
suggest that the central cavity is more strongly ionized than the
eastern rim, as expected. 

While thorough line identification and analysis will be deferred to
forthcoming spectroscopy papers in the series, 
we point out here that a fair number of weaker lines are also detected
in addition to the lines mentioned above, such as those thought to be
the OH 119\,$\mu$m $^{2}\Pi_{3/2}$ J=5/2$^{+}$--3/2$^{-}$ $\Lambda$-doublet
transitions at 119.2 and 119.4\,$\mu$m \citep{melnick87},
OH$^{+}$ lines at 153.0, 290.2, 308.4, and 329.7\,$\mu$m, 
and 
[\ion{C}{I}] at 370.4\,$\mu$m.
Line flux measurements made for these 
lines are summarized in Table\,\ref{linefluxes}.
Among these weaker lines, detection of OH$^{+}$ in emission is particularly
rare.
In fact, the detection of OH$^{+}$ in emission was made in two other PNs of
the HerPlaNS sample, and is the subject of a stand-alone
HerPlaNS paper \citep{aleman13}. 
In addition, the detection of OH$^{+}$ in emission was also made
in two other PNs independently by \citet{etxaluze13} as part of the  
{\sl Herschel} MESS (Mass Loss of Evolved Stars) key program
\citep{mess}. 
As we have already demonstrated, these emission lines contribute
negligibly to broadband flux measurements, and hence, to the SED fitting
analysis of the dust temperature. 

%_____________________________________________________________
% Figure 8: CO rotation diagram
%-------------------------------------------------------------
   \begin{figure}
    \centering
    \includegraphics[width=\hsize]{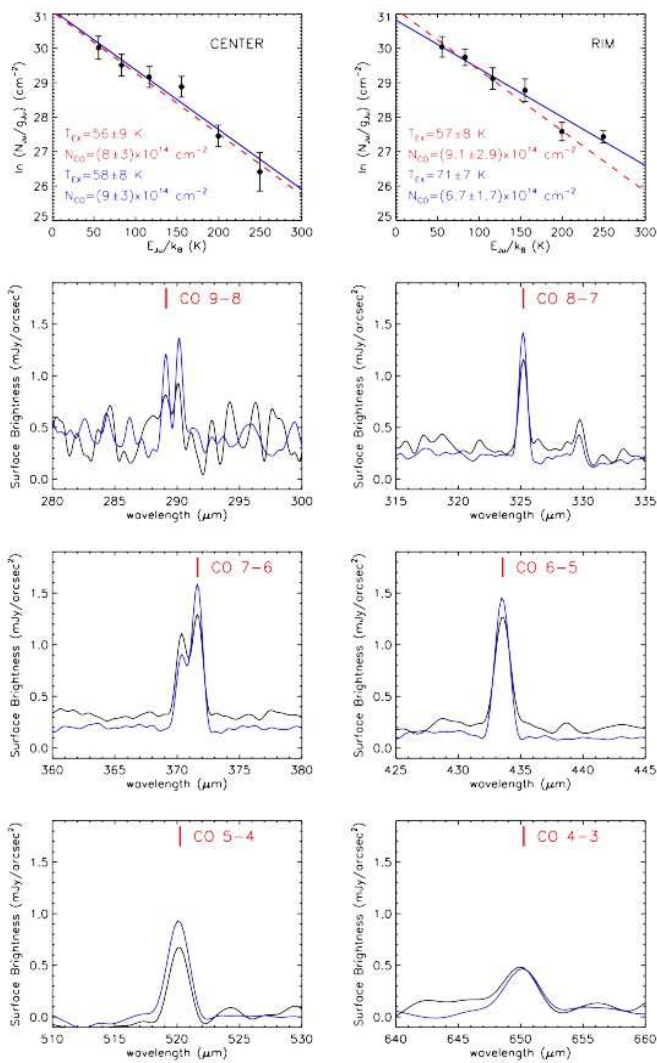}
    \caption{\label{corotation}
    [Top Two Frames] CO rotation diagrams constructed from the lowest six
    of the observed transitions.
    The blue line is a fit using all six transitions and the red line is
    a fit without the lowest two lines with a low S/N.
    The left frame is of the center pointing, while the right is of the
    rim pointing. 
    Here, the beam dilution effect is not considered.
    [Bottom Six Frames] CO line profiles of the six transitions from
    J=9--8 to J=4--3 used in the analysis, extracted from the spectra
    taken at the central bolometer. 
    The black line is of the center pointing, while the blue line is of
    the rim pointing.
    The two transitions not included in the second fit are 
    CO J=9--8 (highly uncertain due to low S/N) and 
    CO J=7--6 (blended with the [\ion{C}{I}] line at 370.3\,$\mu$m).}
   \end{figure}
%_____________________________________________________________

Using measured fluxes of CO lines from J=9--8 to 4--3 transitions
detected above the 2-$\sigma$ signal-to-noise limit
(and ignoring the effects of beam dilution, which has not been
fully calibrated in HIPE),
we calculated the CO excitation temperature, $T_{\rm ex}$, and column
density, $N_{\rm CO}$, by least-squares fitting,  
following the formalism of \citet{gl99} under the optically thin 
assumption.
\citet{bachiller97} reported a $^{12}$CO column density of 
$1.4 \times 10^{16}$\,cm$^{-2}$ towards \object{NGC\,6781} based on 
$^{12}$CO J=2--1 and 1--0 maps. 
Assuming this column density for the upper levels of the transitions
that we detected the optical depth in each line would still be much
less than unity \citep{gl99}.
Hence, our assumption of optically thin CO emission is reasonable.

Fig.\,\ref{corotation} shows the CO excitation diagrams for the center
(top left) and rim (top right) pointings, as well as individual CO
spectra from J=9--8 to 4--3 (the rest of the frames, 
from top left to bottom right).
These calculations yielded $T_{\rm ex} = 56 \pm 9$\,K 
and $N_{\rm CO} = (8\pm3)\times10^{14}$\,cm$^{-2}$
for the center pointing and $57 \pm 8$\,K and
$N_{\rm CO} = (9\pm3)\times10^{14}$\,cm$^{-2}$
for the rim pointing
(blue lines in the top frames of Fig.\,\ref{corotation}). 
Neglecting two lines at J=9--8 (marginal detection at $\sim2$-$\sigma$)
and J=7--6 (blending with the [\ion{C}{I}] line at 370.3\,$\mu$m),
fitting instead resulted in $T_{\rm ex} = 58 \pm 8$\,K 
and $N_{\rm CO} = (9\pm3)\times10^{14}$\,cm$^{-2}$
for the center pointing and 
$71 \pm 7$\,K and $N_{\rm CO} = (7\pm2)\times10^{14}$\,cm$^{-2}$
for the rim pointing (dashed red lines in the top two frames
in Fig.\,\ref{corotation}).
In this excitation diagram fitting, the uncertainties are obtained by
standard error propagation from the uncertainties of the line intensity
measurements.
The values of E(J$_{\rm u}$) and Einstein coefficients are taken from
the HITRAN database.\footnote{http://www.cfa.harvard.edu/hitran/, version 2012.}

The present measurements from higher-J transitions suggest that 
the bulk of CO gas remains at low temperature and preferentially
detected at the lowest-J transitions.
However, the spatial distribution of the CO gas component is very much
restricted to where we see thermal dust continuum emission, indicating
that most of the CO gas is contained within the cylindrical barrel
structure most likely temperature-stratified in the polar directions.

%_____________________________________________________________
% Figure 9: Whole SPIRE Spectra
%-------------------------------------------------------------
   \begin{figure}
    \centering
    \includegraphics[width=\hsize]{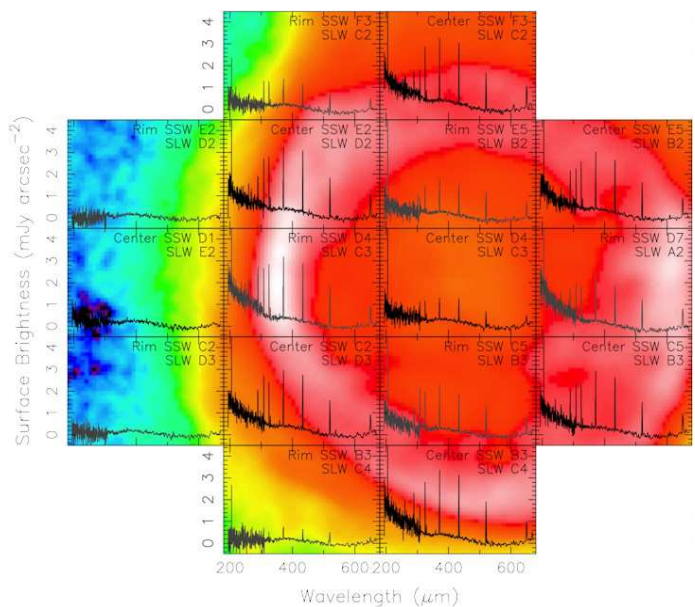}
    \caption{\label{wholespire}
    Spectra of \object{NGC\,6781} over the complete SPIRE spectral
    coverage (194--672\,$\mu$m) at 16 distinct positions, 
    where the SSW and SLW bolometers spatially overlap reasonably well.    
    Bolometer positions are nested between the two pointings, and
    spectra extracted from the center pointing are shown in black and
    those from the rim pointing are shown in dark gray.
    The background PACS 70\,$\mu$m image indicates the approximate
    locations of the corresponding bolometers within the nebula.  
    The flux units are set to the surface brightness (mJy\,arcsec$^{-2}$).
    The measurements are valid roughly within the central 
    $30^{\prime\prime}$ of the specific location of the bolometers.
    The spatial variation of the strength of the CO rotational
    transition lines and thermal dust continuum is clearly detected.}
   \end{figure}
%_____________________________________________________________

%_____________________________________________________________
% Figure 10: Whole PACS Spectra
%-------------------------------------------------------------
   \begin{figure}
    \centering
    \includegraphics[width=\hsize]{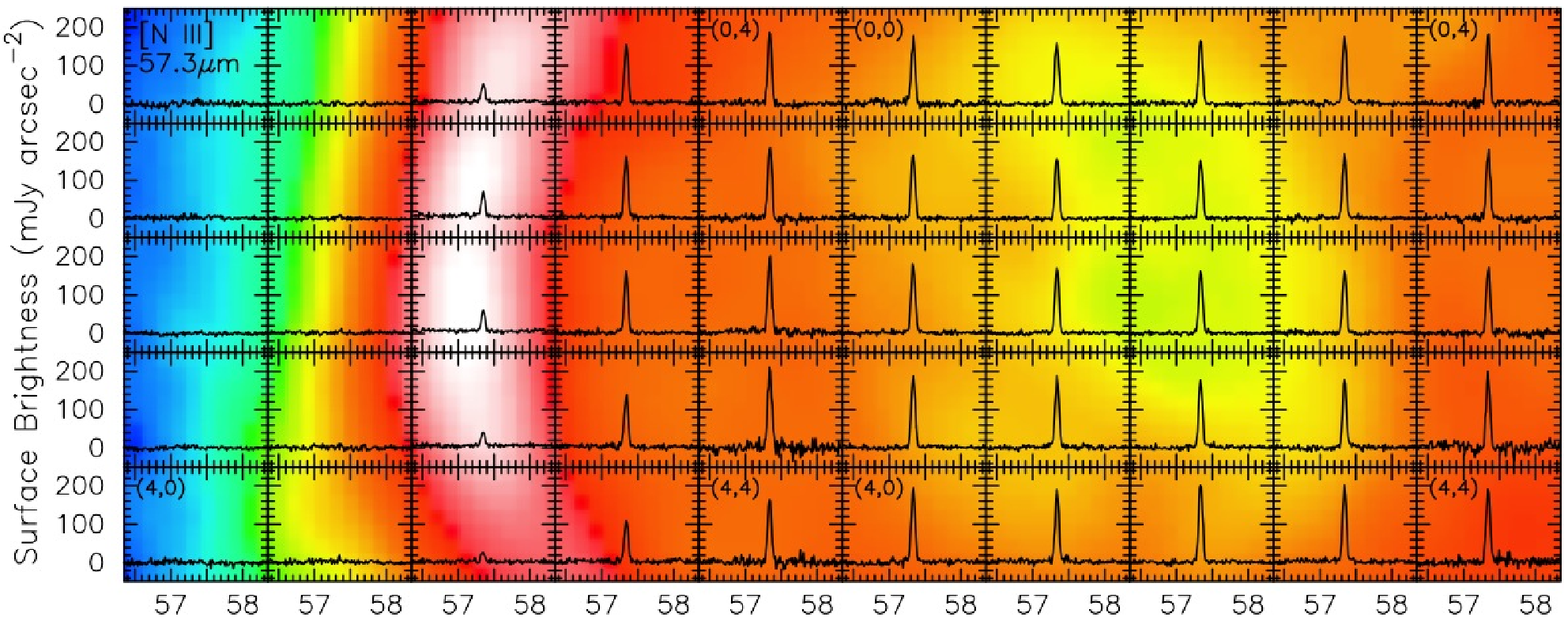}\vspace{5pt}
    \includegraphics[width=\hsize]{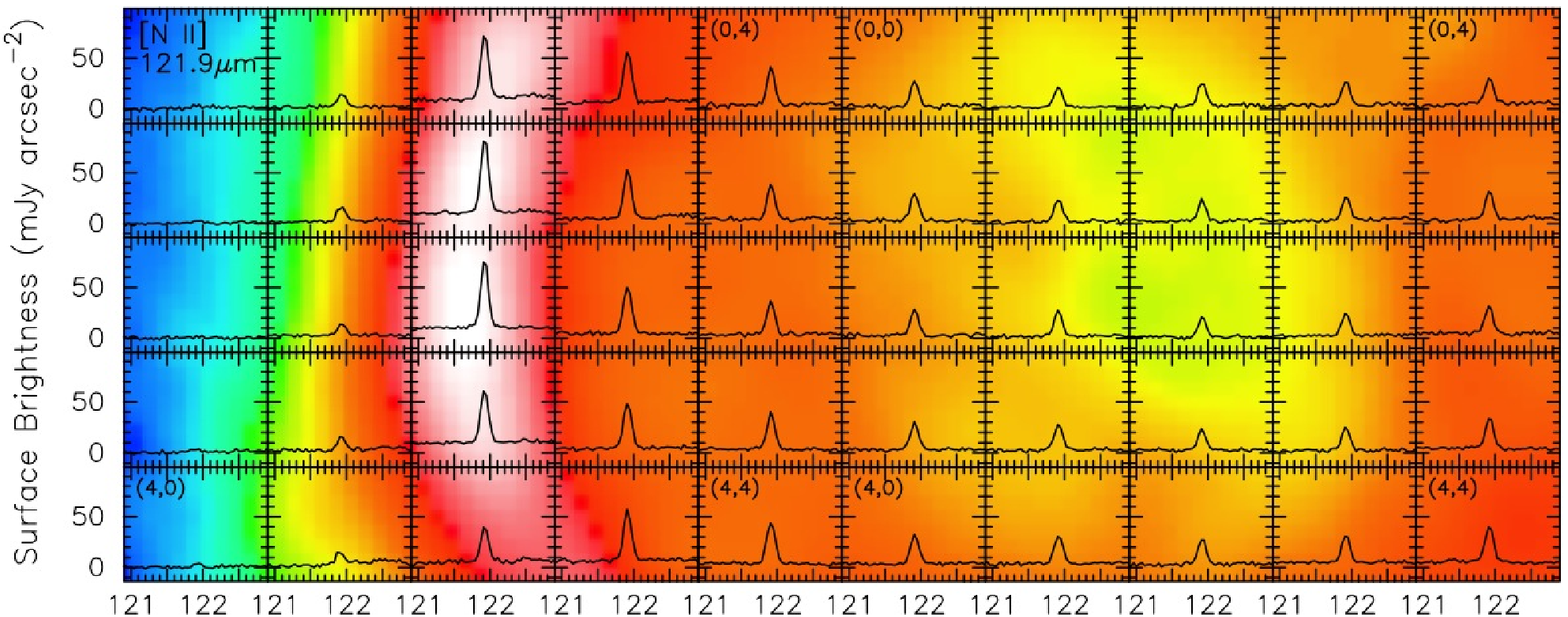}\vspace{5pt}
    \includegraphics[width=\hsize]{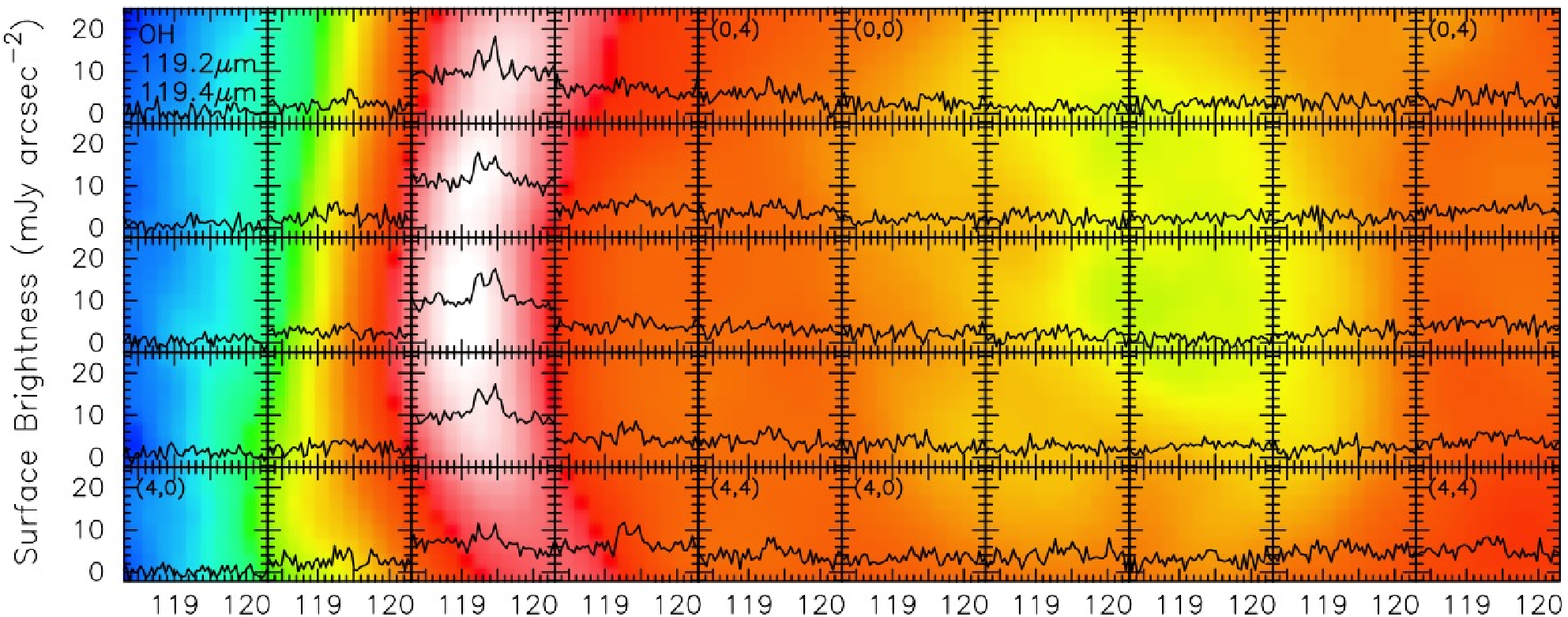}\vspace{5pt}
    \includegraphics[width=\hsize]{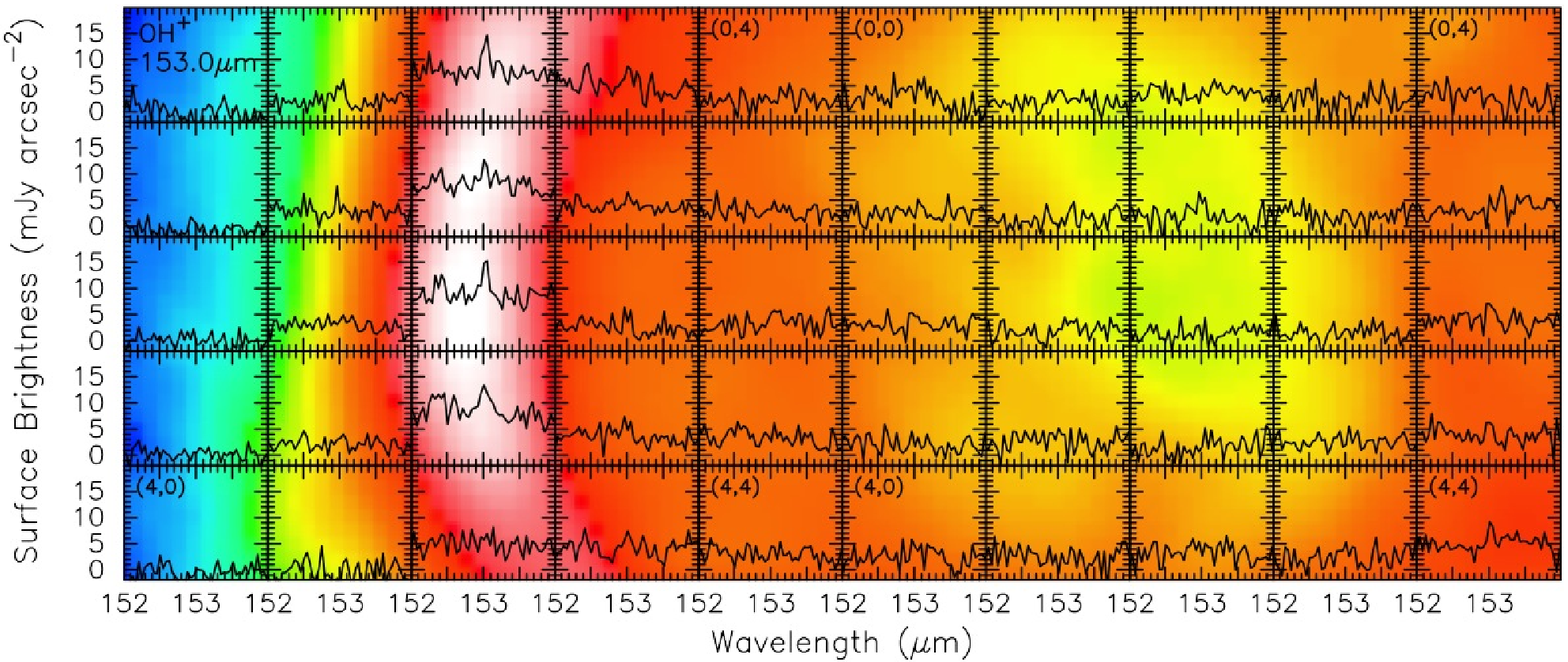}
    \caption{\label{pacsspatial}
    Spectra extracted from individual $5\times5$ PACS spaxels  
    at [\ion{N}{III}]\,57.3\,$\mu$m, [\ion{N}{II}]\,121.9\,$\mu$m,
    OH doublet 119.2/119.4\,$\mu$m, and OH$^{+}$ 153\,$\mu$m
    in each of the two pointings toward
    \object{NGC\,6781} shown side by side: ``center'' on the right and
    ``rim''  on the left.
    To specify the instrument orientation, corner spaxels are
    identified by their identifiers.
    The flux unit is set to the surface brightness (mJy\,arcsec$^{-2}$).
    The background PACS 70\,$\mu$m image indicates  
    the approximate location of each spaxel.
    Ionic lines tend to be strong in the highly-ionized cavity
    of the cylindrical structure, while atomic and molecular lines tend
    to be pronounced in the cylindrical rim of the nebula.
    Note that the footprint of the PACS IFU is not regular as
    implied by the placement of the sub-frames; the slightly irregular
    footprint can be seen in Figs.\ \ref{specposmaps} and 
    \ref{linemaps}.}
   \end{figure}
%_____________________________________________________________

%-----------------------------------------------------------------
% Figure 11: Line Maps 
%-----------------------------------------------------------------
   \begin{figure}
    \centering
    \includegraphics[width=\hsize]{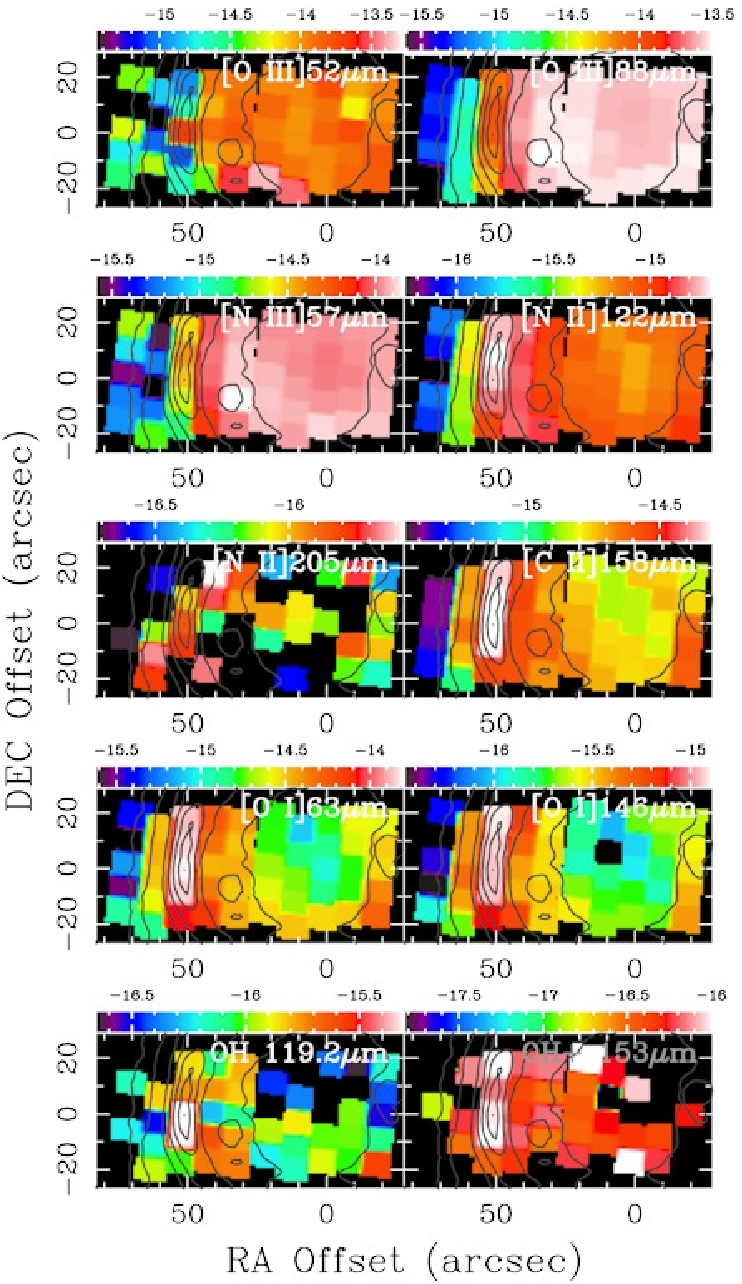}
    \caption{\label{linemaps}
    Line intensity maps covering the central
    $111^{\prime\prime} \times 56^{\prime\prime}$ region of 
    \object{NGC\,6781} at
    [\ion{O}{III}]\,52,88\,$\mu$m,
    [\ion{N}{III}]\,57\,$\mu$m,
    [\ion{N}{II}]\,122,205\,$\mu$m,
    [\ion{C}{II}]\,158\,$\mu$m
    [\ion{O}{I}]\,63,146\,$\mu$m,
    OH doublet\,119.2,119.4\,$\mu$m,
    and
    OH$^{+}$ 153\,$\mu$m.
    Upon integrating the PACS IFU data cube over each line, the
    continuum level was determined using surface brightnesses on both
    sides of the line.
    Pixel values of the maps were not set unless line emission 
    registers more than 3-$\sigma$:
    this is why some spaxels appear to be blank (especially
    in the [\ion{N}{II}]\,205\,$\mu$m map).
    These intensity maps are overlaid with the PACS 70\,$\mu$m contours 
    (as in Fig.\,\ref{broadmaps}).
    These maps are made at
    1\,arcsec\,pix$^{-1}$ so that 
    footprints of the original PACS IFU spaxels of 9\farcs4\,arcsec
    can be seen. 
    The color wedges show the log of the line intensity in units of 
    erg\,s$^{-1}$\,cm$^{-2}$\,arcsec$^{-2}$.
    The [\ion{N}{II}] map at 205\,$\mu$m is scaled to have the total line
    intensity equal to the SPIRE measurements, because PACS flux
    calibration in the 205\,$\mu$m region is uncertain.}   
   \end{figure}
%_________________________________________________________________

Spatial variations of the line strength can be investigated by comparing
individual spectra taken from each PACS spaxel and SPIRE bolometer.
For example, Fig.\,\ref{wholespire} displays 16 spectra covering the
whole SPIRE range (194--672\,$\mu$m) that are recovered from locations
within the nebula at which the SPIRE SSW and SLW bolometers overlap 
(Fig.\,\ref{specposmaps}).
These spectra, presented with the 70\,$\mu$m image in the background,
indicate that both thermal dust continuum and CO line emission are more 
prominent along the barrel wall, hinting at generally colder and denser
conditions within the barrel wall. 

Meanwhile, Fig.\,\ref{pacsspatial} shows spatial variations of
the excitation conditions within the nebula.
The [\ion{N}{III}] line strength distribution (Fig.\,\ref{pacsspatial},
top) reveals uniformly high excitation conditions in the central
cavity, which gradually decreases as the column density increases 
toward the rim over three spaxels ($\equiv 30^{\prime\prime}$).
The opposite trend is seen in the lower-excitation [\ion{N}{II}] line
strength distribution (Fig.\,\ref{pacsspatial}, second from top).
On the other hand, the molecular OH and OH$^{+}$ lines
(Fig.\,\ref{pacsspatial}, bottom two) appear exclusively in the barrel
wall, suggesting that the presence of molecular gas is spatially very
much restricted.
The line strength distribution maps for other lines can be found 
in the Appendix (Figs.\,\ref{pacsspecall1},
\ref{pacsspecall2}, \ref{spirespecall1}, and \ref{spirespecall2}).

\subsection{Spectral Mapping}

\subsubsection{Spatially-Resolved Far-IR Emission Line Maps}

The PACS line strength distribution maps introduced in the previous
section (Fig.\,\ref{pacsspatial}, as well as Figs.\,\ref{pacsspecall1} 
and \ref{pacsspecall2}) are not spatially accurate, as spaxels are 
not exactly aligned on a $5\times5$ square grid due to internal
misalignment.
Therefore, the PACS IFU data cube was rendered into spectral maps 
by taking into account the internal offsets among spaxels.
In the case of \object{NGC\,6781}, the fields of view of two pointings
are adjacent to each other, and hence, cover roughly two-thirds
of the radius of the nebula along the equatorial plane.
Fig.\,\ref{linemaps} shows such mosaicked line emission distribution 
maps of the central $40^{\prime\prime}\times110^{\prime\prime}$ region  
in
[\ion{O}{III}]\,52, 88\,$\mu$m,
[\ion{N}{III}]\,57\,$\mu$m,
[\ion{N}{II}]\,122, 205\,$\mu$m,
[\ion{C}{II}]\,158\,$\mu$m,
[\ion{O}{I}]\,63, 146\,$\mu$m,
OH doublet at 119.2, 119.4\,$\mu$m,
and
OH$^{+}$ at 153\,$\mu$m,
respectively, from top left to bottom right.

These line maps of \object{NGC\,6781} intuitively show that the
distribution of line emission is fairly uniform within the barrel cavity
and tends to vary over a roughly 30$^{\prime\prime}$-wide region across
the barrel wall.  
The high-excitation line maps at [\ion{O}{III}]\,52, 88\,$\mu$m and 
[\ion{N}{III}]\,57\,$\mu$m show stronger emission from within the 
barrel cavity, with the strongest emission tending toward the inner wall 
of the cavity, about 40$^{\prime\prime}$ to the east from the nebula
center. 
On the other hand, the low-excitation and atomic line maps at 
[\ion{N}{II}]\,122, 205\,$\mu$m, 
[\ion{C}{II}]\,158\,$\mu$m, and
[\ion{O}{I}]\,63, 146\,$\mu$m
exhibit concentrations of surface brightness along the barrel wall, 
about 50$^{\prime\prime}$ to the east from the center.

Hence, the barrel wall region is where the gradient of the line
emission strengths tends to become large. 
The spatial coincidence of various emission lines of ionic, neutral,
and molecular nature revealed here shows that 
(1) the temperature gradient is fairly steep across the inner barrel wall,
 and 
(2) the barrel wall is stratified with physically distinct layers. 
The line maps in Fig.\,\ref{linemaps} also indicate that the 
line ratios for a given ionic or atomic species (such as
[\ion{O}{III}]\,52\,$\mu$m/88\,$\mu$m and
[\ion{O}{I}]\,63\,$\mu$m/146\,$\mu$m) vary significantly within the
nebula. 
Therefore, we stress that single-valued line ratios obtained by treating 
PNs as point sources are inadequate for purposes of line diagnostic
investigations to understand their structures. 

Based on the present data augmented with the previous results in the
literature, we propose the following picture of stratification across
the nebula volume of \object{NGC\,6781}.
The inner cavity is highly ionized and the surface of the barrel wall is
mostly ionized, while the wall itself is dense enough to maintain a
large amount of column of molecular species. 
Between the dense molecular/neutral barrel wall and the ionized cavity, 
there should be a layer of photo-dissociation region (PDR) shielding the
barrel from the UV field of the 
central star.
There is also another PDR layer on the outer surface of the barrel wall,
which shields the barrel against the UV field of the 
interstellar radiation field.
These PDRs are likely the origins of various neutral and molecular
lines detected in the far-IR and elsewhere. 
Dense PDR clumps embedded in an otherwise ionized gas in the
central cavity could also provide sites for production of various
emission lines as in NGC\,7293 (e.g., \citealt{speck02}) and NGC\,650
(e.g., \citealt{vanhoof13}). 
This complex cylindrical barrel region (up to about 50$^{\prime\prime}$
radius from the star) is surrounded by a region of cold dust extending
to roughly 100$^{\prime\prime}$ radius (Fig.\,\ref{broadmaps}).   

\subsubsection{Electron Temperature/Density Diagnostics of the \ion{H}{II}/Ionized Regions\label{linediag}}

%_____________________________________________________________
% Table 6: Te, ne 
%_____________________________________________________________
\begin{table*}
\caption{\label{tntable}  
 Spatially-Resolved ($T_{\rm e}$, $n_{\rm e}$) in the \ion{H}{II} region
 of \object{NGC\,6781}}
\centering                   
\begin{tabular}{lcccccc}  
\hline\hline                 
Position\tablefootmark{a} & $\Delta$RA\tablefootmark{b} & 
$n_{\rm e}$[\ion{O}{III}] & 
$n_{\rm e}$[\ion{O}{I}] & 
$n_{\rm e}$[\ion{N}{II}] &
$T_{\rm e}$[\ion{O}{III}] &
$T_{\rm e}$[\ion{N}{II}] \\
& (arcsec) & (cm$^{-3}$) & (cm$^{-3}$) & (cm$^{-3}$) & (K) & (K) \\
\hline                        % inserts single horizontal line
Cen5  &  -20  &  530$\pm$80  &  1,520$\pm$400  &  370$\pm$\phantom{1}90  &  \phantom{1}9,670$\pm$\phantom{1}10  &  \phantom{0}8,300$\pm$\phantom{1}80  \\
Cen4  &  -10  &  320$\pm$30  &  \phantom{1,}820$\pm$420  &  220$\pm$\phantom{1}40  &  \phantom{1}9,690$\pm$\phantom{1}20  &  \phantom{0}7,990$\pm$\phantom{1}70  \\
Cen3  &  \phantom{-0}0  &  350$\pm$40  &  $<600$  &  \phantom{0}80$\pm$\phantom{1}10  &  \phantom{1}9,800$\pm$\phantom{1}10  &  \phantom{0}7,920$\pm$\phantom{1}60  \\
Cen2  &  \phantom{-}10  &  410$\pm$70  &  \dots  &  105$\pm$\phantom{1}20  &  \phantom{1}9,730$\pm$\phantom{1}10  &  \phantom{0}8,240$\pm$\phantom{1}60  \\
Cen1  &  \phantom{-}20  &  350$\pm$40  &  1,920$\pm$670  &  220$\pm$\phantom{1}50  &  \phantom{1}9,780$\pm$\phantom{1}10  &  \phantom{0}8,250$\pm$\phantom{1}70  \\
Rim5  &  \phantom{-}30  &  400$\pm$60  &  \phantom{1,}650$\pm$270  &  960$\pm$430  &  \phantom{1}9,920$\pm$\phantom{1}10  &  \phantom{0}8,730$\pm$\phantom{1}50  \\
Rim4  &  \phantom{-}40  &  270$\pm$40  &  $<300$  &  300$\pm$\phantom{1}80  &  10,000$\pm$\phantom{1}10  &  \phantom{0}9,410$\pm$\phantom{1}40  \\
Rim3  &  \phantom{-}50  &  220$\pm$80  &  1,270$\pm$\phantom{1}80  &  640$\pm$210  &  10,940$\pm$\phantom{1}30  &  11,580$\pm$\phantom{1}60  \\
Rim2  &  \phantom{-}60  &  \dots  &  \dots  &  450$\pm$130  &  \phantom{1}7,400$\pm$\phantom{1}60  &  \phantom{0}6,250$\pm$\phantom{1}50  \\
Rim1  &  \phantom{-}70  &  \dots  &  $<300$  &  \dots  &  \phantom{1}5,700$\pm$100  &  \phantom{0}4,780$\pm$240  \\
\hline                                   %inserts single line
\end{tabular}
\tablefoot{%
\tablefoottext{a}{Position along the RA direction defined by integer
 multiples of $10^{\prime\prime}$ from the nebula center, at which
 the surface line strength is summed over the 5-spaxel column along the
 DEC direction   
 (also indicated at the top of Fig.\,\ref{tnprofiles}).}
\tablefoottext{b}{Relative angular distance in the RA direction from the
 center spaxel of the center pointing.}}
\end{table*}

Far-IR fine-structure line ratios such as 
[\ion{O}{III}]\,52/88\,$\mu$m and
[\ion{N}{II}]\,122/205\,$\mu$m 
are relatively insensitive to the electron temperature 
($T_{\rm e}$), because the fine-structure levels of the
$^{3}$P ground state are close enough in energy: one can derive 
the electron density ($n_{\rm e}$) from a range of 
$T_{\rm e}$ (e.g., \citealt{rubin94,liu01}).
Meanwhile, $T_{\rm e}$ can be inferred from, for example,
optical-to-far-IR line ratios such as  
[\ion{O}{III}]\,$\lambda$5007/88\,$\mu$m and
[\ion{N}{II}]\,$\lambda$6583/122\,$\mu$m, which are relatively  
insensitive to $n_{\rm e}$
(i.e., one can derive $T_{\rm e}$ from a range of $n_{\rm e}$).
By iterative application of the above processes, one can {\sl derive\/}
the optimum ($T_{\rm e}$, $n_{\rm e}$) pair for a given set of line
ratios {\sl without any prior assumption}. 

Now that we obtained spatially-resolved far-IR line maps, 
we can observationally establish the ($T_{\rm e}$, $n_{\rm e}$)
distributions from the central region to the eastern barrel wall of
\object{NGC\,6781} as a function of position.
Because the orientation of the PACS IFU field of view happened to be
almost aligned with the (RA, DEC) coordinates (Fig.\,\ref{linemaps}), 
we collapsed these line maps along the DEC direction to yield a 1-D
surface line strength profile as a function of position in the RA
direction. 
For the present analysis, we resampled the line maps back 
at the nominal $10^{\prime\prime} \times 10^{\prime\prime}$ spaxel
size to verify internal 
consistencies of the line map generation procedure with individual
spaxel measurements.

Using the IRAF/STSDAS {\sl nebular\/} package\footnote{In this analysis
 package, the equation of statistical equilibrium is solved for the
 lowest five excitation levels of a given atom \citet{nebular}.
 Upon using the package, we updated the atomic data following the
 reference list summarized by Otsuka et al.\ (2010; their Table\,7).}
we first derived the $T_{\rm e}$ radial profile 
from the [\ion{O}{III}]\,$\lambda$5007/88\,$\mu$m ratio profile
with an assumed flat $n_{\rm e}$ profile.
Then, the $n_{\rm e}$ radial profile was recovered from the
[\ion{O}{III}]\,52/88\,$\mu$m ratio profile using the derived 
$T_{\rm e}$ profile.
These two steps were repeated until the ($T_{\rm e}$, $n_{\rm e}$)
radial profile pair converged.
Unfortunately, the [\ion{O}{III}]\,52\,$\mu$m line flux calibration was
compromised by spectral leakage between adjacent grating orders. 
Therefore, 
assuming that the relative surface line strength within the
[\ion{O}{III}]\,52\,$\mu$m map was unaffected by the leakage,
we scaled the [\ion{O}{III}]\,52\,$\mu$m map 
by a factor of 3.2 so that the 
[\ion{O}{III}]\,52/88\,$\mu$m ratio averaged over the entire RA profile   
equals 1.45, which is the ratio computed from the respective line
strength measurements ($151/104 = 1.45$) made by \citet{liu01}
in their previous spatially unresolved study with {\sl ISO}.

We also used the [\ion{N}{II}]\,$\lambda$6583/122\,$\mu$m
and [\ion{N}{II}]\,122/205\,$\mu$m ratio profiles
in the same iterative process to derive the optimum
($T_{\rm e}, n_{\rm e}$) radial profiles for lower density regions
(e.g., \citealt{rubin94}).  
Unfortunately, the strength of the [\ion{N}{II}] emission at 205\,$\mu$m
was close to the PACS R1 band detection limit, and hence, many spaxels
in the [\ion{N}{II}]\,205\,$\mu$m map did not yield valid measurements.
In fact, the integrated [\ion{N}{II}] 205\,$\mu$m line strengths measured
from the PACS and SPIRE spectra differed by more than a factor of two
 (after considering the difference in the respective aperture sizes).
Thus, we adopted the more reliable SPIRE measurements and 
scaled the PACS [\ion{N}{II}] 205\,$\mu$m line map so that the integrated
line strength of the map equals with the measured SPIRE line strength.
In our analyses outlined above, the [\ion{O}{III}]\,$\lambda$5007 and 
[\ion{N}{II}]\,$\lambda$6583 maps in the optical were augmented from the 
study by \citet{phillips11} using data obtained at the NOT.

%-----------------------------------------------------------------
% Figure 12: (Te, ne) and abundance profiles
%-----------------------------------------------------------------
   \begin{figure*}
    \centering
    \includegraphics[width=0.49\hsize]{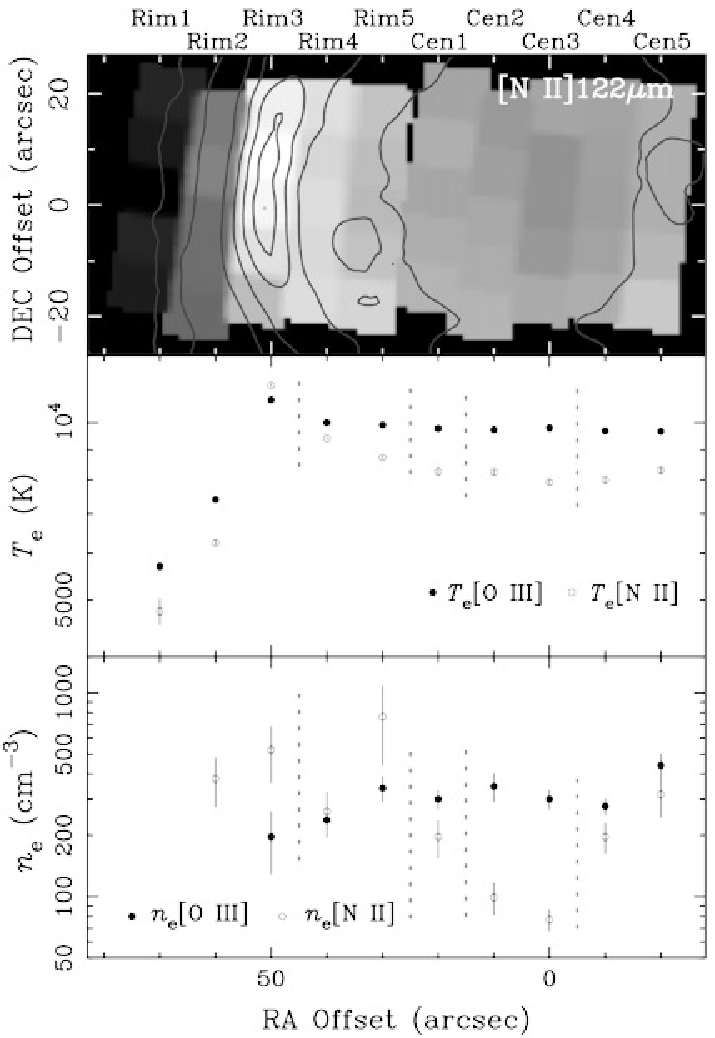}
    \includegraphics[width=0.49\hsize]{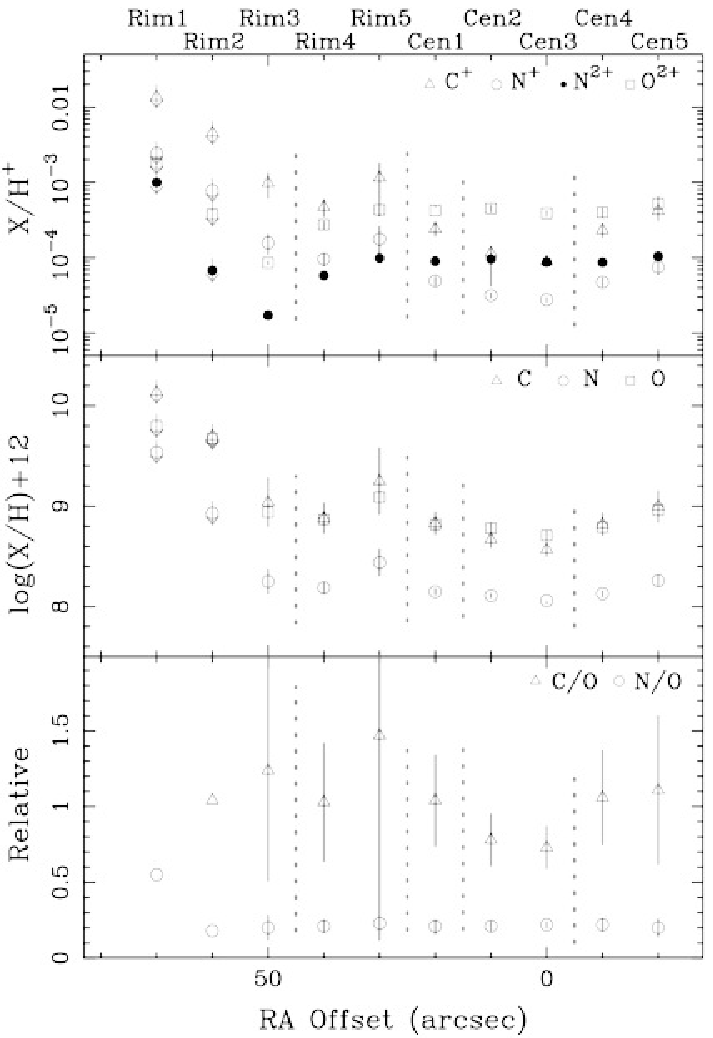}
    \caption{\label{tnprofiles}
    The ($T_{\rm e}$, $n_{\rm e}$) and abundance profiles along the RA
    direction in the observed mid-eastern region of \object{NGC\,6781}, 
    designated as Cen1-Cen5 and Rim1-Rim5 as shown in Table \ref{tntable}.
    [Top Left] The [\ion{N}{II}] 122\,$\mu$m line map from
    Fig.\,\ref{linemaps}.
    [Middle Left] The $\log T_{\rm e}$ profiles.
    [Bottom Left] The $\log n_{\rm e}$ profiles.
    [Top Right] The relative ionic abundance profiles.
    [Middle Right] The relative elemental abundance profiles
    (where $\log N_{\rm H}=12$). 
    [Bottom Right] The C/O and N/O profiles.
    The legend of symbols is given in each frame;
    filled circle: $T_{\rm e}$ or $n_{\rm e}$ based on [\ion{O}{III}],
    open circle: $T_{\rm e}$ or $n_{\rm e}$ based on [\ion{N}{II}],
    triangle: C$^{+}$, $\log({\rm C/H})+12$, and C/O,
    circle: N$^{+}$ (open), N$^{2+}$ (filled), $\log({\rm N/H})+12$, and N/O,
    square: O$^{2+}$ and $\log({\rm O/H})+12$.
    The vertical gray line associated with each symbol represents
    uncertainties, while the vertical down arrows indicate 
    upper limits (all of the values at Rim1 and Rim2).
    Note that these values are exclusively of the \ion{H}{II} region of
    \object{NGC\,6781} (see Sect.\ \ref{pdr}).
    The vertical gray dashed lines indicate boundaries between  
    stratified regions discussed in Sect.\,3.3.5.}   
   \end{figure*}
%_________________________________________________________________

Thus, the above {\sl nebular\/} diagnostics of the line ratio profiles
across the eastern radius of \object{NGC\,6781} yielded the 
($T_{\rm e}$, $n_{\rm e}$) radial profiles, which are summarized in 
Table\,\ref{tntable} and plotted in the left frames of Fig.\,\ref{tnprofiles}.  
The $T_{\rm e}$ radial profiles derived from the [\ion{O}{III}] and
[\ion{N}{II}] line profiles are consistent with each other,
revealing a highly-ionized region of almost constant $T_{\rm e}$ (from 
8,100--9,700\,K) in the barrel cavity, surrounded by the barrel wall at
which $T_{\rm e}$ reaches the maximum ($\sim10,500$\,K; \citealt{mpp01})
and beyond which 
$T_{\rm e}$ tapers off (middle frame of Fig.\,\ref{tnprofiles}).
The $n_{\rm e}$ radial profile derived from the [\ion{O}{III}] line
profile (pluses) shows a generally   
constant distribution ($\sim 400$\,cm$^{-3}$; cf.\ \citealt{mpp01}),
while that derived  
from the [\ion{N}{II}] profile (crosses) hints at a radially increasing
tendency (from 80\,cm$^{-3}$ at the center of the cavity to
$>600$\,cm$^{-3}$ around the barrel wall; bottom frame of
Fig.\,\ref{tnprofiles}).

Such a radially increasing trending of $n_{\rm e}$ is expected from
the line strength surface brightness map in the [\ion{N}{II}] line 
(top left frame of Fig.\,\ref{tnprofiles}).
The lower $n_{\rm e}$[\ion{N}{II}] is probably affected by the presence
of the high excitation region in the very center of the cavity (seen in
the \ion{He}{II}$\lambda$4686, from which [\ion{C}{IV}], [\ion{N}{IV}],
and [\ion{O}{IV}] lines, which were neglected in the present analysis,
probably arise). 
Because various diagnostic lines probe a range of excitations,
$n_{\rm e}$ determined from various $n_{\rm e}$ indicator lines 
will vary depending on which indicator line is used \citep{rubin94}.   
Here, we have demonstrated this indicator-line-dependent nature of line
diagnostics for the first time in the far-IR
by radially resolving the ionization structure.  

In previous far-IR line diagnostics performed by \citet{liu01}, 
$n_{\rm e}$[\ion{O}{III}] was found to be 371.5\,cm$^{-3}$
under the assumption of a constant $T_{\rm e}$ at $10^{4}$\,K.
A follow-up study with optical data in [\ion{O}{III}], [\ion{N}{II}], and
[\ion{O}{II}] later verified that $T_{\rm e}$ is between 10,200 and
10,600\,K \citep{liuy04}. 
Indeed, our iterative, more self-consistent method confirmed 
that $n_{\rm e}$[\ion{O}{III}] in the central cavity 
(i.e., the average of the mid-sections, Cen2, 3, and 4;
Table\,\ref{tntable})
is roughly 360\,cm$^{-3}$, in which $T_{\rm e}$[\ion{O}{III}] is about  
$9,700$\,K.
In the previous single-beam observations, these numbers were
considered representative of the entire nebula.

However, we emphasize here that these values are representative of
just the inner cavity and that details such as the $T_{\rm e}$ peak 
around the barrel wall and the structured $n_{\rm e}$[\ion{N}{II}] 
profile across the nebula would not have been discovered without
spatially-resolved far-IR data.
Hence, it is imperative that spatially-resolved line diagnostics are
performed in future to diagnose the spatially-resolved PN energetics not
only in the far-IR but also in other wavelengths at which relevant
diagnostic lines exist to take full advantage of the maximum spatial
resolving power possible.  

\subsubsection{Physical Conditions of the PDR\label{pdr}}

The low-excitation and atomic line maps at 
[\ion{C}{II}] 158\,$\mu$m and [\ion{O}{I}] 63, 146\,$\mu$m are typically
used to probe the physical conditions of the PDR \citep{tielens10}. 
On one hand, the [\ion{O}{I}] lines at 63 and 146\,$\mu$m are expected to
arise exclusively from the PDR, because the ionization
potential of O$^{0}$ is 13.6\,eV and no O$^{0}$ is expected 
in the \ion{H}{II} regions (\ion{H}{II}Rs). 
On the other hand, the [\ion{C}{II}] line at 158\,$\mu$m can
arise from both the \ion{H}{II}Rs and PDRs,
because the ionization potential of C$^{0}$ is 11.3\,eV and C$^{+}$ can
exist in both \ion{H}{II}Rs and PDRs \citep{malhotra01}.
Therefore, there is a need to separate the \ion{H}{II}R and PDR
components of the [\ion{C}{II}] 158\,$\mu$m line strength distribution.

The intensity of the [\ion{C}{II}] 158\,$\mu$m line emission
arising from \ion{H}{II}Rs along a particular line of sight should
roughly scale with that of the [\ion{N}{II}] 122\,$\mu$m line emission 
along the same line of sight.
Moreover, whenever the [\ion{N}{II}] 122\,$\mu$m emission is detected, 
the observed intensity of the [\ion{C}{II}] 158\,$\mu$m emission along
the same line of sight have to be the sum of contributions arising from
the PDRs and \ion{H}{II}Rs along the line of sight. 
This occurs simply because the ionization potential of N$^{0}$,
14.5\,eV, is close to but larger than that of C$^{0}$, 11.3\,eV.

Hence, assuming that the nebula gas is well-mixed along the line of
sight, we can express the relationship between strengths of the
[\ion{C}{II}] 158\,$\mu$m and [\ion{N}{II}] 122\,$\mu$m lines 
as 
$F_{[\ion{C}{II}]158\mu{\rm m}} =
F_{[\ion{C}{II}]158\mu{\rm m}}^{\ion{H}{II}{\rm R}} +
F_{[\ion{C}{II}]158\mu{\rm m}}^{\rm PDR} =
\alpha \times F_{[\ion{N}{II}]122\mu{\rm m}} + \beta$
for each spaxel, where 
$F_{[\ion{N}{II}]122\mu{\rm m}}$ and $F_{[\ion{C}{II}]158\mu{\rm m}}$ 
are the integrated strengths of the corresponding lines 
{\sl per spaxel\/}, with the superscripts for the 
$F_{[\ion{C}{II}]158\mu{\rm m}}$ line indicating the origins of the
emission. 
If we further assume that the \ion{H}{II}R component of the 
[\ion{C}{II}] 158\,$\mu$m line emission dominates, the scaling factor 
$\alpha$ can be determined by a linear fitting of the
measured line strengths of the 
[\ion{C}{II}] 158\,$\mu$m and [\ion{N}{II}] 122\,$\mu$m lines.
The validity of the last assumption is corroborated by excellent spatial
correspondence between the [\ion{N}{II}] 122\,$\mu$m and [\ion{C}{II}]
158\,$\mu$m line strength distribution maps (Fig.\,\ref{linemaps}, the
second and third frames from the top in the right column, respectively).

%-----------------------------------------------------------------
% Figure 13: C II vs. N II
%-----------------------------------------------------------------
   \begin{figure}
    \centering
    \includegraphics[width=\hsize]{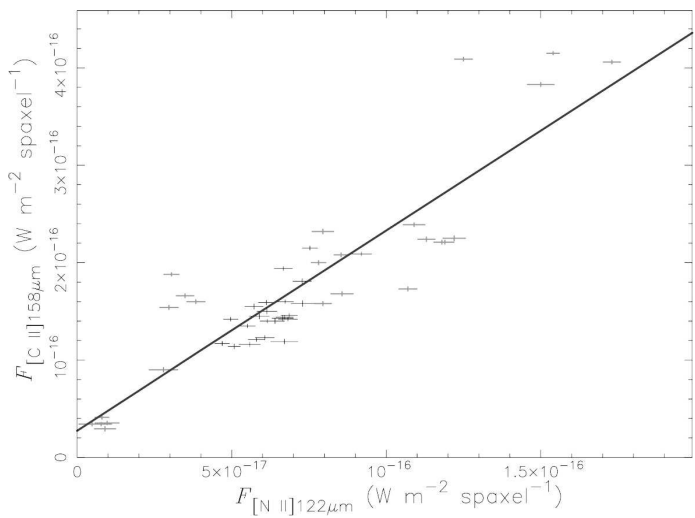}
    \caption{\label{c2vsn2}
    Correlation between integrated line fluxes per spaxel at 
    [\ion{N}{II}] 122\,$\mu$m and [\ion{C}{II}] 158\,$\mu$m for both
    the center (black symbols) and rim (light-gray symbols) pointings.
    The sizes of these symbols indicate uncertainties.
    Measurements from the center pointing are clustered closer together
    than those from the rim pointing, suggesting that 
    the line strength distribution is more or less uniform in the
    central cavity. 
    The dark-gray line is the best-fit to the data points: 
    $F_{[\ion{C}{II}]158\mu{\rm m}} = (2.05 \pm 0.15) \times
    F_{[\ion{N}{II}]122\mu{\rm m}}
    + (2.74 \pm 1.17) \times 10^{-17}$.}
   \end{figure}
%_________________________________________________________________

Fig.\,\ref{c2vsn2} shows the relationship between the integrated line
strengths of the [\ion{C}{II}] 158\,$\mu$m emission against those of the
[\ion{N}{II}] 122\,$\mu$m emission at each spaxel for both the center and
rim pointings. 
The best-fit to the observed linear correlation is determined to be
$F_{[\ion{C}{II}]158\mu{\rm m}} = (2.1 \pm 0.2) \times 
F_{[\ion{N}{II}]122\mu{\rm m}} + (2.7 \pm 1.2) \times 10^{-17}$
W\,m$^{-2}$.
As $F_{[\ion{C}{II}]158\mu{\rm m}} = (1.9 \pm 0.5) \times 10^{15}$\,W\,m$^{-2}$, 
we confirm that the \ion{H}{II}R component of the 
[\ion{C}{II}] 158\,$\mu$m line emission indeed dominates
(which is expected because the observed regions mostly cover the highly 
ionized inner cavity, also revealed by our $T_{\rm e}$ analysis; 
Fig.\,\ref{tnprofiles}, Table\,\ref{tntable}).
By this way, we estimated the surface brightness distributions 
of [\ion{C}{II}] 158\,$\mu$m emission arising exclusively from 
the \ion{H}{II}R by the relation,
$F_{[\ion{C}{II}]158\mu{\rm m}}^{\ion{H}{II}{\rm R}} =
2.1  \times F_{[\ion{N}{II}]122\mu{\rm m}}$,
and 
of [\ion{C}{II}] 158\,$\mu$m arising exclusively from the PDR 
by the difference,
$F_{[\ion{C}{II}]158\mu{\rm m}}^{\rm PDR} = 
F_{[\ion{C}{II}]158\mu{\rm m}} -
2.1 \times F_{[\ion{N}{II}]122\mu{\rm m}}$.

As in the above analysis with the surface brightness maps 
of lines arising exclusively from the \ion{H}{II}R of the nebula, 
we can probe the ($T_{\rm e}$, $n_{\rm e}$) profiles in the PDR of the
nebula using line maps arising exclusively from the PDRs. 
Using the [\ion{O}{I}] 63/146\,$\mu$m ratio profile and the 
$T_{\rm e}$ profile, we can derive the $n_{\rm e}$ profile for the PDR
of the nebula.
Unfortunately, however, there is no suitable optical surface brightness
map in a neutral O line (e.g., an [\ion{O}{I}]\,$\lambda$6300 map) to
be paired with a far-IR line map in order to derive the $T_{\rm e}$
profile for further iteration.
Hence, we simply adopted the $T_{\rm e}$[\ion{N}{II}] profile for the
following analysis of the PDR of the nebula.
The resulting $n_{\rm e}$ profile (Table\,\ref{tntable}) is very
irregular, and even lacks a measured value at 10$^{\prime\prime}$
E of the center.
As one immediately sees from the individual spectra 
(Figs.\,\ref{pacsspecall1} and \ref{pacsspecall2}), the [\ion{O}{I}] line
emission arises almost exclusively from the column density peak of the
barrel wall (i.e., middle of the rim pointing).
Therefore, we adopted the value 
$n_{\rm e}{\rm [\ion{O}{I}]}=1,270$\,cm$^{-3}$ at the barrel wall in the
discussion below.

Following the PDR temperature-density diagnostic scheme elucidated by 
\citet{liu01}, 
the ($T_{\rm PDR}$, $n_{\rm H^{0}}$) pair at the PDR part of the barrel
wall can be estimated.
According to the PDR temperature-density diagnostic diagram
prepared by \citet{vn13} (their Fig.\,10),
the line strength ratios of [\ion{O}{I}] 63\,$\mu$m to 146\,$\mu$m
($13.8$) and of [\ion{O}{I}] 63\,$\mu$m to the PDR part of [\ion{C}{II}]
158\,$\mu$m ($5.2$) would suggest $T_{\rm PDR} \gtrsim 5,000$\,K and 
$n_{\rm H^{0}} \approx 6,300$\,cm$^{-3}$.
We consider these values tentative, however, as the ratios obtained from
the present data are slightly beyond the edge of the parameter 
space explored by \citet{vn13}. 
Hence, further analysis is deferred to the forthcoming spectroscopic
analysis papers of the {\sl HerPlaNS\/} series, in which 
a [\ion{C}{I}] 370\,$\mu$m map is derived from SPIRE spectra
and a custom PDR diagnostic diagram is constructed for a wider range of
parameters. 

\subsubsection{Spatially-Resolved Abundance Analysis}

%_____________________________________________________________
% Table 7: Abundances
%_____________________________________________________________
\begin{table*}
\caption{\label{abundances}
Spatially-Resolved Abundances in \object{NGC\,6781}}
\centering
\begin{tabular}{lcccccccccc}
\hline\hline
Position\tablefootmark{a}&$\Delta$RA\tablefootmark{b}& C$^{+}$/H$^{+}$        &N$^{+}$/H$^{+}$ &N$^{2+}$/H$^{+}$            &O$^{2+}$/H$^{+}$ &C/O         &N/O       &C        &N &O    \\
&($''$)&        (x10$^{-4}$)        &(x10$^{-5}$) &(x10$^{-5}$)    &(x10$^{-4}$)    &        & &        &        &    \\
\hline
Cen5&    -20&    \phantom{1}4.2$\pm$1.1    &\phantom{1}7.6$\pm$1.5 &10.5$\pm$1.3
        &5.3$\pm$1.1    &1.1$\pm$0.5 &0.2$\pm$0.1    &9.0$\pm$0.2    &8.3$\pm$0.1 &9.0$\pm$0.1\\
Cen4&    -10&    \phantom{1}2.3$\pm$0.5    &\phantom{1}4.8$\pm$0.6 &\phantom{1}8.7$\pm$0.6
        &4.0$\pm$0.5    &1.1$\pm$0.3 &0.2$\pm$0.1    &8.8$\pm$0.1    &8.1$\pm$0.1 &8.8$\pm$0.1\\
Cen3&    0&    \phantom{1}0.9$\pm$0.1    &\phantom{1}2.8$\pm$0.2 &\phantom{1}8.8$\pm$0.3
        &3.9$\pm$0.3    &0.7$\pm$0.1 &0.2$\pm$0.1    &8.5$\pm$0.1    &8.1$\pm$0.1 &8.7$\pm$0.1\\
Cen2&    10&    \phantom{1}1.2$\pm$0.2    &\phantom{1}3.2$\pm$0.2 &\phantom{1}9.6$\pm$0.4
        &4.5$\pm$0.4    &0.8$\pm$0.2 &0.2$\pm$0.1    &8.7$\pm$0.1    &8.1$\pm$0.1 &8.8$\pm$0.1\\
Cen1&    20&    \phantom{1}2.4$\pm$0.5    &\phantom{1}5.0$\pm$0.6 &\phantom{1}9.0$\pm$0.6
        &4.2$\pm$0.5    &1.0$\pm$0.3 &0.2$\pm$0.1    &8.8$\pm$0.1    &8.2$\pm$0.1 &8.8$\pm$0.1\\
Rim5&    30&    11.5$\pm$6.2&17.8$\pm$8.2 &\phantom{1}9.9$\pm$1.1
        &4.3$\pm$0.8    &1.5$\pm$1.3 &0.2$\pm$0.1    &9.3$\pm$0.3    &8.4$\pm$0.1 &9.1$\pm$0.2\\
Rim4&    40&    \phantom{1}4.7$\pm$1.2    &\phantom{1}9.7$\pm$1.8 &\phantom{1}5.8$\pm$0.4
        &2.8$\pm$0.3    &1.0$\pm$0.4 &0.2$\pm$0.1    &8.9$\pm$0.2    &8.2$\pm$0.1 &8.9$\pm$0.1\\
Rim3&    50&    \phantom{1}9.8$\pm$3.4    &15.9$\pm$4.7 &\phantom{1}1.7$\pm$0.1
        &0.9$\pm$0.1    &1.2$\pm$0.8 &0.2$\pm$0.1    &9.0$\pm$0.2    &8.3$\pm$0.1 &8.9$\pm$0.1\\
Rim2&    60&    $<$46.    &$<$78. &$<$6.8         &$<$3.8         &$<$1.0      &$<$0.2         &$<$9.7 &$<$8.9         &$<$9.7    \\
Rim1&    70&    $<$140.    &$<$240. &$<$100.         &$<$18.&$<$2.2 &$<$0.6         &$<$10.        &$<$9.5              &$<$9.8    \\

\hline
\end{tabular}
\tablefoot{%
\tablefoottext{a}{As designated in Fig.\,\ref{tnprofiles}} 
\tablefoottext{b}{Relative angular distance in the RA direction from the
center spaxel of the center pointing.}}
\end{table*}
%_____________________________________________________________

Using the IRAF/STSDAS {\sl nebular\/} package,
we translated the ($T_{\rm e}$, $n_{\rm e}$) profiles 
of the \ion{H}{II}R component of the object into
spatially-resolved 
ionic abundance profiles of
C$^{+}$, N$^{+}$, N$^{2+}$/H$^{+}$, and O$^{2+}$ relative to H$^{+}$,
abundance ratio profiles of C/O and N/O, and
elemental abundance profiles of C, N, and O relative to H
(Fig.\,\ref{tnprofiles}, right frames; Table\,\ref{abundances}).

To do so, the number density of N relative to H was estimated via  
\begin{eqnarray}
 \frac{{\rm N}}{{\rm H}} = 
\frac{{\rm <}{\rm N}^{+}{\rm >}+{\rm <}{\rm N}^{2+}{\rm >}}{{\rm <}{\rm H}^{+}{\rm >}}
=
\frac{F_{[\ion{N}{II}]122}/\epsilon_{[\ion{N}{II}]122}+
F_{[\ion{N}{III}]57}/\epsilon_{[\ion{N}{III}]57}}{F_{H\beta}/\epsilon_{H\beta}},
\label{n/h}
\end{eqnarray}
where the
$F$'s are the integrated line strengths per spaxel of the corresponding
lines in W\,m$^{-2}$ and $\epsilon$'s are the volume emissivities for the
corresponding lines.
These $\epsilon$'s were computed based on the ($T_{\rm e}$, $n_{\rm e}$) 
values determined for each spaxel: 
T$_{\rm e}$[\ion{N}{II}] and n$_{\rm e}$[\ion{N}{II}] were adopted to
represent physical conditions of the low-excitation regions, 
while 
T$_{\rm e}$[\ion{O}{III}] and n$_{\rm e}$[\ion{O}{III}] were adopted to
represent physical conditions of the high-excitation regions. 
For this formulation to work, it was assumed that there
existed at most N$^{2+}$ in the \ion{H}{II}R of the nebula close to 
the barrel wall.

In formulating the above N/H equation, we initially assumed that there
were only insignificant 
number of ${\rm N}^{3+}$ and higher charge magnitude ions.
However, the archived {\sl ISO\/} SWS spectrum of \object{NGC\,6781} shows
the [\ion{O}{IV}] 25.9\,$\mu$m line with the peak flux of 64\,Jy. 
The ionization potential of O$^{2+}$ is 54.9\,eV, which is higher than 
that of N$^{2+}$, 47.4\,eV.
Hence, N$^{3+}$ ions are expected to be present in the nebula. 
However, the {\sl ISO\/} SWS aperture 
($14^{\prime\prime}\times27^{\prime\prime}$) is much smaller than the
extended structure of \object{NGC\,6781}, and there is no way for us to
know the spatial distribution of the [\ion{O}{IV}] and [\ion{N}{IV}]
emission. 
According to the \ion{He}{II}$\lambda$4686 map (the ionization potential
of He$^{+}$ is 54.4\,eV) presented by \citet{mpp01}, there appears to be
a region of higher excitation well within the inner cavity of the
cylindrical barrel. 
Therefore, we assumed further that ${\rm N}^{3+}$ (and ${\rm C}^{3+}$
for that matter) would be restricted to this central high excitation 
region and would not significantly affect the subsequent line diagnostic
analysis for the bulk of the nebula beyond this centrally restricted
region. 

Thus, ionic abundance profiles, N$^{2+}$/H$^{+}$ and N$^{+}/$H$^{+}$
(Fig.\,\ref{tnprofiles}, top-right; Table \ref{abundances}), were
determined based on the 
[\ion{N}{III}] and [\ion{N}{II}] line strengths relative to the 
${\rm H}\beta$ line strength. 
For the present analysis, the H$\beta$ map was synthesized by scaling
the H$\alpha$ map of \citet{phillips11} with the extinction-corrected
H$\alpha$/H$\beta$ line ratio.
We computed the H$\alpha$/H$\beta$ line ratio ourselves from the
archival optical spectrum of the object taken with the William Herschel
Telescope Intermediate dispersion Spectrograph and Imaging System, which
was originally presented partially by \citet{liuy04}. 
Finally, the elemental abundance profile, N/H, was obtained by combining
N$^{2+}$/H$^{+}$ and N$^{+}/$H$^{+}$ as formulated above in units of
$\log({\rm X/H})$, for which the H abundance is set to 
$\log(N_{\rm H})=12$ (Fig.\,\ref{tnprofiles}, middle-right; 
Table \ref{abundances}).  

As the ionization potentials of N$^{+}$ and O$^{+}$ are similar (29.60 and
35.1\,eV, respectively) and that of O$^{0}$ is 13.6\,eV (i.e., no
O$^{+}$ remained in the \ion{H}{II}R of the nebula), the number
density of N relative to O in the \ion{H}{II}R of the nebula was
determined from  
\begin{eqnarray}
 \frac{{\rm N}}{{\rm O}}
=\frac{{\rm <}{\rm N}^{2+}{\rm >}}{{\rm <}{\rm O}^{2+}{\rm >}}=
  \frac{F_{[\ion{N}{III}]57}/\epsilon_{[\ion{N}{III}]57}}{F_{[\ion{O}{III}]88}/\epsilon_{[\ion{O}{III}]88}}
\label{n/o}
\end{eqnarray}
(Fig.\,\ref{tnprofiles}, bottom-right; Table \ref{abundances}),
and this ratio allowed us to obtain the O$^{2+}$ ionic abundance profile.
Similarly, the number density of C relative to N in the \ion{H}{II}R of
the nebula was computed from
\begin{eqnarray}
 \frac{{\rm C}}{{\rm N}} 
= \frac{{\rm <}{\rm C}^{+}{\rm >}}{{\rm <}{\rm N}^{+}{\rm >}}
= \frac{F_{[\ion{C}{II}]158}^{\ion{H}{II}}/\epsilon_{[\ion{C}{II}]158}}{F_{[\ion{N}{II}]122}/\epsilon_{[\ion{N}{II}]122}},
\label{c/n}
\end{eqnarray}
which followed from
\begin{eqnarray}
 {\rm C} = {\rm <}{\rm C}^{+}{\rm >} + {\rm <}{\rm C}^{2+}{\rm >}
= {\rm <}{\rm C}^{+}{\rm >} + {\rm C}\times\frac{{\rm <}{\rm N}^{2+}{\rm >}}{{\rm N}}.
\end{eqnarray}
Here, we assumed the second term of the last relation would hold
because the ionization potentials of N$^{+}$ and C$^{+}$ are similar (29.6 and
24.4\,eV, respectively) and the relative ionic abundances are similar
for both N and C.

\subsubsection{Physical Stratification across the Nebula Volume}

The resulting ionic abundance profiles (Fig.\,\ref{tnprofiles}, top
right; Table\,\ref{abundances}) revealed the physical
stratification of the nebula across the cylindrical cavity and barrel
wall along the plane of the sky. 
Based on the relative ionic abundances obtained from the analysis, aided
by the literature data of various optical line emission (e.g.,
\citealt{mpp01}), the nebula structure 
was split into four stratified zones:
(1) the very highly-ionized centrally-restricted region close to the
center of the cavity, in which N$^{2+}$, C$^{2+}$, and O$^{2+}$ are
ionized (47.4, 47.9, and 54.9\,eV, respectively), corresponding to
where \ion{He}{II}$\lambda4687$ emission is detected 
($\lesssim20^{\prime\prime}$),\footnote{\citet{mpp01} states that the
\ion{He}{II} emission region is $\sim38^{\prime\prime}$ radius;
however this appears to have been confused with diameter given the
relative sizes of various emission regions (their Fig.\,1).} 
(2) the highly-ionized cavity roughly between
$20^{\prime\prime}$ and $30^{\prime\prime}$, in
which the bulk of photons carry enough energy to ionize N$^{+}$ and O$^{+}$
(29.6 and 35.1\,eV, respectively), corresponding to where
[\ion{O}{III}]$\lambda5007$ emission is seen,
(3) the inner surface of the barrel wall roughly between
$30^{\prime\prime}$ and $50^{\prime\prime}$ at which photons
still energetic enough to ionize N$^{0}$ and C$^{+}$(14.5 and
24.4\,eV, respectively) impinge on the rising density gradient of the
cylindrical barrel wall, corresponding to the [\ion{N}{II}]$\lambda6584$ 
emission region,
and 
(4) the barrel wall density peak and beyond about $50^{\prime\prime}$,
where only less energetic photons are left to produce O$^{0}$
(13.6\,eV), corresponding to where H$\alpha$ emission is found.

It is worth emphasizing that these ionic abundance transitions take
place at two radial locations in the cavity within the barrel wall at
around $30^{\prime\prime}$ and $50^{\prime\prime}$ from the center.
The first transition at around $30^{\prime\prime}$
occurs well within the ionized cavity region.
Were it not for the [\ion{O}{III}] and [\ion{N}{III}] profiles
(Fig.\,\ref{pacsspatial}, top; Figs.\,\ref{pacsspecall1}, top and bottom), 
this transition zone at around $30^{\prime\prime}$ would not have been
recognized. 

The CNO relative elemental abundances are nearly the same 
within the barrel wall ($\lesssim50^{\prime\prime}$;
Fig.\,\ref{tnprofiles}, middle-right; Table \ref{abundances}).
The average elemental abundances relative to H are 
8.9 for C,
8.2 for N, and 
8.9 for O,
while 
the corresponding solar values are 8.4, 7.8, and 8.7, respectively
\citep{grevesse10}. 
Hence, in \object{NGC\,6781} these elements are 0.5, 0.4, and 0.2 dex
more abundant, respectively, with respect to the solar values.
The N/O relative abundance is especially uniform and relatively low 
across the entire nebula (${\rm N/O} = 0.2$), which lead us to conclude
that \object{NGC\,6781} did not have a high enough N abundance to
be of Peimbert Type I (${\rm N/O} \gtrsim 0.5$; \citealt{peimbert83}).
This is consistent with the previous determination by \citet{liu04} 
based on the nebular He abundance, which barely satisfies the Type I
criterion, N$_{\rm He}$/N$_{\rm H} \ge 0.125$ (\citealt{peimbert83}). 
The above observation of \object{NGC\,6781} not being a Peimbert Type I
PN suggests an upper limit of approximately 2\,M$_{\odot}$ for the 
initial mass of the progenitor star.
While the C/O relative abundance is rather uncertain 
due to uncertainties in separating [\ion{C}{II}]\,158$\mu$m 
line emission into \ion{H}{II}R and PDR components,
the median C/O is $1.1 \pm 0.2$.
This means that the ionized gas is marginally carbon-rich 
(solar ${\rm C/O} = 0.55$; \citealt{grevesse10}).
At any rate, there is no clear indication of an increasing
C/O abundance toward the center of the nebula in the present data, as
one might expect if 
the progenitor star was massive enough to have driven carbon-rich winds 
in the recent past.

In previous spatially-unresolved line diagnostics by 
\citet{liu01} and \citet{liuy04,liu04}, these authors determined
the following ionic abundances:
N$^{+}/{\rm H}^{+}=$\,(5.6--6.8)\,$\times 10^{-5}$,
N$^{2+}/{\rm H}^{+}=1.7\times10^{-4}$, and 
O$^{2+}/{\rm H}^{+}=$\,(2.7--6.2)\,$\times10^{-4}$, 
assuming $T_{\rm e} = 10^{4}$\,K and
$n_{\rm e} = 240$\,cm$^{-3}$.
Our spatially-resolved measurements corroborated the use of uniform
$T_{\rm e}$ and $n_{\rm e}$ values at least in the barrel cavity for the
moderately ionized gas by yielding the averaged ionic abundances of 
$6.3\times10^{-5}$, $8.9\times10^{-5}$, and $4.1\times10^{-4}$,
for N$^{+}$, N$^{2+}$, and O$^{+}$, respectively.

In terms of elemental abundances, previous nebular studies 
reported a dichotomy between abundance determinations (e.g., 
\citealt{gd01,liu04,tsamis04}), in which heavy
element abundances relative to H calculated from weak optical
recombination lines (ORLs) are systematically higher than those derived
from collisionally excited lines (CELs).
Presently, this discrepancy between abundances derived from ORLs and
CELs is suspected to be a strong function of nebular evolution.
For example, \citet{liu04} assessed that the discrepancy was the most
prominent among large, evolved, low-excitation PNs.

The range of elemental abundances that we resolved in different
parts of  
\object{NGC\,6781} actually overlaps with the range of abundances
obtained from ORLs and CELs. 
In previous studies, irrespective of the wavelengths employed, elemental
abundance derivations were almost always performed in a
spatially-integrated manner.
Therefore, the present ORL-CEL discrepancies of elemental abundances
could well be due simply to spatial resolution effects, i.e.,  
each of the ORL and CEL methods may be sensitive to a particular and 
distinct physical condition of the nebula along the unresolved line of   
sight.

The analysis outlined above is very specific to ionized \ion{H}{II}Rs. 
Non-ionized PDRs can be probed similarly to investigate physical
conditions in colder regions. 
For example, surface brightness maps in neutral N
([\ion{N}{I}]\,$\lambda$5199,5202) and neutral H (\ion{H}{I} at 
21\,cm; e.g., \citealt{rgw02}) would allow investigations into neutral
regions.
While such spatially-resolved analyses will resolve the degeneracy in
the plane of the sky, degeneracy along the line of sight will still
remain.  
This remaining degeneracy will have to be addressed by radiation
transfer modeling including all the necessary ingredients -- ionized,
atomic, and molecular gas components and dust grains. 
Such modeling will be a topic of one of the follow-up HerPlaNS papers
(e.g., Otsuka et al.\ {\sl in prep}).

\subsection{Empirical Gas-to-Dust Mass Ratio Distribution}

One of the goals of HerPlaNS is to empirically obtain 
spatially-resolved gas-to-dust mass ratio distribution maps by deriving
both the dust and gas column mass distribution maps directly from
observational data. 
As we already obtained the dust column mass map for 
\object{NGC\,6781} from thermal continuum emission maps in \S\,\ref{bbandimg}
(Fig.\,\ref{tempmaps}), here we outline how the gas column 
map for the nebula was derived. 

The H$^{+}$ column mass map for \object{NGC\,6781} was obtained by
equating the H$\alpha$ emissivity of the nebula\footnote{The
coefficients of this analytic expression were interpolated 
from the results by \citet{b71} under the assumption of an optically
thin, Case B nebula \citet{bm38}.},
$\epsilon_{{\rm H}\alpha} = (3.86 \times 10^{-25}) n_{{\rm H}^{+}}
n_{\rm e} (T_{\rm e}/10^{4})^{-1.077}$\,erg\,s$^{-1}$\,cm$^{-3}$,
to the H$\alpha$ image taken at the NOT \citep{phillips11}.
Upon solving for $n_{{\rm H}^{+}}$, we determined the 
($T_{\rm e}$, $n_{\rm e}$) radial profiles up to $60^{\prime\prime}$
from the central star by interpolating the profiles calculated from the
[\ion{N}{II}] line ratio profiles (Fig.\,\ref{tnprofiles}) through
polynomial fitting. 
Also, the H$\alpha$ map was made background-source free by removing 
all point sources within the nebula with empirically-defined PSFs using
the IRAF {\sl daophot} package.
Then, the cleaned H$\alpha$ map was properly scaled to have the 
de-reddened total H$\alpha$ flux of 
$3.88 \times 10^{-10}$\,erg\,s$^{-1}$\,cm$^{-2}$ as explained in the
previous section.
Moreover, we assumed a constant filling factor of 0.5 and a
distance of 950\,pc.

The resulting ionized H$^{+}$ gas column map of a $60^{\prime\prime}$
radius was determined to contain 0.37\,M$_{\odot}$.
By scaling this map by 1.452 
($= 1 + 4 \times n_{{\rm He}^{+}}/n_{{\rm H}^{+}}$, 
where $n_{{\rm He}^{+}}/n_{{\rm H}^{+}} = 0.113$; \citealt{liuy04}) to
account for the ionized He component, we obtained the total ionized gas
column map with a total mass of 0.53\,M$_{\odot}$. 
Based on the neutral to ionized gas mass ratio of 0.23 as estimated from
the H column density, the total mass of atomic gas amounts to 
0.12\,M$_{\odot}$.
The total atomic gas column map was then synthesized by scaling the
total dust column mass map, assuming that the lower-temperature atomic
gas was distributed as the dust component, following the physical
stratification discussed in the previous section.

Moreover, we synthesized the total molecular gas column map based on the
total H$_2$ mass estimate of 0.2\,M$_{\odot}$ derived via rotational
diagram analysis using H$_{2}$ emission line measurements of
v=0--0 S(2) to S(7) transitions \citep{phillips11}. 
Again, we assumed that H$_2$ was distributed in the same way 
as the low-temperature dust component.
Finally, the total gas column mass map was obtained by summing these
different gas phase maps and resampling at the same scale as the
dust column mass map.
By taking the ratio between the total gas and dust column maps, 
we thus empirically derived the gas-to-dust mass ratio map.

These column mass maps of the gas and dust components suggest that
the total mass of the observed shell of \object{NGC\,6781} is about 
0.86\,M$_{\odot}$.
Assuming the mass of the central star at present is roughly 
0.6\,M$_{\odot}$ 
(the average mass of a white dwarf; e.g., \citealt{kepler07}),
we conclude that the initial mass of the progenitor 
is approximately 1.5\,M$_{\odot}$.
Based on theoretical evolutionary tracks by \citet{vw94}, the
central star appears to be passing its maximum $T_{\rm eff}$ point
as it evolves onto the white dwarf cooling track.

%-----------------------------------------------------------------
% Figure 14 dust/gas profiles
%-----------------------------------------------------------------
   \begin{figure}
    \centering
    \includegraphics[width=\hsize]{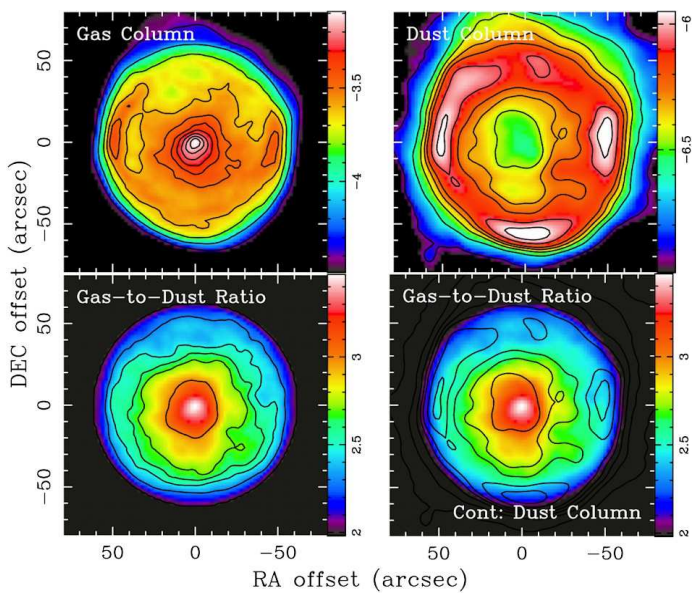}
    \caption{\label{g/d}
    [Top Left] The total gas column mass maps of 
    \object{NGC\,6781} in the log scale of
    M$_{\odot}$
    with contours from 90\% to 10\% (with a 10\% increment) 
    and 5\% of the peak. 
    [Top Right] The dust column mass map of \object{NGC\,6781} 
    in the log scale of M$_{\odot}$
    (the same as the bottom-right frame of Fig.\,\ref{tempmaps})
    with contours at the same levels as the gas column map.
    [Bottom Left] The gas-to-dust mass ratio distribution map in the log 
    scale with contours at the linear ratio of 500, 400, 300, 200,
    100, and 50. 
    [Bottom Right] The gas-to-dust mass ratio distribution map in the log
    scale overlaid with the dust column mass map contours, to indicate
    typical gas-to-dust mass ratio values in the dust barrel structures.}
   \end{figure}
%_________________________________________________________________

Fig.\,\ref{g/d} shows both the synthesized total gas column map and
observationally derived dust column mass map (top frames), 
as well as the gas-to-dust mass ratio distribution map (bottom frames
with different contours).  
The gas column map is peaked at the center, with a slight enhancement in
the dust barrel structure.
The high gas column in the middle of the barrel cavity is due to 
low $n_{\rm e}$[\ion{N}{II}] values:
the gas column distribution would look more uniform if we instead adopted  
$n_{\rm e}$[\ion{O}{III}] in this analysis.
However, the gas column in the middle of the central cavity is
relatively unimportant in terms of the gas-to-dust ratio, because the
bulk of the dust grains resides in the barrel structure.

The gas-to-dust mass ratio map is peaked at the center (with a peak
value of about 3000, owing to lack of dust grains in the cavity), with a
generally radially 
decreasing profile reflecting the centrally-concentrated
gas column mass distribution.
The overall median gas-to-dust mass ratio is 335 with a standard
deviation of 378.
However, the gas-to-dust mass ratio over the central cavity is enhanced
because the amount of dust present in the highly ionized barrel cavity
is small. 
If we restrict ourselves to the region where the dust column registers
more than 50\% of the peak, the gas-to-dust ratio comes down to
$\sim500$ at most and the median gas-to-dust mass ratio is 195 with a
standard deviation of 110.

These findings are in general consistent with a typical ballpark
gas-to-dust mass 
ratio of roughly 160 for oxygen-rich and 400 for carbon-rich AGB winds
\citep{knapp85}, and 100 for interstellar dust \citep{knapp74}.
However, we note that the gas-to-dust mass ratio varies radially from
over 550 to 100 within the cylindrical barrel structure.
Here, we have to exercise caution for the present derivation of the
gas-to-dust mass ratio, especially for the lower limit, because 
(1) the lower temperature atomic and molecular gas components
may be more preferentially distributed in the barrel than the ionized
gas component
and
(2) there is only so much gas that can condense into dust grains. 
For a solar metallicity gas, the condensable mass fraction is about
(5--6)\,$\times10^{-3}$, which gives a minimum gas-to-dust mass ratio of
about 180.

For carbon stars, all excess carbon can potentially condense.
While the gas-to-dust mass ratio depends critically on the C/O
ratio, the dependence is poorly understood.
The C/O ratio probably ranges from 1.1--2.0 at solar metallicity.
Hence, the gas-to-dust mass ratio is expected to vary correspondingly
from 220 to 100 (e.g., \citealt{matsu07}).
At any rate, we may want to reconsider the presently rampant 
``one-value-fits-all'' approach of extrapolating the unknown amount of
gas or dust component by scaling the empirically determined amount of
the other component with a single gas-to-dust mass ratio.
The HerPlaNS data will allow us to investigate how such variations of
the gas-to-dust ratio within PNs change among the target sources or 
show a particular trend.

In the present analysis, the atomic and molecular gas column maps 
were simply scaled from the derived ionized gas column map under the
assumption that all these components of various gas phases have
similar spatial distributions. 
By performing the same line diagnostic analyses for the atomic and 
molecular components of the target nebulae with appropriate maps in
atomic and molecular lines, it is possible to further constrain atomic
and molecular gas column maps observationally so as to derive more
accurate gas-to-dust mass ratio maps that are genuinely
observationally-established. 
Such an analysis will be a topic in the forthcoming HerPlaNS papers
(e.g., Exter et al.\ {\sl in prep}).

\section{Summary\label{sum}}

Using the {\sl Herschel Space Observatory}, we have collected a rich
far-IR imaging and spectroscopic data set for a group of 11\,PNs under
the framework of the {\sl Herschel} Planetary Nebula Survey (HerPlaNS).
In this survey, we used all available observational modes of the PACS and
SPIRE instruments aboard {\sl Herschel} to investigate the far-IR
characteristics of both the dust and gas components of the circumstellar
nebulae of the target sources. 

We obtained 
(1) broadband maps of the target sources at five far-IR bands, 
70, 160, 250, 350, and 500\,$\mu$m, 
with rms sensitivities of 0.01--0.1\,mJy\,arcsec$^{-2}$ 
(0.4--4 MJy\,sr$^{-1}$);
(2) $5 \times 5$ IFU spectral cubes of 51--220\,$\mu$m 
covering a $\sim50^{\prime\prime}\times50^{\prime\prime}$ field
at multiple positions in the target sources,
with rms sensitivities of 0.1--1\,mJy\,arcsec$^{-2}$ 
(4--40 MJy\,sr$^{-1}$) per wavelength bin;
and
(3) sparsely sampled spectral array of 194--672\,$\mu$m 
covering a $\sim3^{\prime}$ field
at multiple positions in the target sources, 
with rms sensitivities of 0.001--0.1\,mJy\,arcsec$^{-2}$ 
(0.04--4 MJy\,sr$^{-1}$) per wavelength bin.

In this first part of the HerPlaNS series, we described the data
acquisition and processing and illustrated the potential of the HerPlaNS
data using \object{NGC\,6781}, a dusty molecular-rich bipolar PN oriented
at nearly pole-on, as an example. 
Broadband images unveiled the surface brightness distribution of thermal
continuum emission from \object{NGC\,6781}.
Spatially resolved was the object's signature ring structure of
a $40^{\prime\prime}$ radius with a $20^{\prime\prime}$ width, 
embedded in an extended halo of about 100$^{\prime\prime}$ radius.
This far-IR ring represents a nearly pole-on bipolar/cylindrical barrel
structure, containing at least 
$M_{\rm dust}=4 \times 10^{-3}$\,M$_{\odot}$, at the adopted 
distance of $950\pm143$\,pc (\citealt{sm06};
all distance-dependent quantities were based on this value and
subject to its 15\% uncertainty).

Spectral fitting of the broadband images indicated that dust grains are 
composed mostly of amorphous-carbon based material (i.e., the power-law
emissivity index of $\beta \approx 1$) of the temperature between 26 and
40\,K. 
In the past, far-IR SED fitting with broadband fluxes were performed 
under the assumption of negligible line contamination.
With the {\sl HerPlaNS} data, we verified that the degree of line
contamination is approximately 8--20\% and does not significantly
affect the fitting results.

The {\sl Herschel} spectra obtained at various locations within
\object{NGC\,6781} revealed both the physical and chemical nature 
of the nebula. 
The spectra revealed a number of ionic and atomic lines such as
[\ion{O}{III}]\,52, 88\,$\mu$m,
[\ion{N}{III}]\,57\,$\mu$m,
[\ion{N}{II}]\,122, 205\,$\mu$m, 
[\ion{C}{II}]\,158\,$\mu$m, and 
[\ion{O}{I}]\,63, 146\,$\mu$m,
as well as various molecular lines, in particular, 
high-J CO rotational transitions, OH, and 
OH$^{+}$ emission lines: 
see \citet{aleman13} and \citet{etxaluze13} for more details on the
discovery of OH$^{+}$ emission in PNs. 
Thermal dust continuum emission was also detected in most bands
in these deep exposure spectra. 
Moreover, spectra taken at multiple spatial locations elucidated
the spatial variations of the line and continuum emission,
which reflect changes of the physical conditions within the nebula
projected onto the plane of sky.
On average, the relative distributions of emission lines of various
nature suggested 
that the barrel cavity is uniformly highly ionized, with a region of 
lower ionization delineating the inner surface of the barrel wall, 
and 
that the least ionic and atomic gas, molecular, and dust
species are concentrated in the cylindrical barrel structure.

The CO rotation diagram diagnostics yielded 
$T_{\rm ex} \approx$\,60--70\,K and 
$N_{\rm CO}\approx 10^{15}$\,cm$^{-2}$.
Compared with the previous CO measurements and diagnostics 
by \citet{bachiller93}, the present observations and analysis with
higher-J transitions sampled much warmer CO gas component in the
cylindrical barrel structure, 
probably located closer to the equatorial region along the line of sight. 
However, the amount of this warm component was determined to be
an order of magnitude smaller than the cold component.

Based on the PACS IFU spectral cube data, we derived line maps in the
detected ionic and atomic fine-structure lines.
Then, diagnostics of the electron temperature and density using line
ratios such as [\ion{O}{III}]  
52/88\,$\mu$m and [\ion{N}{II}] 122/205\,$\mu$m resulted in 
($T_{\rm e}$, $n_{\rm e}$) and ionic/elemental/relative
abundance profiles for the first time in the far-IR for any PN.
The derived $T_{\rm e}$ profile substantiated the typical assumption of
uniform $T_{\rm e} = 10^4$\,K in the main ionized region, while showing
an interesting increase in the barrel wall up to 11,000\,K, followed by a
sudden tapering off toward the halo region. 
The $n_{\rm e}$ profile of high-excitation species is nearly flat at 
$\sim400$\,cm$^{-3}$ across the inner cavity of the nebula, 
whereas the $n_{\rm e}$ profile of low-excitation species exhibits a
radially increasing tendency from 
80\,cm$^{-3}$ to $>600$\,cm$^{-3}$ with a somewhat complex variation
around the barrel wall. 

In fact, this $n_{\rm e}$[\ion{N}{II}] profile is reflected in the
physical stratification of the nebula revealed by the ionic/elemental 
abundance analysis. 
We found
(1) a very highly ionized, centrally restricted \ion{H}{II}R 
within $\sim20^{\prime\prime}$,
(2) a highly ionized \ion{H}{II}R for 20--30$^{\prime\prime}$
of the center marked by a high relative abundance of N$^{2+}$ and
O$^{2+}$, 
(3) a moderately ionized \ion{H}{II}R within the inner surface of the
barrel wall (for 30--50$^{\prime\prime}$) marked by a high relative
abundance of N$^{+}$ and C$^{+}$, 
and
(4) a least ionized \ion{H}{II}R transitioning into a PDR on the barrel
wall (beyond $\sim50^{\prime\prime}$) marked by the presence of
molecular and dust species. 
The detected stratification is consistent with the previous inferences
made from the past optical imaging observations in various emission
lines of varying levels of excitation.

The derived relative elemental abundance profiles showed uniformly low
N and C abundances, confirming the low initial mass ($<2$\,M$_{\odot}$) 
and marginally carbon-rich nature of the central star.
However, the profiles did not appear to reveal variations reflecting the
evolutionary change of the central star, such as a radially increasing
carbon abundance.
Nevertheless, the range of relative elemental abundances measured for
spatially resolved observations of \object{NGC\,6781} overlaps with the
range of abundances obtained from spatially-unresolved measurements of
ORLs and CELs. 
This may indicate that the issue of dichotomy between abundance
measurements made from ORLs and CELs is simply due to the spatial
resolution effects.
Therefore, it is interesting to revisit spatially-resolved diagnostics
using optical line maps as performed here using far-IR line maps.  

Direct comparison between the dust column mass map derived from the
HerPlaNS broadband thermal dust data and the gas column mass map derived
from the HerPlaNS fine-structure line mapping data augmented with
literature data in other wavelengths yielded an empirical gas-to-dust
mass ratio distribution map for \object{NGC\,6781}.
The resulting empirical gas-to-dust mass ratio map showed a range of
ratios within the cylindrical barrel structure, in general radially
decreasing roughly from 550 to 100.
The average gas-to-dust mass ratio in the dust barrel was determined to
be $195\pm110$, and hence, is generally consistent with the typical
spatially-unresolved ratio of 100--400 widely used in the literature
for the case of PNs and AGB stars.
The HerPlaNS data would therefore allow further investigations of the
distribution of gas-to-dust mass ratios across the target PNs in a
spatially resolved manner. 

The derivation of column mass distribution maps for various components
of \object{NGC\,6781}, 
based on the empirically-established distribution of the ionized
gas component, 
yielded an estimate
for the total mass of the shell of 0.86\,M$_{\odot}$, consisting of 
0.54\,M$_{\odot}$ of ionized gas,
0.12\,M$_{\odot}$ of atomic gas,
0.2\,M$_{\odot}$ of molecular gas, and
$4\times10^{-3}$\,M$_{\odot}$ of dust grains.
Provided that the present core mass is 0.6\,M$_{\odot}$, 
we concluded that the progenitor star had an initial mass of
1.5\,M$_{\odot}$. 
Then, theoretical evolutionary tracks of this 1.5\,M$_{\odot}$ star would
suggest that the star is nearing to the end of its PN evolution,
transitioning onto the white dwarf cooling track. 

In the forthcoming papers of the HerPlaNS series, we will focus on
separate analyses of the broadband maps (Ladjal et al.\ {\sl in prep})  
and spectra (Exter et al.\ {\sl in prep})
using the wealth of the entire HerPlaNS data set to present more
in-depth results of the analyses outlined above.
In addition, we will report more on the energetics of the entire
gas-dust system as a function of location in the nebulae, emphasizing
the results' statistical implications. 

   \begin{acknowledgements}
    This work is based on observations made with the Herschel Space
    Observatory, a European Space Agency (ESA) Cornerstone Mission with 
    significant participation by NASA. 
    Support for this work was provided by 
    NASA through an award issued by Jet Propulsion Laboratory,
    Caltech (Ueta, Ladjal, Kastner, Sahai), 
    the Japan Society of the Promotion of Science (JSPS) through  
    a FY2013 long-term invitation fellowship program (Ueta),
    the Belgian Federal Science Policy Office
    via the PRODEX Programme of ESA (Exter, van Hoof),
    the Polish NCN through a grant 2011/01/B/ST9/02031 (Szczerba,
    Si\'{o}dmiak), and
    the European Research Council via the advanced-ERC grant 246976
    and the Dutch Science Agency (NWO) via the Dutch Astrochemistry
    Network and the Spinoza prize (Aleman, Tielens).
    The authors thank M.~A.~Guerrero for sharing the NOT optical
    images of \object{NGC\,6781} with us.
    Also, H.~Monteiro's generosity is appreciated for the reproduction
    of one of his figures (Fig.\,3 of \citealt{sm06}).
    Sahai acknowledges that his contribution to the research
    described here was carried out at the JPL/Caltech, under a contract
    with NASA.    
    Finally, Ueta also acknowledges the hospitality of the members of
    the Laboratory of Infrared Astrophysics at ISAS/JAXA during his
    sabbatical stay as a JSPS invitation fellow.  
   \end{acknowledgements}

\Online

\begin{appendix}

 \section{PACS SED Range Spectroscopy Spectral Maps\label{allpacsspec}}

\begin{figure*}
    \centering
    \includegraphics[width=\hsize]{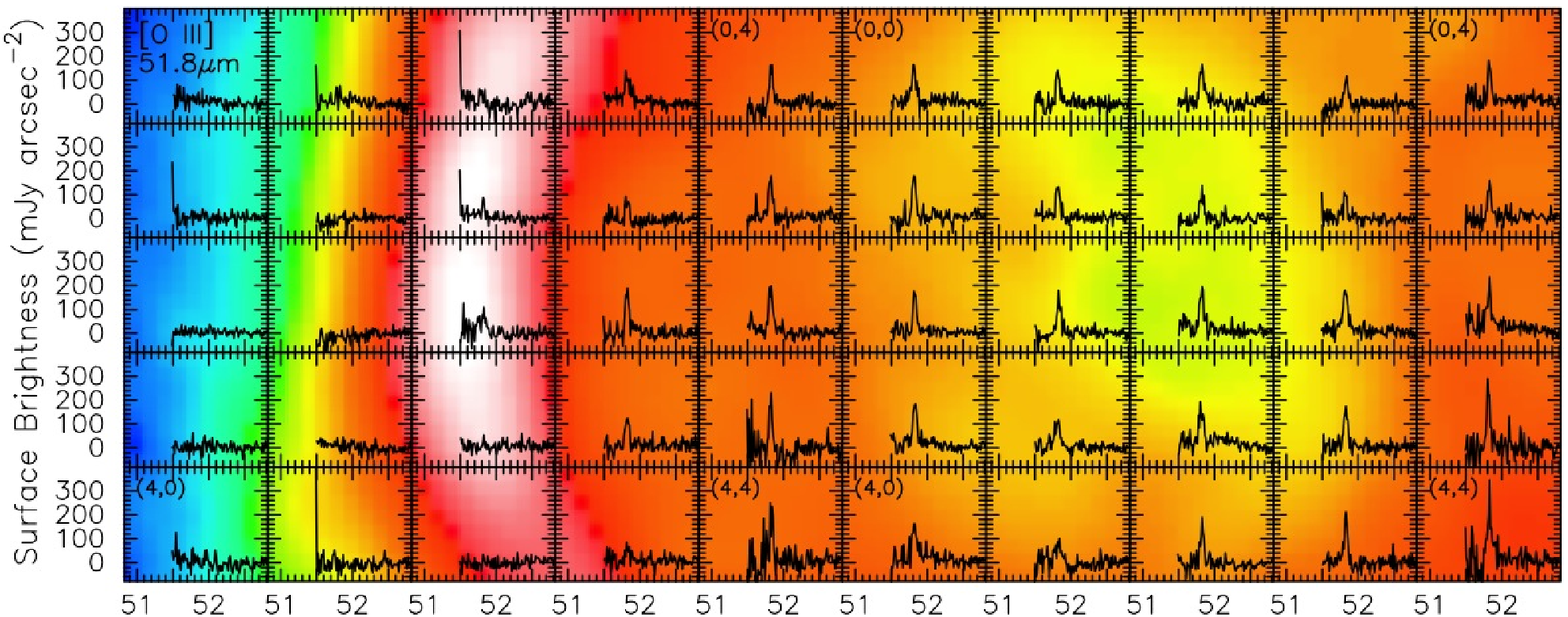}\vspace{5pt}
    \includegraphics[width=\hsize]{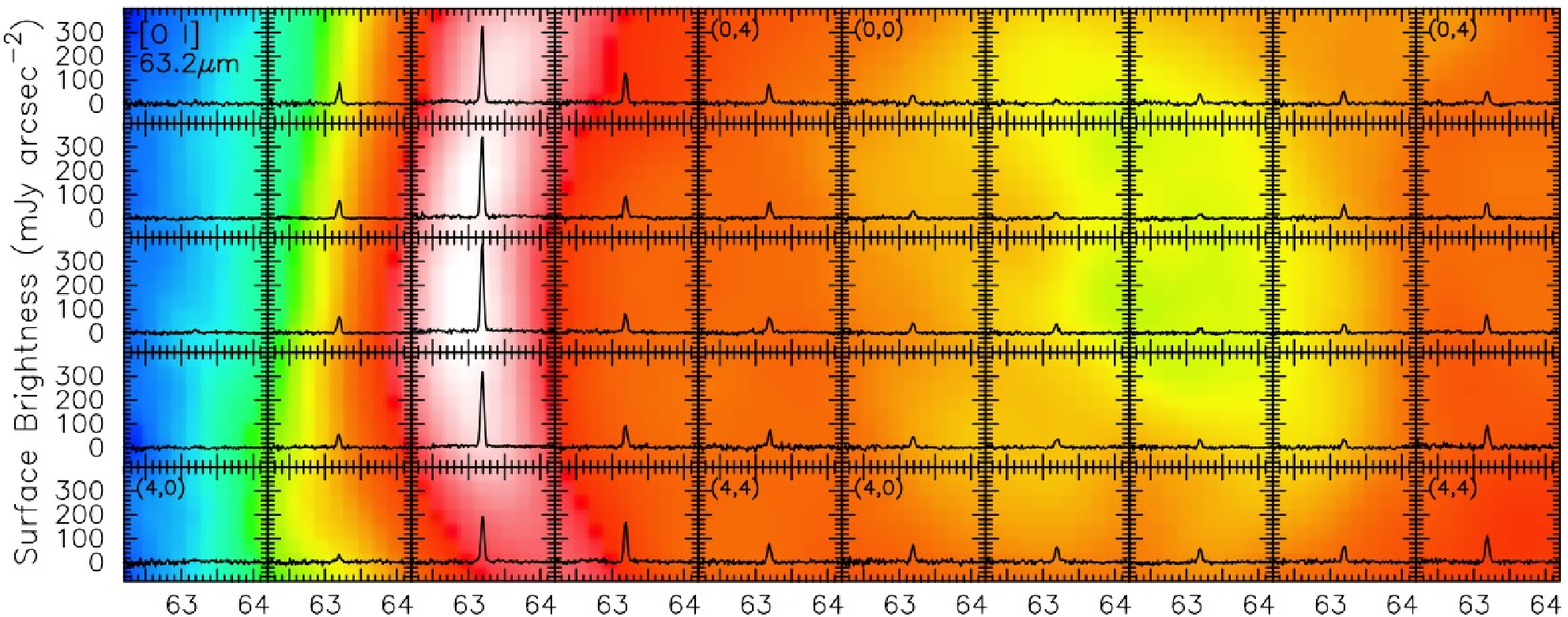}\vspace{5pt}
    \includegraphics[width=\hsize]{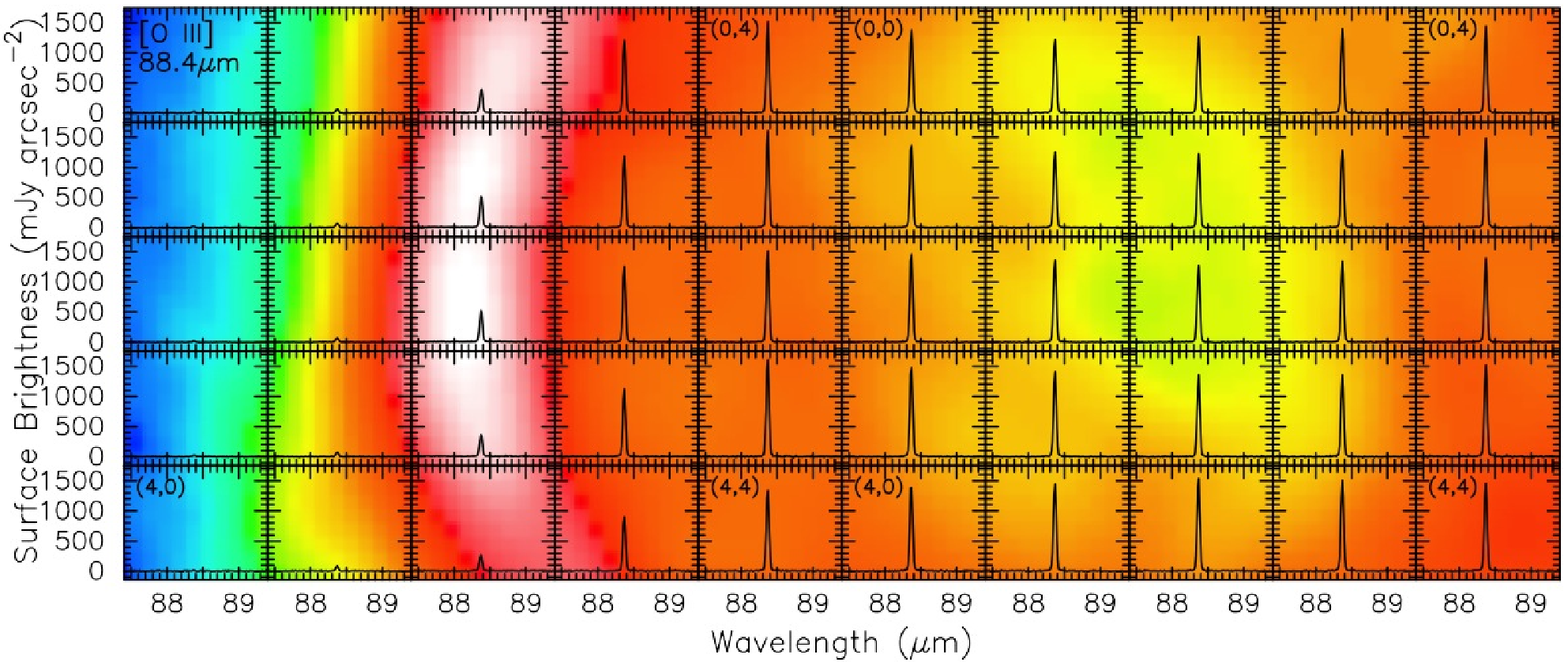}
    \caption{\label{pacsspecall1}
    Spatially-varying line emission of NGC\,6781 at 
    [\ion{O}{III}] 51.8\,$\mu$m, [\ion{O}{I}] 63.2\,$\mu$m, and 
    [\ion{O}{III}] 88.4\,$\mu$m
    in each of the two pointings toward
    \object{NGC\,6781} shown side by side: ``center'' on the right and
    ``rim''  on the left.
    To specify the instrument orientation, corner spaxels are
    identified by their identifiers.
    The flux unit is set to the surface brightness (mJy\,arcsec$^{-2}$).
    The background PACS 70\,$\mu$m image indicates  
    the approximate location of each spaxel.
    Ionic lines tend to be strong in the highly-ionized cavity
    of the cylindrical structure, while atomic and molecular lines tend
    to be pronounced in the cylindrical rim of the nebula.
    Note that the footprint of the PACS IFU is not a regular grid as
    implied by the placement of the sub-plots; the slightly irregular
    footprint can be seen in Figs.\ \ref{specposmaps} and 
    \ref{linemaps}.}
   \end{figure*}

\begin{figure*}
    \centering
    \includegraphics[width=\hsize]{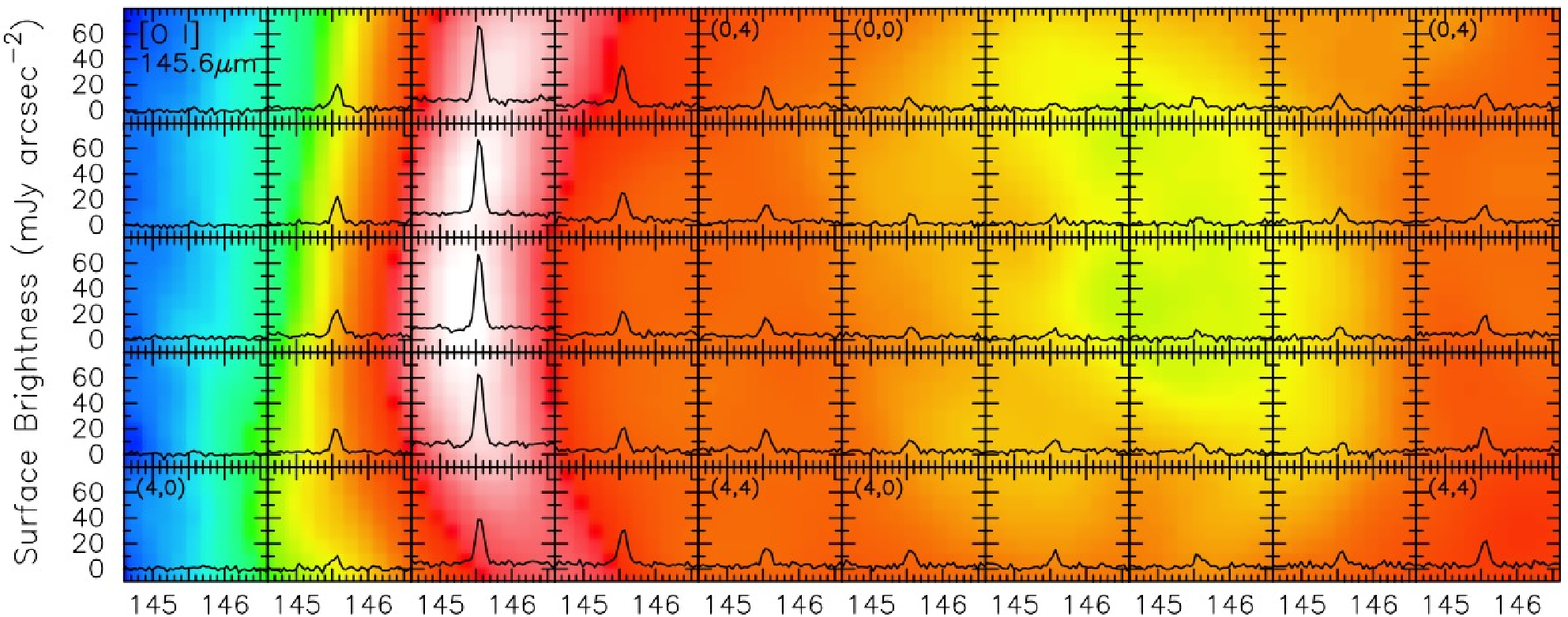}\vspace{5pt}
    \includegraphics[width=\hsize]{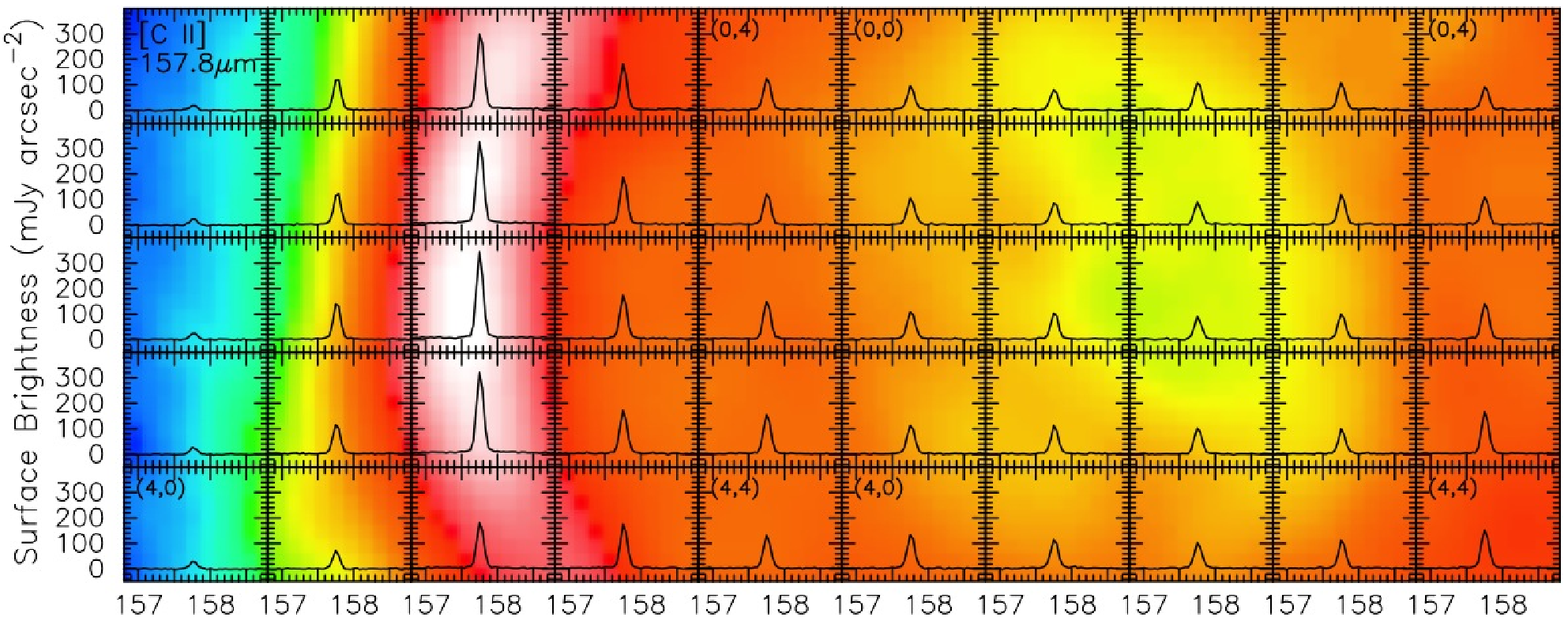}\vspace{5pt}
    \includegraphics[width=\hsize]{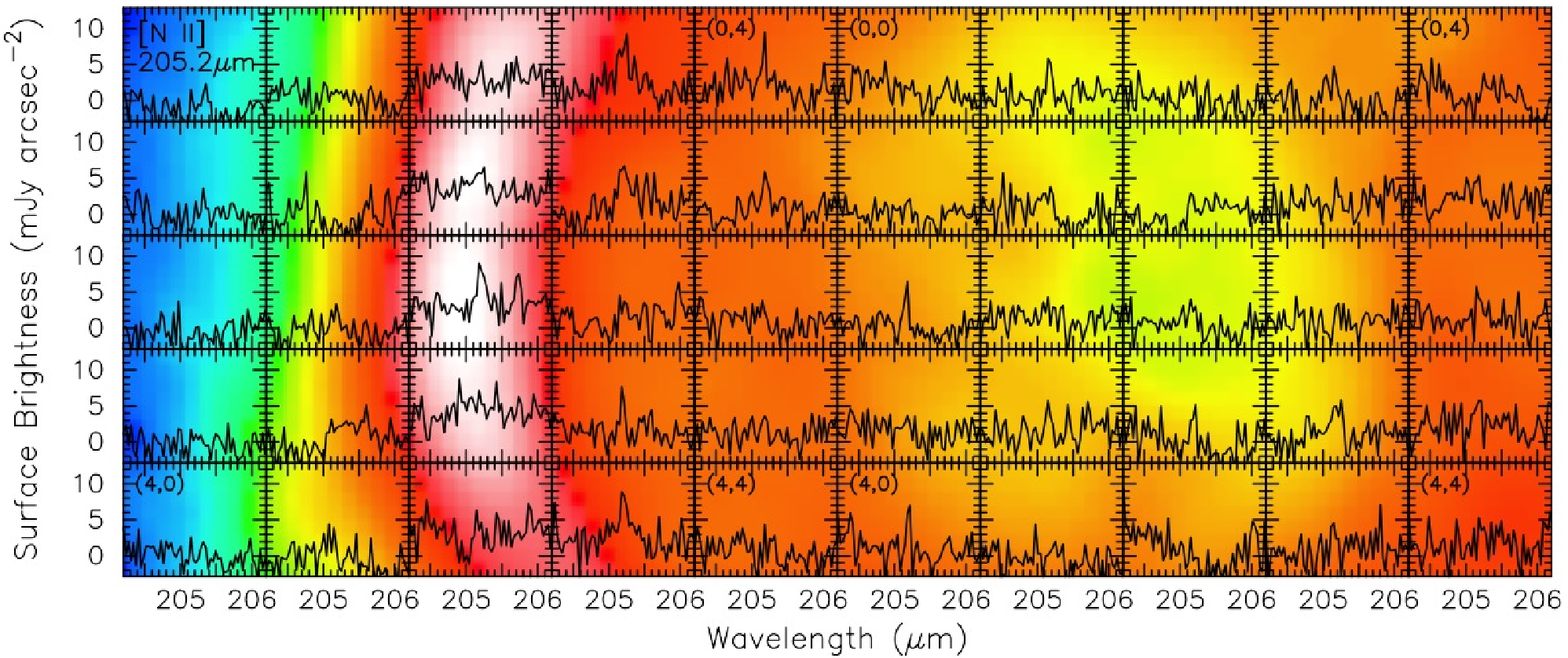}
    \caption{\label{pacsspecall2}
    Cont'd from the previous figure.
    Spatially-varying line emission of NGC\,6781 at 
    [\ion{O}{I}] 145.6\,$\mu$m, [\ion{C}{II}] 157.8\,$\mu$m, and 
    [\ion{N}{II}] 205.2\,$\mu$m.}
   \end{figure*}

 Fig.\,\ref{pacsspecall1} and \ref{pacsspecall2} show spatially-varying
 emission lines of \object{NGC\,6781} other than those already presented
 in Fig.\,\ref{pacsspatial}. 
 Displayed are all 50 spectra extracted from each of the $5 \times 5$
 spaxels at each of the two spatial pointings on the eastern rim (left)
 and at the center (right). 
 These spectra show that the relative strengths of
 these ionic, atomic, and molecular lines 
 change depending on the spaxel position along each line of sight within
 the PACS IFU aperture.
 The approximate spatial correspondence of the 50 spaxels to the
 70\,$\mu$m broadband map is shown by the color-scale surface brightness
 map in the background.
 Within the central ionized region, ionic lines ([\ion{O}{III}] and
 [\ion{N}{III}]) are strong (seen in the spaxels of the western half of
 the ``rim'' pointing and of the ``center'' pointing).
 While moving laterally away from the center of the nebula to outer
 regions along the plane of the sky via the eastern rim, the largest
 change of the relative strengths of lines occurs when going across the
 rim at which the ionic lines become weaker and atomic lines
 ([\ion{O}{I}], [\ion{N}{II}], and [\ion{C}{II}]) become stronger. 
 This transition region, however, is physically very restricted as the
 change is seen across nearly only one-spaxel width.  

%_____________________________________________________________
% Figure A:
%-------------------------------------------------------------
%_____________________________________________________________

 \section{SPIRE FTS High-Spectral Resolution Spectral Maps}

\begin{figure*}
   \centering
   \includegraphics[width=\hsize]{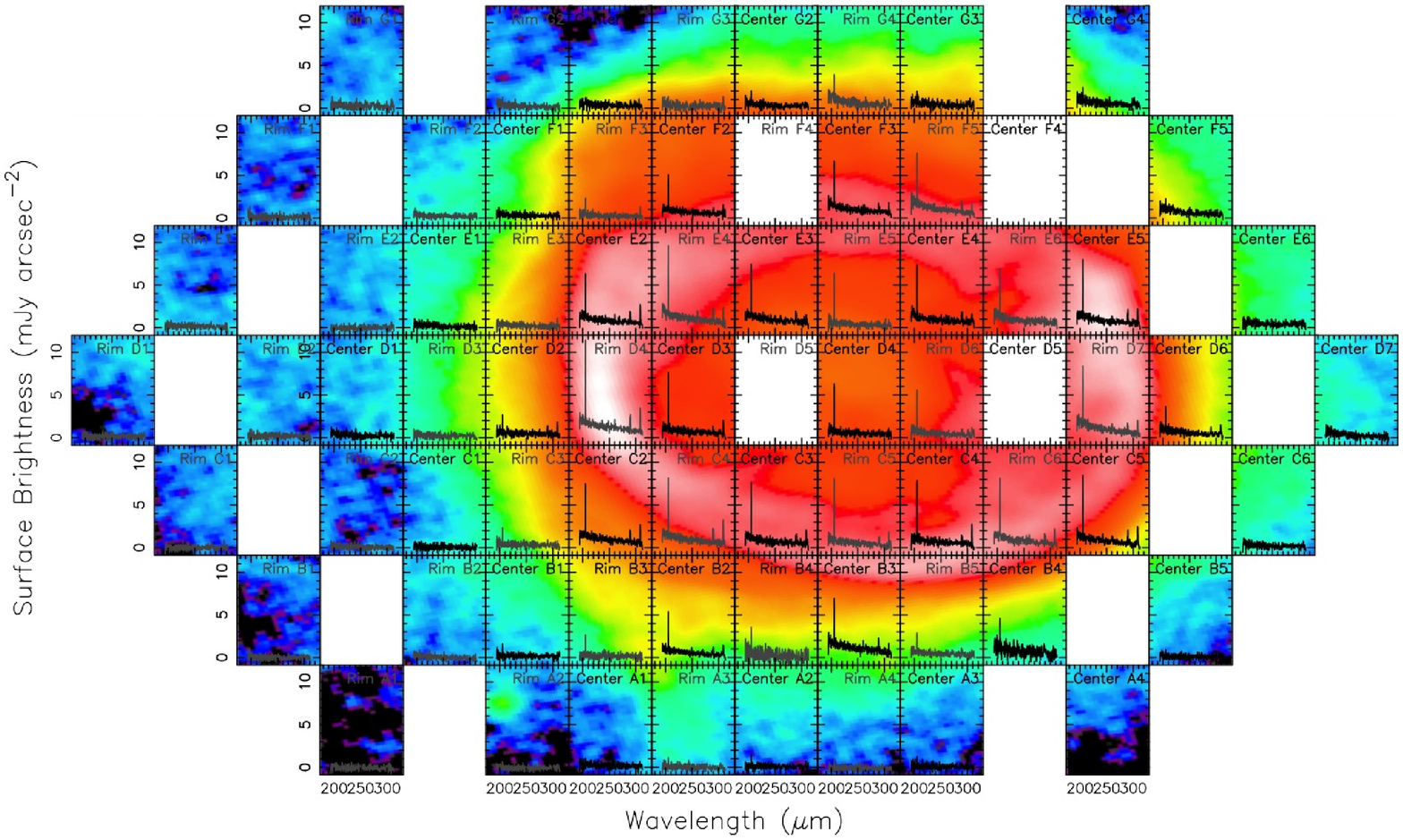}
   \caption{\label{spirespecall1}
   Pseudo-mosaic SPIRE spectral maps constructed with individual
   spectra from each SSW/SLW bolometer:
   All 70 SPIRE/SSW spectra covering 194--342\,$\mu$m extracted
   from each of the 35 bolometers (two bolometers are blind) at two
   pointings. 
   The background PACS 70\,$\mu$m map is shown to indicate the
   approximate spatial
   coverage of each SPIRE bolometer at each of the two pointings.
   The image is 
   rotated by $90^{\circ}$ to make individual these frames sufficiently
   large for printing.}
  \end{figure*}

\begin{figure*}
    \includegraphics[width=\hsize]{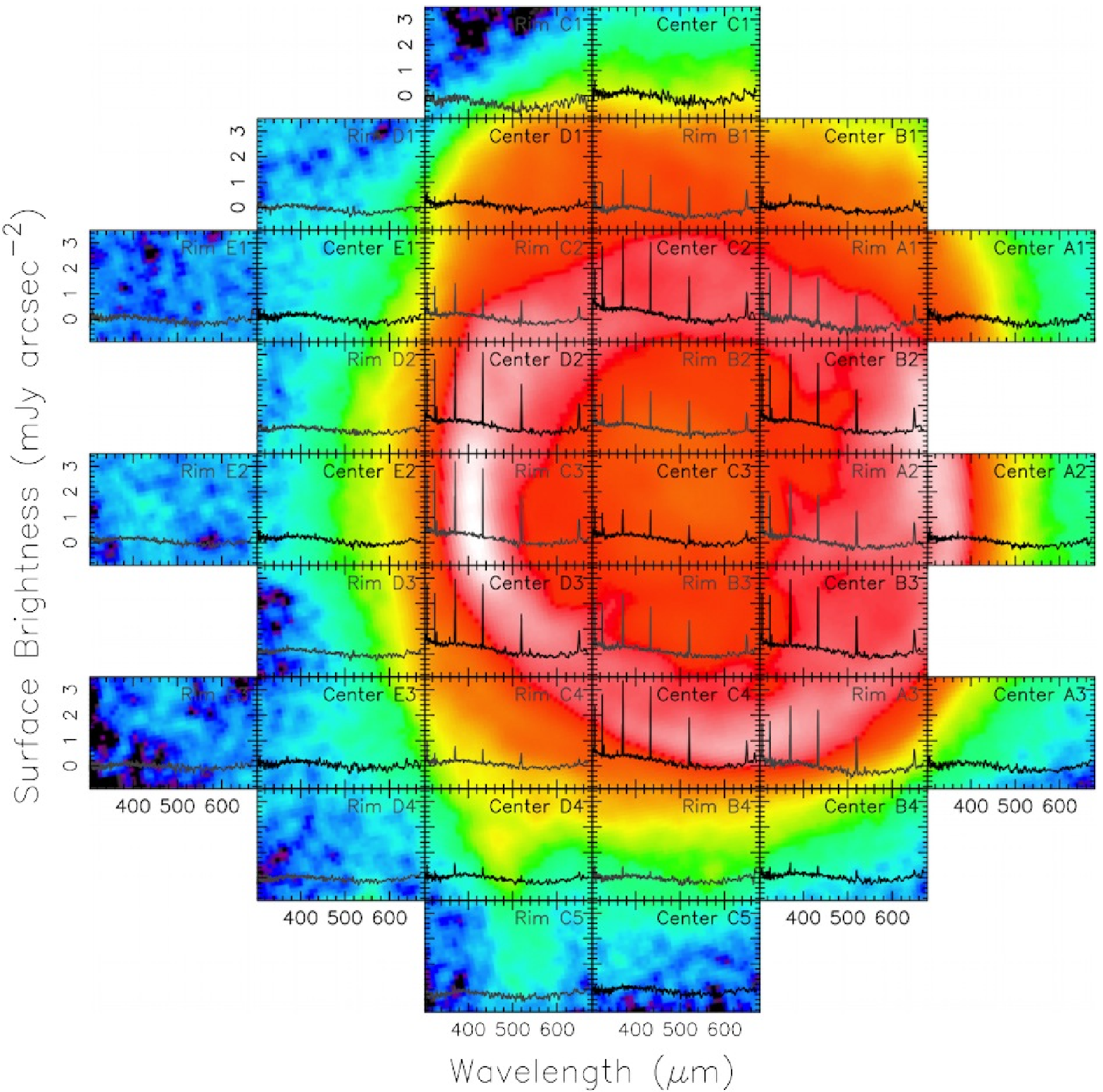}
    \caption{\label{spirespecall2}
    Pseudo-mosaic SPIRE spectral maps constructed with individual
    spectra from each SSW/SLW bolometer:
    All 38 SPIRE/SLW spectra covering 316--672\,$\mu$m extracted
    from each of the 19 bolometers at two pointings. 
    The background PACS 70\,$\mu$m map is shown to indicate the
    approximate spatial
    coverage of each SPIRE bolometer at each of the two pointings.}
   \end{figure*}

 Figs.\,\ref{spirespecall1} and \ref{spirespecall2} show all 70 spectra
 extracted from each of the two pointings of the 35-bolometer SSW array
 (2 bolometers are blind -- they are left blank in the map), covering
 194--342\,$\mu$m, and all 38 spectra extracted from each of the two
 pointings of the 19-bolometer SLW array, covering 316--672\,$\mu$m,
 respectively. 
 The entire SPIRE wavelength coverage from 194--672\,$\mu$m is achieved 
 in at least 16 locations (which are shown as Fig.\,\ref{wholespire} in
 the main text). 
 These spectra show that the relative strengths of the ionic 
 [\ion{N}{II}] line 
 in the SSW spectral range and CO rotational lines in the SLW spectral
 range, vary depending on the bolometer position within the target
 nebula. 
 Within the central ionized region, both atomic and molecular lines
 are weaker than in the ring structure. 
 In both SSW and SLW spectral ranges, however, the line strengths
 suddenly decrease once the line of sight goes beyond the central
 ionized region,
 suggesting that the presence of the gas component in the nebula is
 fairly spatially restricted -- not much beyond the central ionized
 region and the cylindrical barrel structure. 

%_____________________________________________________________
% Figure B:
%-------------------------------------------------------------
 
%_____________________________________________________________

\end{appendix}


\begin{thebibliography}{}

 \bibitem[Aleman et al.(2013)]{aleman13}
   Aleman, I., Ueta, T., Ladjal. D., et al.
%		 Exter, K.\ M., Kastner, J.\ H.,
%		 Montez, R., Tielens, A.\ G.\ G.\ M., Chu, Y.-H.,
%		 Izumiura, H., McDonald, I., Sahai, R., Si\'{o}dmiak,
%		 N., Szczerba, R., van Hoof, P.\ A.\ M., Villaver, E.,
%		 Vlemmings, W., Wittkowski, M., \& Zijlstra, A.\ A.
   2013, \aap, submitted

 \bibitem[Baker \& Menzel(1938)]{bm38}
   Baker, J.\ G., \& Menzel, D.\ H.
   1938, \apj, 88, 52

 \bibitem[Bachiller et al.(1993)]{bachiller93}
   Bachiller, R., Huggins, P.\ J., Cox, P., \& Forveille, T. 
   1993, \aap, 267, 177

 \bibitem[Bachiller et al.(1993)]{bachiller97}
   Bachiller, R., Forveille, T., Huggins, P.\ J., \& Cox, P.
   1997, \aap, 324, 1123

 \bibitem[Bohren \& Huffman(1983)]{bh83}
   Bohren, C.\ F., \& Huffman, D.\ R.
   1983, Absorption and scattering of light by small particles
   (New York: Wiley)

 \bibitem[Boyer et al.(2012)]{boyer12}
   Boyer, M.\ L., Srinivasan, S., Riebel, D., et al.
%		 McDonald, I., van
%   Loon, J.\ Th., Clayton, G.\ C., Gordon, K.\ D., Meixner, M.,
%   Sargent, B.\ A., \& Sloan, G.\ C.
   2012, \apj, 748, 40
   
 \bibitem[Brocklehurst(1971)]{b71}
   Brocklehurst, M.
   1971, \mnras, 153, 471

 \bibitem[Cantalupo et al.(2010)]{madmap}
   Cantalupo, C.\ M., Borill, J.\ D., Jaffe, A.\ H., \& Stompor, R. 
   2010, \apjs, 187, 212

 \bibitem[Castro-Carrizo et al.(2001)]{cc01}
   Castro-Carrizo, A., Bujarrabal, V., Fong, D., et al.
%		 Meixner, M., Tielens,
%		 A.\ G.\ G.\ M., Latter, W.\ B., \& Barlow, M.\ J.
   2001, /aap, 367, 674

 \bibitem[Corradi et al.(2003)]{corradi03}
   Corradi, R.\ L.\ M., Sch\"{o}nberner, D., Steffen, M.,
   \& Perinotto, M. 
   2003, \mnras, 340, 417

 \bibitem[Cox et al.(2011)]{cox11}
   Cox, N.\ L.\ J., Garc\'{\i}a-Hern\'{a}ndez, D.\ A.,
   Garc\'{\i}a-Lario, P., \& Manchado, A.
   2011, \aj, 141, 111

 \bibitem[Draine \& Lee(1984)]{dl84}
   Draine, B.\ T., \& Lee, H.\ M.
   1984, \apj, 285, 89

 \bibitem[Etxaluze et al.(2013)]{etxaluze13}
   Etxaluze, M., Cernicharo, J., Goicoechea, J.\ R., et al.
%		 van Hoof, P.\ A.\
%		 M., Swinyard, B.\ M., Barlow, M.\ J., Groenenwegen, M.\
%		 A.\ T., Kerschbaum, F., Lim, T.\ J., Matsuura, M.,
%		 Polehampton, E.\ T., Royer, P., Ueta, T., \& Van de
%		 Steene, G.\ C.
   2013, \aap, submitted

 \bibitem[Frew(2008)]{frew08}
   Frew, D.\ J.
   2008, Ph.D.\ Thesis, Macquarie Univ., NSW, Australia

 \bibitem[Fong et al.(2001)]{fong01}
   Fong, D., Meixner, M., Castro-Carrizo, A., et al.
%		 Bujarrabal, V., Latter,
%	   W.\ B., Tielens, A.\ G.\ G.\ M., Kelly, D.\ M., \& Sutton,
%	   E.\ C.
   2001, \aap, 367, 652

  \bibitem[Garnett \& Dinerstein(2001)]{gd01}
   Garnett, D.\ R., \& Dinerstein, H.\ L.
   2001, Rev.\ Mex.\ Astron.\ Astrofis.\ Ser.\ Conf., 10, 13

  \bibitem[Gledhill et al.(2002)]{gby02}
   Gledhill, T.\ M., Bains, I., \& Yates, J.\ A.
   2002, \mnras, 332, L55

  \bibitem[Goldsmith \& Langer(1999)]{gl99}
   Goldsmith, P.\ F., \& Langer, W.\ D.
   1999, \apj, 517, 209
		 
 \bibitem[Grevesse et al.(2010)]{grevesse10}
   Grevesse, N., Asplund, M., Sauval, A.\ J., \& Scott, P.
   2010, \apss, 328, 179

 \bibitem[Griffin et al.(2010)]{spire}
   Griffin, M.\ J., Abergel, A., Abreu, A., et al.
%		 Ade, P.\ A.\ R., Andr\'{e},
%   P., Augueres, J.-L., Babbedge, T., Bae, Y., Baillie, T.,
%   Baluteau, J.-P., Barlow, M.\ J., Bendo, G., Benielli, D., Bock,
%   J.\ J., Bonhomme, P., Brisbin, D., Brockley-Blatt, C.,
%   Caldwell, M., Cara, C., Castro-Rodriguez, N., Cerulli, R.,
%   Chanial, P., Chen, S., Clark, E., Clements, D.\ L., Clerc, L.,
%   Coker, J., Communal, D., Conversi, L., Cox, P., Crumb, D.,
%   Cunningham, C., Daly, F., Davis, G.\ R., de Antoni, P.,
%   Delderfield, J., Devin, N., di Giorgio, A., Didschuns, I.,
%   Dohlen, K., Donati, M., Dowell, A., Dowell, C.\ D., Duband, L.,
%   Dumaye, L., Emery, R.\ J., Ferlet, M., Ferrand, D., Fontignie,
%   J., Fox, M., Franceschini, A., Frerking, M., Fulton, T.,
%   Garcia, J., Gastaud, R., Gear, W.\ K., Glenn, J., Goizel, A.,
%   Griffin, D.\ K., Grundy, T., Guest, S., Guillemet, L.,
%   Hargrave, P.\ C., Harwit, M., Hastings, P., Hatziminaoglou, E.,
%   Herman, M., Hinde, B., Hristov, V., Huang, M., Imhof, P.,
%   Isaak, K.\ J., Israelsson, U., Ivison, R.\ J., Jennings, D.,
%   Kiernan, B., King, K.\ J., Lange, A.\ E., Latter, W., Laurent,
%   G., Laurent, P., Leeks, S.\ J., Lellouch, E., Levenson, L., Li,
%   B., Li, J., Lilienthal, J., Lim, T., Liu, S.\ J., Lu, N.,
%   Madden, S., Mainetti, G., Marliani, P., McKay, D., Mercier,
%   K., Molinari, S., Morris, H., Moseley, H., Mulder, J., Mur,
%   M., Naylor, D.\ A., Nguyen, H., O'Halloran, B., Oliver, S.,
%   Olofsson, G., Olofsson, H.-G., Orfei, R., Page, M.\ J., Pain,
%   I., Panuzzo, P., Papageorgiou, A., Parks, G., Parr-Burman, P.,
%   Pearce, A., Pearson, C., P\'{e}rez-Fournon, I., Pinsard, F.,
%   Pisano, G., Podosek, J., Pohlen, M., Polehampton, E.\ T.,
%   Pouliquen, D., Rigopoulou, D., Rizzo, D., Roseboom, I.\ G.,
%   Roussel, H., Rowan-Robinson, M., Rownd, B., Saraceno, P.,
%   Sauvage, M., Savage, R., Savini, G., Sawyer, E., Scharmberg,
%   C., Schmitt, D., Schneider, N., Schulz, B., Schwartz, A.,
%   Shafer, R., Shupe, D.\ L., Sibthorpe, B., Sidher, S., Smith,
%   A., Smith, A.\ J., Smith, D., Spencer, L., Stobie, B.,
%   Sudiwala, R., Sukhatme, K., Surace, C., Stevens, J.\ A.,
%   Swinyard, B.\ M., Trichas, M., Tourette, T., Triou, H., Tseng,
%   S., Tucker, C., Turner, A., Vaccari, M., Valtchanov, I.,
%   Vigroux, L., Virique, E., Voellmer, G., Walker, H., Ward, R.,
%   Waskett, T., Weilert, M., Wesson, R., White, G.\ J.,
%   Whitehouse, N., Wilson, C.\ D., Winter, B., Woodcraft, A.\ L.,
%   Wright, G.\ S., Xu, C.\ K., Zavagno, A., Zemcov, M., Zhang, L.,
%   \& Zonca, E.
   2010, \aap, 518, L3

 \bibitem[Groenewegen et al.(2011)]{mess}
  Groenewegen, M.\ A.\ T., Waelkens, C., Barlow, M.\ J., et al.
%		 Kerschbaum, F.,
%		 Garcia-Lario, P., Cernicharo, J., Blommaert, J.\ A.\
%		 D.\ L., Bouwman, J., Cohen, M., Cox, N., Decin, L.,
%		 Exter, K., Gear, W.\ K., Gomez, H.\ L., Hargrave, P.\
%		 C., Henning, Th., Hutsem\'{e}kers, D., Ivison, R.\ J.,
%		 Jorissen, A., Krause, O., Ladjal, D., Leeks, S.\ J.,
%		 Lim, T.\ L., Matsuura, M., Naz\'{e}, Y., Olofsson, G.,
%		 Ottensamer, R., Polehampton, E., Posch, T., Rauw, G.,
%		 Royer, P., Sibthorpe, B., Swinyard, B.\ M., Ueta, T.,
%		 Vamvatira-Nakou, C., Vandenbussche, B., Van de Steene,
%		 G.\ C., Van Eck, S., van Hoof, P.\ A.\ M., Van Winckel,
%		 H., Verdugo, E., \& Wesson, R.  
  2011, \aap, 526, A162 

 \bibitem[Gruendl et al.(2006)]{ggcw06}
   Gruendl, R.\ A., Guerrero, M.\ A., Chu, Y.-H., 
	   \& Williams, R.\ M.
   2006, \apj, 653, 339

 \bibitem[Gry et al.(2003)]{isohb}
   Gry, C., Swinyard, B., Harwood, A., et al.
%		 Trams, N., Leeks, S., Lim, T.,
%		 Sidher, S., Lloyd, C., Pezzuto, S., Molinari, S.,
%		 Lorente, R., Caux, E., Polehampton, E., Chan, J.,
%		 Hutchinson, G., M\"{u}ller, T., Burgdorf, M., \& Grundy, T.
   2003, The ISO Handbook, Volume III -- LWS -- The Long Wavelength Spectrometer
		 SAI-99-077/Dc, Version 2.1 (Vilspa: ESA)

 \bibitem[Guerrero \& De Marco(2013)]{gdm13}
   Guerrero, M.\ A., \& De Marco, O.
   2013, A\&A, 553, 126

 \bibitem[Hiriart(2005)]{hiriart05}
   Hiriart, D.
   2005, \aap, 187, 181

 \bibitem[Hora, Latter, \& Deutsch(1999)]{hora99}
   Hora, J.\ L., Latter, W.\ B., \& Deutsch, L.\ K.
   1999, \apjs, 124, 195 

 \bibitem[Kastner et al.(2003)]{kastner03}
   Kastner, J.\ H., Balick, B., Blackman, E.\ G., et al.
%		 Frank, A., Soker, N.,
%	   Vrt\'{\i}lek, S.\ D., \& Li, J.
   2003, \apjl, 591, L37

 \bibitem[Kastner et al.(2012)]{chanplans}
   Kastner, J.\ H., Montez, R., Jr., Balick, B., et al.
%		 Frew, D.\
%   J., Miszalski, B., Sahai, R., Blackman, E., Chu, Y.-H.,
%   De Marco, O., Frank, A., Guerrero, M.\ A., Lopez, J.\
%   A., Rapson, V., Zijlstra, A., Behar, E., Bujarrabal,
%   V., Corradi, R.\ L.\ M., Nordhaus, J., Parker, Q.\ A.,
%   Sandin, C., Sch\"{o}nberner, D., Soker, N., Sokoloski, J.\
%   L., Steffen, M., Ueta, T., \& Villaver, E. 
   2012, \aj, 144, 58

 \bibitem[Kepler et al.(2007)]{kepler07}
   Kepler, S.\ O., Kleinman, S.\ J., Nitta, A., et al.
		 %Koester, D.; Castanheira, B. G.; Giovannini, O.; Costa, A. F. M.; Althaus, L.
		 2007, \mnras, 375, 1315

 \bibitem[Kerber et al.(2003)]{kerber03}
   Kerber, F., Mignani, R.\ P., Guglielmetti, F., \& Wicenec, A.
   2003, \aap, 408, 1029

 \bibitem[Kessler et al.(1996)]{iso}
   Kessler, M.\ F., Steinz, J.\ A., Anderegg, M.\ E., et al.
%		 Clavel, J., Drechsel,
%   G., Estaria, P., Faelker, J., Riedinger, J.\ R., Robson, A.,
%   Taylor, B.\ G., \& Ximenez de Ferran, S.\ 
   1996, \aap, 315, L27

 \bibitem[Knapp(1985)]{knapp85}
   Knapp, G.\ R.
   1985, \apj, 293, 273      

 \bibitem[Knapp et al.(1994)]{kbyp94}
   Knapp, G.\ R., Bowers, P.\ F., Young, K., \& Phillips, T. G.
   1994, \apjl, 429, 33

 \bibitem[Knapp \& Kerr(1974)]{knapp74}
   Knapp, G.\ R., \& Kerr, F.\ J.
   1974, \apj, 35, 361      

 \bibitem[Knapp et al.(1993)]{ksr93}
   Knapp, G.\ R., Sandell, G., \& Robson, E.\ I. 
   1993, \apjs, 88, 173

 \bibitem[Kwok(2000)]{kwokbook}
   Kwok, S. 
   2000, The Origin and Evolution of Planetary Nebulae
   (Campridge: CUP) 

% \bibitem[Leal-Ferreira et al.(2011)]{ferreira11}
%   Leal-Ferreira, M.\ L., Gon\c{c}alves, D.\ R., Monteiro, H., \& Richards, J.\ W.
%   2011, \mnras, 411, 1395

 \bibitem[Liu et al.(2001)]{liu01}
   Liu, X.-W., Barlow, M.\ J., Cohen, M., et al.
%		 Danziger, I.\
%   J., Luo, S.-G., Baluteau, J.\ P., Cox, P., Emery, R.\
%   J., Lim, T., \& P\'{e}quignot, D.
   2001, \mnras, 323, 343  

 \bibitem[Liu et al.(2004a)]{liuy04}
   Liu, Y., Liu, X.-W., Luo, S.-G., \& Barlow, M. J.
   2004a, \mnras, 353, 1231

 \bibitem[Liu et al.(2004b)]{liu04}
   Liu, Y., Liu, X.-W., Barlow, M. J., \& Luo, S.-G.
   2004b, \mnras, 353, 1251

 \bibitem[Malhotra et al.(2001)]{malhotra01}
   Malhotra, S., Kaufman, M.\ J., Hollenbach, D., et al.
%		 Helou, G., 
%   Rubin, R.\ H., Brauher, J., Dale, D., Lu, N.\ Y.,
%   Lord, S., Stacey, G., Contursi, A., Hunter, D.\ A., \&
%   Dinerstein, H.
   2001, \apj, 561, 766

 \bibitem[Mathis, Rumpl, \& Nordsieck(1977)]{mrn}
   Mathis, J.\ S., Rumpl, W., \& Nordsieck, K.\ H.
   1977, \apj, 217, 425

 \bibitem[Matsuura et al.(2009)]{matsuura09}
   Matsuura, M., Barlow, M.\ J., Zijlstra, A.\ A., et al.
%		 Whitelock, P.\ A.,
%   Cioni, M.-R.\ L., Groenewegen, M.\ A.\ T., Volk, K., Kemper, F.,
%   Kodama, T., Lagadec, E., Meixner, M., Sloan, G.\ C., \& Srinivasan, S.
   2009, \mnras, 396, 918

 \bibitem[Matsuura et al.(2007)]{matsu07}
   Matsuura, M., Zijlstra, A.\ A., Bernard-Salas, J., et al.
%		 Menzies, J.\ W., Sloan,
%		 G.\ C., Whitelock, P.\ A., Wood, P.\ R., Cioni, M.-R.\ L.,
%		 Feast, M.\ W., Lagadec, E., van Loon, J.\ Th.,
%		 Groenewegen, M.\ A.\ T., \& Harris, G.\ J.
   2007, \mnras, 382, 1889

 \bibitem[Mavromatakis et al.(2001)]{mpp01}
   Mavromatakis, F., Papamastorakis, J., \& Paleologou, E.\ V.
   2001, \aap, 374, 280

 \bibitem[Melnick et al.(1987)]{melnick87}
   Melnick, G.\ J., Genzel, R., \& Lugten, J.\ B.
   1987, \apj, 321, 530

 \bibitem[Mennella et al.(1995)]{mcb95}
   Mennella, V., Colangeli, L., \& Bussoletti, E.
   1995, \aap, 295, 165

 \bibitem[Milanova \& Kholtygin(2009)]{mk09}
   Milanona, Y.\ V., \& Kholtygin, A.\ F.
   2009, Astron.\ Lett., 35, 518

 \bibitem[Murakami et al.(2007)]{murakami07}
   Murakami, H., Baba, H., Barthel, P., et al.
%		 Clements, D.\ L., Cohen, M.,
%   Doi, Y., Enya, K., Figueredo, E., Fujishiro, N., Fujiwara, H.,
%   Fujiwara, M., Garcia-Lario, P., Goto, T., Hasegawa, S., Hibi,
%   Y., Hirao, T., Hiromoto, N., Hong, S.\ S., Imai, K., Ishigaki,
%   M., Ishiguro, M., Ishihara, D., Ita, Y., Jeong, W.-S., Jeong,
%   K.\ S., Kaneda, H., Kataza, H., Kawada, M., Kawai, T., Kawamura,
%   A., Kessler, M.\ F., Kester, D., Kii, T., Kim, D.\ Chan, Kim, W.,
%   Kobayashi, H., Koo, B.\ C., Kwon, S.\ M., Lee, H.\ M., Lorente, R.,
%   Makiuti, S., Matsuhara, H., Matsumoto, T., Matsuo, H., Matsuura,
%   S., M\"{u}ller, T.\ G., Murakami, N., Nagata, H., Nakagawa, T.,
%   Naoi, T., Narita, M., Noda, M., Oh, S.\ H., Ohnishi, A., Ohyama,
%   Y., Okada, Y., Okuda, H., Oliver, S., Onaka, T., Ootsubo, T.,
%   Oyabu, S., Pak, S., Park, Y.-S., Pearson, C.\ P., Rowan-Robinson,
%   M., Saito, T., Sakon, I., Salama, A., Sato, S., Savage, R.\ S.,
%   Serjeant, S., Shibai, H., Shirahata, M., Sohn, J., Suzuki, T.,
%   Takagi, T., Takahashi, H., Tanab\'{e}, T., Takeuchi, T.\ T.,
%   Takita, S., Thomson, M., Uemizu, K., Ueno, M., Usui, F.,
%   Verdugo, E., Wada, T., Wang, L., Watabe, T., Watarai, H., White,
%   G.\ J., Yamamura, I., Yamauchi, C., \& Yasuda, A.
   2007, \pasj, 59, S369

 \bibitem[Ott(2010)]{hipe}
		 Ott, S.
		 2010, ASP Conf.\ Ser.\ 434, 
		 Astronomical Data Analysis Software and Systems XIX, 
		 eds.\ Y.\ Mizumoto, K.\ Morita, \& M.\ Ohishi
		 (San Francisco: ASP), 139

 \bibitem[Peimbert \& Torres-Peimbert(1983)]{peimbert83}
   Peimbert, M., \& Torres-Peimbert, S.
   1983, in Planetary nebulae: Proceedings of the Symposium
	   (Dordrecht: D.\ Reidel Publishing), p.233

 \bibitem[Phillips et al.(2011)]{phillips11}
   Phillips, J.\ P., Ramos-Larios, G., \& Guerrero, M.\ A.
   2011, \mnras, 415, 513

 \bibitem[Pilbratt et al.(2010)]{pilbratt10}
   Pilbratt, G.\ L., Riedinger, J.\ R., Passvogel, T., et al.
%		 Crone, G.,
%   Doyle, D., Gageur, U., Heras, A.\ M., Jewell, C., Metcalfe, L.,
%   Ott, S., \& Schmidt, M.
   2010, \aap, 518, L1
 
 \bibitem[Poglitsch et al.(2010)]{pacs}  
   Poglitsch, A., Waelkens, C., Geis, N., et al.
%		 Feuchtgruber,
%   H., Vandenbussche, B., Rodriguez, L., Krause, O.,
%   Renotte, E., van Hoof, C., Saraceno, P., Cepa, J.,
%   Kerschbaum, F., Agn\`{e}se, P., Ali, B., Altieri, B.,
%   Andreani, P., Augueres, J.-L., Balog, Z., Barl, L.,
%   Bauer, O.\ H., Belbachir, N., Benedettini, M., Billot,
%   N., Boulade, O., Bischof, H., Blommaert, J., Callut,
%   E., Cara, C., Cerulli, R., Cesarsky, D., Contursi, A.,
%   Creten, Y., De Meester, W., Doublier, V., Doumayrou,
%   E., Duband, L., Exter, K., Genzel, R., Gillis, J.-M.,
%   Gr\"{o}zinger, U., Henning, T., Herreros, J., Huygen, R.,
%   Inguscio, M., Jakob, G., Jamar, C., Jean, C., de Jong,
%   J., Katterloher, R., Kiss, C., Klaas, U., Lemke, D.,
%   Lutz, D., Madden, S., Marquet, B., Martignac, J., Mazy,
%   A., Merken, P., Montfort, F., Morbidelli, L., M\"{u}ller,
%   T., Nielbock, M., Okumura, K., Orfei, R., Ottensamer,
%   R., Pezzuto, S., Popesso, P., Putzeys, J., Regibo, S.,
%   Reveret, V., Royer, P., Sauvage, M., Schreiber, J.,
%   Stegmaier, J., Schmitt, D., Schubert, J., Sturm, E.,
%   Thiel, M., Tofani, G., Vavrek, R., Wetzstein, M.,
%   Wieprecht, E., \& Wiezorrek, E. 
   2010, \aap, 518, L2

 \bibitem[Pottasch et al.(1984)]{pottasch84}
   Pottasch, S.\ R., Baud, B., Beintema, D., et al.
%		 Emerson, J., Harris, S.,
%		 Habing, H.\ J., Houck, J., Jennings, R., \& Marsden, P.
   1984, A\&A, 138, 10

 \bibitem[Rodr\'{\i}guez et al.(2002)]{rgw02}
   Rodr\'{\i}guez, L.\ F., Goss, W.\ M., \& Williams, R.
   2002, \apj, 574, 179

 \bibitem[Rouleau \& Martin(1991)]{rm91}
   Rouleau, F. \& Martin, P.\ G. 
   1991, \apj, 377, 526

 \bibitem[Roussel(2013)]{scana}
   Roussel, H.
   2013, \pasp, 125, 1126

 \bibitem[Rubin et al.(1994)]{rubin94}
   Rubin, R.\ H., Simpson, J.\ P., Lord, S.\ D., et al.
%		 Colgan, S.\ W.\ J., 
%   Erickson, E.\ F., \& Haas, M.\ R.
   1994, \apj, 420, 772

 \bibitem[Sabin et al.(2010)]{sabin10}
   Sabin, L., Zijlstra, A.\ A., Wareing, C., et al.
%   Corradi, R.\ L.\ M.. Mampaso, A.,
%   Viironen, K., Wright, N.\ J., \&
%   Parker, Q.\ A.
   2010, \pasa, 27, 166 
     
 \bibitem[Sahai et al.(2011)]{sahai11}
   Sahai, R., Morris, M.\ R., \& Villar, G.\ G.
   2011, \aj, 141, 134

 \bibitem[Sandin et al.(2008)]{sandin08}
   Sandin, C., Sch\"{o}nberner, D., Roth, M.\ M., et al.
%		 Steffen, M., B\"{o}hm,
%	   P., \& Monreal-Ibero, A.
   2008, \aap, 486, 545

 \bibitem[Sch\"{o}nberner et al.(2005)]{schoen05}
   Sch\"{o}nberner, D., Jacob, R., Steffen, M., et al.
		 %Perinotto, M.; Corradi, R. L. M.; Acker, A.
		 2005, \aap, 431, 963

 \bibitem[Schwarz \& Monteiro(2006)]{sm06}
   Schwarz, H.\ E., \& Monteiro, H.
   2006, \apj, 648, 430

 \bibitem[Shaw \& Dufour(1995)]{nebular}
   Shaw, R.\ A., \& Dufour, R.\ J.
   1995, \pasp, 107, 896

 \bibitem[Si\'{o}dmiak \& Tylenda(2001)]{st01}
   Si\'{o}dmiak, N., \& Tylenda, R.
   2001, \aap, 373, 1032

 \bibitem[Smith(2003)]{smith03} 
   Smith, N.
   2003, \mnras, 342, 383

 \bibitem[Speck et al.(2002)]{speck02}
   Speck, A.\ K., Meixner, M., Fong, D., et al.
%		 McCullough, P.\ R., Moser, D.\
%		 E., \& Ueta, T. 
   2002, \aj, 123, 346

 \bibitem[Su et al.(2007)]{su07}
   Su, K.\ Y.\ L., Chu, Y.-H., Rieke, G.\ H., et al.
%		 Huggins, P.\ J., 
%   Gruendl, R., Napiwotzki, R., Rauch, T., Latter, W.\ B.,
%   \& Volk, K.
   2007, \aj, 657, L41

 \bibitem[Tielens(2010)]{tielens10}
   Tielens, A.\ G.\ G.\ M.
   2010, The Physics and Chemistry of the Interstellar Medium
  (Cambridge: CUP)

 \bibitem[Tsamis et al.(2004)]{tsamis04}
   Tsamis, Y.\ G., Barlow, M.\ J., Liu, X.-W., et al.
%		 Storey, P.\ J., \& 
%		 Danziger, I.\ J.
   2004, \mnras, 353, 953

 \bibitem[Ueta(2006)]{ueta06}
   Ueta, T. 2006, \apj, 650, 228

 \bibitem[Vamvatira-Nakou et al.(2013)]{vn13}
   Vamvatira-Nakou, C., Hutsem\'{e}kers, D., Royer, P., et al.
%		 Naz\'{e}, Y.,
%	   Magain, P., Exter, K., Waelkens, C., \&
%	   Groenewegen, M.\ A.\ T.
	   2013, \aap, 557, A20

 \bibitem[van Hoof et al.(2010)]{vanhoof10}
   van Hoof, P.\ A.\ M., van de Steene, G.\ C., Barlow, M.\ J., et al.
%		 Exter,
%   K.\ M., Sibthorpe, B., Ueta, T., Peris, V., Groenewegen,
%   M.\ A.\ T., Blommaert, J.\ A.\ D.\ L., Cohen, M., De Meester,
%   W., Ferland, G.\ J., Gear, W.\ K., Gomez, H.\ L., Hargrave,
%   P.\ C., Huygen, E., Ivison, R.\ J., Jean, C., Leeks,
%   S.\ J., Lim, T.\ L., Olofsson, G., Polehampton, E.\ T.,
%   Regibo, S., Royer, P., Swinyard, B.\ M., Vandenbussche,
%   B., van Winckel, H., Waelkens, C., Walker, H.\ J.,
%   \& Wesson, R.
   2010, \aap, 518, L137
   
 \bibitem[van Hoof et al.(2013)]{vanhoof13}
   van Hoof, P.\ A.\ M., van de Steene, G.\ C., Exter, K.\ M., et al.
%   Barlow, M.\ J., Ueta, T., Groenewegen, M.\ A.\ T., Gear, W.\ K., 
%   Gomez, H.\ L., Hargrave, P.\ C., Ivison, R.\ J., Leeks, S.\ J., 
%   Lim, T.\ L., Olofsson, G., Polehampton, E.\ T.,
%   Swinyard, B.\ M., van Winckel, H., Waelkens, C., \& Wesson, R.
   2013, \aap, 560, 7

 \bibitem[Vassiliadis \& Wood(1994)]{vw94} 
   Vassiliadis, E., \& Wood, P.\ R. 
   1994, \apjs, 92, 125 

 \bibitem[Villaver, Manchado, \& Garc\'{\i}a-Segura(2002)]{villaver02}
   Villaver, E., Manchado, A., \& Garc\'{\i}a-Segura, G.
   2002, \apj, 581, 1204	

% \bibitem[Villaver et al.(2003)]{villaver03}
%   Villaver, E., Garc\'{\i}a-Segura, G., \& Manchado, A.
%   2003, \apj, 585, L49 
      
% \bibitem[Villaver et al.(2012)]{villaver12}
%   Villaver, E., Manchado, A., \& Garc{\i}a-Segura, G.
%   2012, \apj, 748, 94

 \bibitem[Volk \& Kwok(1988)]{vk88}
   Volk, K., \& Kwok, S.
   1988, \apj, 331, 435

% \bibitem[Wareing et al.(2007)]{wareing07}
%   Wareing, C.\ J., Zijlstra, A.\ A., \& O'Brien, T.\ J.
%   2007, \mnras, 382, 1233

 \bibitem[Wareing et al.(2006)]{wareing06} 
   Wareing, C.\ J., O'Brien, T.\ J., Zijlstra, A.\ A., et al.
%   Kwitter, K.\ B., Irwin, J., Wright, N., Greimel, R.,
%   \& Drew, J.\ E.
   2006, \mnras, 366, 387

 \bibitem[Weinberger(1989)]{weinberger98}
   Weinberger, R.
   1989, \aaps, 78, 301

 \bibitem[Weisskopf et al.(2002)]{cxc}
   Weisskopf, M.\ C., Brinkman, B., Canizares, C., et al.
%   Garmire, G., Murray, S., \& Van Speybroeck, L.\ P.
   2002,\pasp. 114, 1
  
 \bibitem[Werner et al.(2004)]{werner04}
   Werner, M.\ W., Roellig, T.\ L., Low, F.\ J., et al.
%		 Rieke, G.\ H., Rieke, M.,
%   Hoffmann, W.\ F., Young, E., Houck, J.\ R., Brandl, B., Fazio,
%   G.\ G., Hora, J.\ L., Gehrz, R.\ D., Helou, G., Soifer, B.\ T.,
%   Stauffer, J., Keene, J., Eisenhardt, P., Gallagher, D., Gautier,
%   T.\ N., Irace, W., Lawrence, C.\ R., Simmons, L., Van Cleve,
%   J.\ E., Jura, M., Wright, E.\ L., \& Cruikshank, D.\ P.
   2004, \apjs, 154, 1

 \bibitem[Zhang et al.(2012)]{zhang12}
   Zhang, Y,\ Hsia, C.-H., \& Kwok, S.
   2012, \apj, 755, 53 

 \bibitem[Zijlstra, Pottasch, \& Bignell(1989)]{z89}
   Zijlstra, A.\ A., Pottasch, S.\ R., \& Bignell, C.
   1989, \aaps, 70, 329

 \bibitem[Zuckerman et al.(1990)]{z90}   
   Zuckerman, B, Kastner, J.\ H., Balick, B., \& Gatley, I.
   1990, \apjl, 356, L59
\end{thebibliography}
\end{document}